\documentclass[10pt]{IEEEtran}
\IEEEoverridecommandlockouts
\usepackage{cite}
\usepackage{amsmath,amssymb,amsfonts,amsthm}
\usepackage{algorithm}
\usepackage{algorithmic}
\usepackage{graphicx}
\usepackage{textcomp}
\usepackage[dvipsnames]{xcolor}
\usepackage{bbm}
\usepackage{nccmath}
\usepackage{booktabs}
\usepackage{mathtools}
\usepackage{subcaption}

\def\BibTeX{{\rm B\kern-.05em{\sc i\kern-.025em b}\kern-.08em
    T\kern-.1667em\lower.7ex\hbox{E}\kern-.125emX}}

\usepackage{tikz}
\usetikzlibrary{shapes.arrows}
\usetikzlibrary{plotmarks}
\usetikzlibrary{patterns}
\usetikzlibrary{arrows.meta}
\usetikzlibrary{positioning}
\usetikzlibrary{spy}

\usepackage{pgfplots}
\usepgfplotslibrary{colorbrewer}
\usepgfplotslibrary{fillbetween}
\usepackage{adjustbox}

\usepackage{gincltex}

\pgfplotsset{compat=1.15}
\tikzset{>=latex,  every pin edge/.append style={<-, very thin, solid, black, opacity=1}
    }

\usepackage[acronym]{glossaries}

\newacronym{aoi}{AoI}{Age of Information}
\newacronym{paoi}{PAoI}{Peak Age of Information}
\newacronym{sic}{SIC}{successive interference cancellation}
\newacronym{bs}{BS}{base station}
\newacronym{pmf}{pmf}{probability mass function}
\newacronym{cdf}{CDF}{Cumulative Distribution Function}
\newacronym{irsa}{IRSA}{irregular repetition slotted ALOHA}
\newacronym{sinr}{SINR}{signal-to-interference-plus-noise ratio}
\newacronym{snr}{SNR}{signal-to-noise ratio}
\newacronym{kpi}{KPI}{Key Performance Indicator}
\newacronym{plr}{PLR}{packet loss rate}
\newacronym{csi}{CSI}{channel state information}
\newacronym{ofdma}{OFDMA}{Orthogonal Frequency-Division Multiple Access}
\newacronym{oma}{OMA}{Orthogonal Multiple Access}
\newacronym{noma}{NOMA}{Non-Orthogonal Multiple Access}
\newacronym{pnoma}{PNOMA}{Partial Non-Orthogonal Multiple Access}
\newacronym{rlnc}{RLNC}{Random Linear Network Coding}
\newacronym{mds}{MDS}{Maximum Distance Separable}
\newacronym{urllc}{URLLC}{ultra-reliable low-latency communications}
\newacronym{embb}{eMBB}{enhanced mobile broadband}
\newacronym{mmtc}{mMTC}{massive machine-type communications}
\newacronym{bec}{BEC}{binary erasure channel}
\newacronym{ran}{RAN}{radio access network}
\newacronym{csma}{CSMA}{Carrier Sense Multiple Access}
\newacronym{awgn}{AWGN}{additive white Gaussian noise}
\newacronym{tdma}{TDMA}{time division multiple access}
\newacronym{cdma}{CDMA}{code division multiple access}
\newacronym{lr}{LR}{latency-reliability}
\newacronym{rv}{RV}{random variable}
\newacronym{nr}{NR}{New Radio}
\newacronym{fdma}{FDMA}{frequency division multiple access}

\def \fwidth{0.9\linewidth}
\def \fheight {0.55\linewidth}

\definecolor{color2}{HTML}{7BCCC4}
\definecolor{color4}{HTML}{43A2CA}
\definecolor{color6}{HTML}{0868AC}
\definecolor{color3}{HTML}{DF65B0}
\definecolor{color5}{HTML}{DD1C77}
\definecolor{color7}{HTML}{980043}
\definecolor{color0}{HTML}{FD8D3C}
\definecolor{color1}{HTML}{F03B20}
\definecolor{color8}{HTML}{BD0026}

\theoremstyle{definition}
\newtheorem{definition}{Definition}

\newlength{\hatchspread}
\newlength{\hatchthickness}
\newlength{\hatchshift}
\newcommand{\hatchcolor}{}
\tikzset{hatchspread/.code={\setlength{\hatchspread}{#1}},
         hatchthickness/.code={\setlength{\hatchthickness}{#1}},
         hatchshift/.code={\setlength{\hatchshift}{#1}},
         hatchcolor/.code={\renewcommand{\hatchcolor}{#1}}}
\tikzset{hatchspread=3pt,
         hatchthickness=0.4pt,
         hatchshift=0pt,
         hatchcolor=black}
\pgfdeclarepatternformonly[\hatchspread,\hatchthickness,\hatchshift,\hatchcolor]
   {custom north west lines}
   {\pgfqpoint{\dimexpr-2\hatchthickness}{\dimexpr-2\hatchthickness}}
   {\pgfqpoint{\dimexpr\hatchspread+2\hatchthickness}{\dimexpr\hatchspread+2\hatchthickness}}
   {\pgfqpoint{\dimexpr\hatchspread}{\dimexpr\hatchspread}}
   {
    \pgfsetlinewidth{\hatchthickness}
    \pgfpathmoveto{\pgfqpoint{0pt}{\dimexpr\hatchspread+\hatchshift}}
    \pgfpathlineto{\pgfqpoint{\dimexpr\hatchspread+0.15pt+\hatchshift}{-0.15pt}}
    \ifdim \hatchshift > 0pt
      \pgfpathmoveto{\pgfqpoint{0pt}{\hatchshift}}
      \pgfpathlineto{\pgfqpoint{\dimexpr0.15pt+\hatchshift}{-0.15pt}}
    \fi
    \pgfsetstrokecolor{\hatchcolor}
    \pgfusepath{stroke}
   }
\title{RAN Slicing Performance Trade-offs: Timing versus Throughput Requirements}

\author{Federico Chiariotti,~\IEEEmembership{Member,~IEEE,} Israel Leyva-Mayorga,~\IEEEmembership{Member,~IEEE,} \v Cedomir Stefanovi\' c,~\IEEEmembership{Senior Member,~IEEE,} Anders E. Kal{\o}r,~\IEEEmembership{Student Member,~IEEE,}  and Petar Popovski,~\IEEEmembership{Fellow,~IEEE}%
\thanks{The authors are with the Department of Electronic Systems, Aalborg University, 9220 Aalborg, Denmark (e-mail: fchi@es.aau.dk; ilm@es.aau.dk; cs@es.aau.dk; aek@es.aau.dk; petarp@es.aau.dk).}%
}

\begin{document}

\maketitle
\thispagestyle{plain}
\pagestyle{plain}
\begin{abstract}
The coexistence of diverse services with heterogeneous requirements is a fundamental feature of 5G. This necessitates efficient \emph{\gls{ran} slicing}, defined as sharing of the wireless resources among diverse services while guaranteeing their respective throughput, timing, and/or reliability requirements.
In this paper, we investigate \gls{ran} slicing for an uplink scenario in the form of multiple access schemes for two user types: (1) broadband users with throughput requirements and (2) intermittently active users with timing requirements, expressed as either \gls{lr} or \gls{paoi}.
Broadband users transmit data continuously, hence, are allocated non-overlapping parts of the spectrum. We evaluate the trade-offs between the achievable throughput of a broadband user and the timing requirements of an intermittent user under \gls{oma} and \gls{noma}, considering capture. Our analysis shows that \gls{noma}, in combination with packet-level coding, is a superior strategy in most cases for both \gls{lr} and \gls{paoi}, achieving a similar \gls{lr} with only slight $2$\% decrease in throughput with respect to the upper bound in performance. However, there are extreme cases where \gls{oma} achieves a slightly greater throughput than \gls{noma} at the expense of an increased \gls{paoi}.
\end{abstract}
\glsresetall


\section{Introduction}
The fifth generation of mobile networks (5G) aims to support three main service categories
: \gls{embb}, \gls{urllc}, and \gls{mmtc}~\cite{3GPPTR38913}.
\gls{embb} is the direct evolution of the 4G mobile broadband service with higher data rates, along with greater spectral and spatial efficiency. Even though \gls{embb} use cases mainly occur in the downlink (e.g., video streaming or file download) novel \gls{embb} use cases in the uplink are quickly gaining relevance. These include remote driving and live-video streaming in, for example, tactile internet applications and sport and cultural events.
\gls{urllc} services, on the other hand, usually involve exchange of small amounts of data, but require latency in the order of a few milliseconds and high reliability guarantees, e.g., a packet loss ratio below $10^{-5}$.
Finally, \gls{mmtc} also involve transmissions of small amounts of data per device, but consist of hundreds or thousands of devices in the service area.
The main challenge in \gls{mmtc} is to design access networking mechanisms that maximize the success probability while maintaining an adequate \emph{timing} in data delivery and resource efficiency.

The main strategy for service co-existence adopted by 3GPP~\cite{rel15,rel16} is \emph{network slicing}, which refers to the allocation of subsets of the network resources to the active services.
The idea is to provide performance guarantees by limiting the mutual impact among services and/or service categories
~\cite{rost2017network}.
Arguably, in the context of the \gls{ran} the most natural form of slicing is \gls{fdma}, which provides a high degree of isolation between services in different frequency bands. However, while \gls{fdma} slicing works well for high throughput services that are constantly demanding resources, it yields low utilization for intermittent services that are active only a fraction of the time. This motivates the search for alternative slicing schemes that are more suitable for heterogeneous services types.

In general, \gls{ran} slicing can be implemented in the form of \gls{oma} and \gls{noma}. \gls{oma} schemes, such as \gls{fdma}, \gls{tdma}, and \gls{cdma}, have been extensively studied and implemented in commercial and cellular systems.
Moreover, \gls{oma} seems to be the approach preferred by 3GPP for 5G and beyond 5G systems, contextualized in the concept of bandwidth part~\cite{rel15}.
On the other hand, in \gls{noma} the same time-frequency resources are assigned to multiple services or users. This allows, for example, to increase the number of served users with the available resources and/or the spectral efficiency of the system~\cite{Vaezi2019}. To enable communication in shared time-frequency resources, \gls{noma} is usually accompanied by multi-user detection techniques, like separation of the users in the code domain, or in the power domain accompanied with \gls{sic} where the individual signals are in turn decoded and subtracted from the received composite signal~\cite{Song2017, Islam2017, Vaezi2019, Dai2018}.

\begin{figure}[t!]
	\centering
	\input{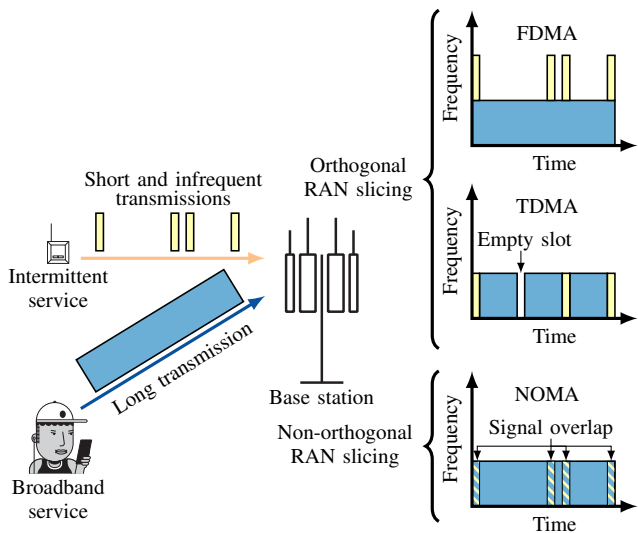}
 \caption{Orthogonal (i.e., \gls{fdma} and \gls{tdma}) and non-orthogonal \gls{ran} slicing (i.e., \gls{noma}) between broadband and intermittent services.}
 \label{fig:access}
\end{figure}

The difference between \gls{oma} and \gls{noma} slicing is illustrated in Fig.~\ref{fig:access}. Here, it can be seen that \gls{tdma} and \gls{noma} achieve higher resource utilization than \gls{fdma} when the intermittent service transmits infrequently, while the difference will be less pronounced when the intermittent service transmits frequently. In the latter case, \gls{noma} may even be inefficient due to the high amount of interference introduced to the broadband service. Further, \gls{tdma} faces the challenge of carefully dimensioning the allocated resources to the intermittent service, which is essential to achieve an acceptably low amount of idle resources (i.e., empty slots) and waiting time for transmission.

Despite this apparent trade-off between efficiency and timeliness of the slicing schemes, only a few studies have compared the performance between \gls{tdma} and \gls{noma} with heterogeneous services~\cite{Richart2016, Popovski2018, chiariotti2021non}. On the other hand, \gls{oma} and \gls{noma} slicing has been widely studied in the presence of multiple users of the same service type~\cite{Maatouk2019, Dai2015, Wu2018, Song2017}. For instance, the trade-offs in achievable data rates for \gls{embb} services are 
characterized in an \gls{awgn} channel with \gls{oma} and power-domain \gls{noma}~\cite{Dai2015, Wu2018}. In our previous work, we derived the performance trade-offs with heterogeneous services with \gls{tdma} and \gls{noma} with packet-level coding in a simplified collision channel model, which provides conservative results for \gls{noma}~\cite{Chiariotti2021, chiariotti2021non}. Some results with capture, obtained by simulation, were provided in \cite{chiariotti2021non}, which served as one of the main motivations for this study, as these illustrated the potential gains of \gls{noma}.
The aim of this work is to provide an extensive and exact analytical treatment of \gls{oma} and \gls{noma} slicing in an uplink scenario with two different service types: broadband and intermittent users with throughput and timing requirements, respectively. Because the trade-off between \gls{tdma} and \gls{noma} is least apparent, this combination is our primary focus and include \gls{fdma} as benchmark that neglects resource efficiency. Furthermore, to ensure a realistic evaluation, we consider a channel model with capture, which inevitably occurs in practice.


As mentioned, we assume that broadband users transmit data continuously and are primarily interested in achieving a high throughput.
In contrast, intermittent users transmit short packets sporadically and are primarily interested in the \emph{timeliness} of their data, according to the underlying application.
To consider packet-level and flow-level timing requirements, we have selected two different \glspl{kpi}. 
The first \gls{kpi}, which reflects flow-level requirements, is \gls{paoi}, relevant for users that send updates of an ongoing process in which the freshness of information is the most important objective.
\gls{paoi} measures the time elapsed since the generation of the last received update until a new update is received~\cite{Costa2014}, and it is therefore determined by the transmission latency, reliability and the update generation pattern. 
\gls{paoi}-focused applications can tolerate individual packet losses, as there are no strict reliability requirements and new updates can supersede old ones.
The second \gls{kpi}, which reflects packet-level requirements, is denoted by \emph{\gls{lr}} and
captures the probability of delivering individual packets within a given latency threshold\cite{Nielsen2018}. For this, we use the distribution of latency where lost packets are defined to have infinite latency. \Gls{lr} captures, for example, \gls{urllc} traffic with strict constraints on the reliability of communication within a maximum latency.
Our specific focus will be on computing high percentiles of \gls{lr} and \gls{paoi}, which can be used to design systems with probabilistic reliability guarantees.

\begin{figure}[t]
\centering
\begin{tikzpicture}[>=latex, scale=1,font=\small]

\pgfmathsetmacro{\pone}{0.3}
\pgfmathsetmacro{\ptwo}{0.7}
\pgfmathsetmacro{\pthree}{1.2}
\pgfmathsetmacro{\pfour}{2.7}
\pgfmathsetmacro{\pw}{0.1}
\pgfmathsetmacro{\ph}{0.2}
\pgfmathsetmacro{\paone}{0.3}
\pgfmathsetmacro{\patwo}{0.3}
\pgfmathsetmacro{\pathree}{0.4}
\pgfmathsetmacro{\pafour}{0.5}

\filldraw[semithick,fill=Set3-A] (\pone-\pw,0) rectangle (\pone,\ph);
\filldraw[semithick,fill=Reds-K] (\ptwo-\pw,0) rectangle (\ptwo,\ph);
\filldraw[semithick,fill=Set3-A] (\pthree-\pw,0) rectangle (\pthree,\ph);
\filldraw[semithick,fill=Set3-A] (\pfour-\pw,0) rectangle (\pfour,\ph);

\draw[->](\pone-\paone,0.3)--++(0,-0.3);
\draw[->](\ptwo-\patwo,0.3)--++(0,-0.3);
\draw[->](\pthree-\pathree,0.3)--++(0,-0.3);
\draw[->](\pfour-\pafour,0.3)--++(0,-0.3);

\draw[densely dashed, semithick, gray](\pone-\paone,0)--++(\paone,\paone);
\draw[densely dashed, semithick, gray](\ptwo-\patwo,0)--++(\pthree-\ptwo+\patwo,\pthree+\patwo-\ptwo);
\draw[densely dashed, semithick, gray](\pthree-\pathree,0)--++(\pathree,\pathree);
\draw[densely dashed, semithick, gray](\pfour-\pafour,0)--++(\pafour,\pafour);

\draw[thick,->] (0,0)node[anchor=east, inner sep=1pt]{$\dotsc$}--(\pfour+0.5,0)node[below]{Time}; 

\draw[thick, PuBu-J] (0,0.5)--(\pone,\pone+0.5)--(\pone,\paone)--++(\pthree-\pone,\pthree-\pone)--(\pthree,\pathree)--++(\pfour-\pthree,\pfour-\pthree)--(\pfour,\pafour)--++(0.2,0.2)node[right, inner sep=1pt, yshift=2pt]{AoI};

\draw[|-|] (\pone-\paone,-0.1)--++(\paone,0)node[below, midway, ]{$\ell_1$};
\draw[|-|] (\pthree-\pathree,-0.1)--++(\pathree,0)node[below, midway, ]{$\ell_3$};
\draw[|-|] (\pfour-\pafour,-0.1)--++(\pafour,0)node[below, midway, ]{$\ell_4$};


\pgfmathsetmacro{\ly}{2.5}
\pgfmathsetmacro{\ldx}{1.4}
\pgfmathsetmacro{\xd}{-0.3}

\draw[->](\xd,\ly)--++(0.3,0)node[anchor=west,]{Arrival};
\filldraw[semithick,fill=Set3-A] (\ldx+\xd,\ly-0.5*\ph) rectangle (\ldx+\pw+\xd,\ly+0.5*\ph);
\node[anchor=west, align=center] at (\ldx+\pw+\xd,\ly){Success};
\filldraw[semithick,fill=Reds-K] (2*\ldx+\xd,\ly-0.5*\ph) rectangle (2*\ldx+\pw+\xd,\ly+0.5*\ph);
\node[anchor=west, align=center] at (2*\ldx+\pw+\xd,\ly){Error};

\draw[semithick] (\xd-0.1,\ly-0.2) rectangle (\pfour+0.85,\ly+0.2);

\begin{scope}[xshift=4.5cm,scale=0.666]
\pgfmathsetmacro{\xmx}{0.7}
\begin{axis}[
scale=1.5,
axis lines*=left,
height=3.5cm,
width=3.5cm,
ymin=0,
ymax=1,
ytick={0,0.25,0.5,0.75,1},
yticklabels={$0$,$1/4$,$1/2$,$3/4$,$1$},
xmin=0,
xmax=\xmx,
xtick={0,\paone,\pathree,\pafour,\xmx},
xticklabels={$0$,$\ell_1$,$\ell_3$,$\ell_4$, Time},
ylabel={Latency-reliability (LR)},
ylabel shift = -5 pt,
clip mode=individual,
ymajorgrids
]
\addplot[PuBu-J, const plot, thick] coordinates{(0,0)(\paone,0.25)(\pathree,0.5)(\pafour,0.75)(\xmx+1,0.75)};
\draw[|-|](\xmx+0.02,0.75)--(\xmx+0.02,1)node[midway,right, align=center]{Error\\[-0.4em]probability};
\end{axis}
\end{scope}
\end{tikzpicture}
\caption{Exemplary diagram of the \gls{aoi} and latency-reliability \glspl{kpi} in a period with four packet transmissions. The latency $\ell_i$ for packets transmitted with errors is set to $\infty$.}\label{fig:aoi_vs_li_diag}
\end{figure}
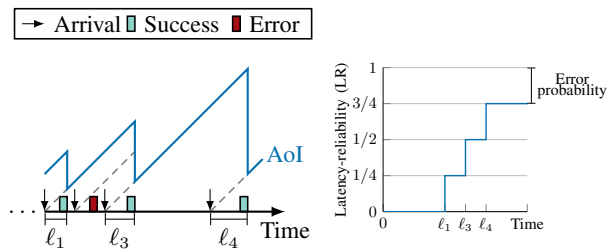

The scenario assumed in this paper comprises a single frequency band (i.e., bandwidth part in 5G \gls{nr} terminology) sliced in time to accommodate one broadband and one intermittent user.\footnote{The scenario is inspired by the latest non-orthogonal multiplexing approaches in the uplink studied by the 3GPP~\cite{3GPPTR38812}.}.
We study \gls{oma}, focusing on \gls{tdma} but also including \gls{fdma}, and \gls{noma} with \gls{sic}.
In the latter case, \gls{sic} can be applied (1) in conjunction with the capture effect, such that the colliding packets are immediately resolved, and (2) coupled with the packet-level coding, such that after decoding of the broadband user block, the interference is removed and past packets from the intermittent user can be recovered. 
We analyze the achievable performance and the inherent trade-offs, providing closed-form expressions for throughput of the broadband user and timing of the intermittent user.
The derivations are contextualized for a simple fading-based channel model, however, the elaborated approach is general and easily transferable to other settings.

In particular, the main contributions of this paper are the following.
\begin{itemize}
    \item We provide exact expressions that allowed us to analyze the operating regions and trade-offs with \gls{oma} and \gls{noma} with a realistic channel model that includes the possibility of capture. The results with this model show fundamental differences when compared to the analyses carried out in our previous work and confirm the trends observed through simulations~\cite{Chiariotti2021,chiariotti2021non}. 
    \item We analyze the impact of the wireless conditions of the intermittent user, including distance from the \gls{bs} and path loss, in the performance of \gls{noma}.
    \item We provide design guidelines for selecting the multiple access scheme and its parameters, depending on:
    \begin{enumerate}
        \item The requirements and features of the different types of services in the system.
        \item The available bandwidth.
        \item The wireless conditions of the intermittent user.
    \end{enumerate}
\end{itemize} 

We observe that, while \gls{fdma} provides the upper bound in performance, \gls{noma} schemes offer significant benefits w.r.t. \gls{oma} when the target \gls{kpi} for the intermittent user is \gls{lr}. Specifically, \gls{noma} can achieve similar performance trade-offs as \gls{fdma} but with a much higher resource utilization.
On the other hand, the potential gains of \gls{noma} w.r.t. \gls{tdma} decrease when the target \gls{kpi} is \gls{paoi}, with \gls{tdma} outperforming \gls{noma} in extreme cases where throughput is maximized in exchange for a longer \gls{paoi}.


The rest of the paper is organized as follows.
Sec.~\ref{sec:related} presents the related work on \gls{aoi} and slicing-based access for heterogeneous user classes.
The system model and \glspl{kpi} are specified in Sec.~\ref{sec:system_model}. 
We then derive the analytical distributions of those metrics for \gls{oma} and \gls{noma} in Sec.~\ref{sec:oma} and Sec.~\ref{sec:noma}, respectively.
Sec.~\ref{sec:results} presents simulation results and discussion of the performance of the different access schemes.
Finally, Sec.~\ref{sec:concl} concludes the paper. 

\section{Related work}\label{sec:related}

Non-orthogonal slicing, in the form of \gls{noma}, offers the possibility of increasing the spectral efficiency and the number of supported users with respect to \gls{oma} in exchange for a greater decoding complexity at the receiver to perform \gls{sic}~\cite{Vaezi2019, Liu2017} or other multi-user detection techniques. Hence, \gls{noma} has been widely studied in the literature in systems with a single service type~\cite{Maatouk2019, Dai2015, Wu2018, Song2017, Vaezi2019}. \gls{noma} often assumes user separation in the power domain such that the benefits of \gls{sic} can be fully exploited.
However, different performance gains have been observed for \gls{noma} in the uplink and in the downlink.
In particular, the effect of power control in the uplink can be eclipsed by the channel conditions of the users in combination with imperfect channel state information~\cite{Liu2017}.

A particularly interesting approach towards heterogeneous service coexistence with \gls{noma} is presented in~\cite{Song2017}, which emphasizes the importance of power control in \gls{noma} and formulates resource allocation as a non-cooperative game and as a matching problem. While not addressing them directly, this non-cooperative game may be able to handle heterogeneous service types, as each user defines and attempts to maximize its own utility.

One of the first studies that addresses the coexistence of heterogeneous services in \gls{oma} and \gls{noma} was presented in~\cite{Popovski2018}, considering different combinations of 5G services in an uplink scenario. Specifically, \gls{embb} users are allocated orthogonal resources between them; these coexist with either one \gls{urllc} user or with  \gls{mmtc} traffic, which follows a Poisson distribution.
It was observed that \gls{noma} may offer benefits with respect to \gls{oma} depending on the rate of the \gls{embb} users and on the type of coexisting traffic. Specifically, these benefits were evaluated in terms of the achievable rates for \gls{embb} and \gls{urllc} traffic and the achievable \gls{embb} rates as a function of the arrival rate of \gls{mmtc} packets.
This work was later extended to a multi-cell scenario with strict latency guarantees for \gls{urllc} traffic~\cite{Kassab2019}, where it was observed that \gls{noma} leads to a greater spectral efficiency w.r.t. \gls{oma}. This same conclusion was drawn by Maatouk et al.~\cite{Maatouk2019} in an uplink scenario with two users with and one service type. The aim of the latter study was to minimize the average \gls{aoi}, however, it was also observed that a greater spectral efficiency does not directly translate in a lower average \gls{aoi}.

\gls{aoi} is a relatively new performance metric, but it has been rapidly adopted due to its relevance in remote control tasks~\cite{Kaul2012}. Most papers in the literature have examined it in the context of queuing theory, often in ideal systems with Markovian service~\cite{kosta2019age}, because of the relative simplicity of the analysis, but a few have considered the effect of physical layer issues and medium access schemes on it. Recent works compute the average \gls{aoi} in \gls{csma}~\cite{maatouk2020age}, ALOHA~\cite{yates2020age} and slotted ALOHA~\cite{chen2020age} networks, considering the impact of the different medium access policies on the age.

Another important missing piece in the \gls{aoi} literature is the worst-case performance analysis: while studies on average \gls{aoi} are common, the tail of its distribution is rarely considered~\cite{yates2020agenet}, limiting the relevance of the existing body of work for reliability-oriented applications. The analytical complexity of deriving the complete distribution of the age is a daunting obstacle; only recently, advances have been made in this line. A recent work~\cite{champati2019statistical} uses the Chernoff bound to derive an upper bound of the quantile function of the \gls{aoi} for two queues in tandem with deterministic arrivals. Using a more analytical approach, the \gls{paoi} distribution was computed over a single-hop link with fading and retransmissions in~\cite{devassy2019reliable}.
We also mention the work in~\cite{Yates2019isit}, where different service classes are defined and the system is modeled as an M/G/1/1 clocking queue with hyperexponential service time. However, in the latter, only the service rate is different among classes. Then, the classes can adapt the arrival rate to minimize the \gls{aoi}.
finally, for a more detailed overview of the literature on \gls{aoi}, we refer the reader to~\cite{yates2020agesurvey}.

\section{System model}
\label{sec:system_model}

We consider an uplink scenario with a set of users $\mathcal{U}$ transmitting data to a \gls{bs} through an \gls{ofdma} system whose time-frequency resources are divided into time slots and bandwidth parts as in 5G \gls{nr}~\cite{3GPP38211}. 
A bandwidth part is defined as a set of contiguous resource blocks in the frequency domain. 
We consider the case where the users transmit up to one packet per time slot, occupying the whole bandwidth part.
Herein, we consider the case where two heterogeneous users must be allocated resources. The options for the \gls{bs} are 1) allocate the users in the same bandwidth part and define how the resources should be shared among them or 2) allocate a different bandwidth part for each of the users using \gls{fdma}.

User 1 is a broadband user following the \gls{embb} model~\cite{3GPPTR38913}: it is a full-buffer user, that is, it always has data to transmit, and maintains an infinite transmission queue. To counteract potential packet losses due to fading and noise, the broadband user implements a packet-level coding scheme, where blocks of $K$ (source) packets are encoded to generate a \emph{frame} of $N$ coded packets.
The coded packets are transmitted in the same bandwidth part, one after the other, and have a zero probability of linear dependence, which can be achieved, for example, with \gls{mds} codes. Hence, the block of packets is decoded when any $K$ packets from the same frame are received without errors. 

User 2 is an intermittent user that, with (a relatively low) probability $\alpha$, may generate a short packet at each slot. Each generated packet is transmitted just once at the next available time slot.
User 2 maintains up to $Q$ of the generated packets in a transmission queue.
We denote by $q_t\leq Q$ the length of the queue at time slot $t$.
If a new packet is generated when $q_t=Q$, user 2 discards the oldest buffered packet and adds the newly generated one at the end of the queue. 

When both users are allocated to the same bandwidth part, the \gls{bs} must allocate the time slots that are available for the transmission of each of the two users.
For this, we define the resource allocation set $\mathcal{A}_{t}\subseteq\mathcal{U}$ as the subset of users that can access the bandwidth part at slot $t$.
We define the following three types of slot allocations.
\begin{enumerate}
\item\emph{Broadband:} The slot is reserved for the broadband user. Hence, $\mathcal{A}_{t}=1$. 

    \item\emph{Intermittent:} The slot is reserved for the intermittent user and may use it if it has one or more packets in its queue. Hence, $\mathcal{A}_{t}=2$. 
    
    \item\emph{Mixed:} Both users are allowed to access the slot, implying that the signals will overlap if the intermittent user transmits. Hence, $\mathcal{A}_{t}=\{1,2\}$.
\end{enumerate}
Building on these, we define the following different access schemes. 
\begin{enumerate}
\item  \emph{\gls{tdma}}: Both users are allocated resources in a single bandwidth part with separate broadband and intermittent slots. Specifically, the intermittent user has a reserved intermittent slot once every $T_\text{int}$ slots, while the rest of the slots are reserved for the broadband user. As such, this is a non-orthogonal slicing in the frequency domain but orthogonal in the time domain where there is no interference among the users.

\item \emph{\Gls{noma}}: Both users are allocated resources in a single bandwidth part with only mixed slots. Hence, the intermittent user may transmit at any slot. The two users interfere any time the intermittent user transmits, but the packets can be recovered through \gls{sic} by decoding one of the signals immediately or at a later time slot. More details on the operation of \gls{sic} are given in Section~\ref{sec:channel}.

\item \emph{\Gls{fdma}}: The users are allocated resources in different and non-overlapping bandwidth parts. Hence, one of the bandwidth parts contains only broadband slots and the other bandwidth part contains only intermittent slots.  
\end{enumerate}

\begin{figure}[t!]
	\centering
	\resizebox{0.9\columnwidth}{!}{
        \includegraphics{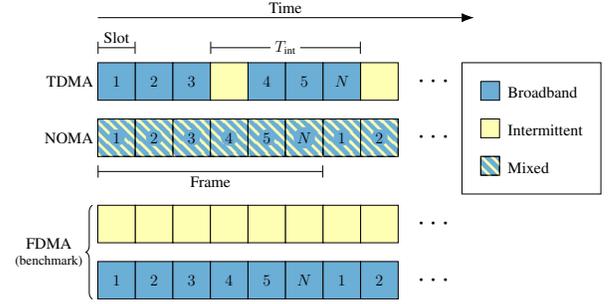}}	
 \caption{Frame structure for the \gls{tdma}, \gls{noma}, and \gls{fdma} schemes with $K=4$ and $N=6$.}
 \label{fig_system}
\end{figure}

The frame structures for these access schemes are illustrated in Fig.~\ref{fig_system}. Naturally, \gls{fdma} can only take place when there are two bandwidth parts available.
Needles to say, the bandwidth part allocated to the intermittent user with \gls{fdma} is likely to be under-utilized, as $\alpha$ is relatively small. Hence, this approach results in a low resource efficiency, and is used here only as a benchmark scheme in which the performance of the users is fully independent of each other. 

\subsection{Channel model}
\label{sec:channel}

We consider a block fading channel, where the received signal by the \gls{bs} at time slot $t$ is given as
\begin{equation}
    y_{t}=\sum_{u\in\{1,2\}} h_{u,t} a_{u,t} x_{u,t} +z_{t}
\end{equation}
where $h_{u,t}\in\mathbb{C}$ is the random fading coefficient for user $u$ and $z_{t}$ is the circularly-symmetric Gaussian noise with power $\sigma^2$. The variable $a_{u,t}$ is an activity indicator, equal to 1 if the user is active in slot $t$ and 0 otherwise. A user is active at time $t$ if and only if $u\in\mathcal{A}_t$ and if its packet queue $q_{u,t}$ is not empty
\begin{equation}
    a_{u,t}=I(u\in\mathcal{A}_t)I(q_{u,t}>0),
\end{equation}
where $I(\cdot)$ is the indicator function, equal to 1 if the condition is true and 0 otherwise. 
Let $P_u\leq P_\text{max}$ be the selected (i.e., fixed) transmission power for user $u$, where $P_\text{max}$ is the maximum transmission power.
The \gls{snr} of user $u$ is given as
\begin{equation}
    \text{SNR}_{u,t}=\frac{|h_{u,t}|^2P_{u}a_{u,t}}{\sigma^2}=\frac{|h'_{u,t}|^2P_{u}a_{u,t}}{\ell_u\sigma^2},
    \label{eq:snr}
\end{equation}
where $\ell_u$ is the constant large-scale fading, including path loss, and $|h'_{u,t}|$ is the envelope of the channel coefficient due to fast fading.
The path loss is a function of the distance of user $u$ to the \gls{bs} $r_u$, the carrier frequency $f_c$, and a path loss exponent $\eta$.
We assume the standard path loss model
\begin{equation}
    \ell_u=\frac{(4\pi f_c)^2r_u^\eta}{c^2},
\end{equation}
where $c$ is the speed of light.

The expected \gls{snr} for a transmission by user $u$ is
\begin{equation}
    \overline{\text{SNR}}_u=\frac{\mathbb{E}\left[|h_{u,t}|^2\right]P_{u}}{\sigma^2}=\frac{\mathbb{E}\left[|h'_{u,t}|^2\right]P_{u}}{\ell_u\sigma^2},
    \label{eq:mean_snr}
\end{equation}

By using the standard assumption of treating the interfering signal as \gls{awgn} noise, the \gls{sinr} for user $u$ in the considered scenario is
\begin{equation}
    \text{SINR}_{u,t}=\frac{|h_{u,t}|^2P_{u}a_{u,t}}{\sigma^2+ |h_{v,t}|^2P_{v}a_{v,t}}=\frac{\text{SNR}_{u,t}}{1+\text{SNR}_{v,t}},\quad \text{s.t. } v\neq u.\label{eq:SINR}
\end{equation}

\subsection{Reception model}

Let $X$ be the \gls{rv} of the number of packets from user 1 that belong to the same block and are received without errors. The success probability of user 1, denoted as $p_{s,1}$, is defined as the probability of receiving $K$ or more packets out of the $N$ that comprise the block. That is,
\begin{equation}
    p_{s,1}\triangleq \Pr\left[X\geq K\middle| N\right].
\end{equation}
We define $\gamma_u$ as the threshold in the \gls{sinr} to decode a packet transmitted by user $u$.
In practice, the threshold is mainly a function of the modulation and coding scheme and the receiver sensitivity.
In the following, we consider the case in which the fading envelope $|h_{u,t}|$ is Rayleigh distributed and define the erasure probabilities for the two users.

\emph{Erasure probability for the broadband user:} The \gls{bs} has collected sufficient \gls{csi} about the broadband user so that the appropriate transmission power $P_1\leq P_\text{max}$, blocklength, and data rate (i.e., modulation and coding) to achieve a target erasure probability $\varepsilon_1$ are signaled back to the broadband user. Therefore, user 1 transmits with power
\begin{equation}
   P_1\leq P_\text{max}:\Pr\left[\text{SNR}_{1,t}<\gamma_1\right]=\varepsilon_1
\end{equation}

\emph{Erasure probability for the intermittent user:} Due to the infrequent transmissions, the \gls{csi} of this user at the \gls{bs} is insufficient to perform a precise selection of parameters as done for the broadband user. Instead, the user always transmits at $P_2=P_\text{max}$ and its erasure probability $\varepsilon_2$ is determined by its path loss $\ell_2$ and by $\gamma_2$. Hence, the erasure probability for user 2 is calculated from~\eqref{eq:mean_snr} as
\begin{equation}
    \varepsilon_2=\Pr\left[\text{SNR}_{2,t}<\gamma_2\right]=1-e^{\frac{-\gamma}{\overline{\text{SNR}_{2}}}}=1-e^{\frac{-\gamma \ell_2\sigma^2}{P_{u}}}
\end{equation}
since $\mathbb{E}\left[|h'_{u,t}|^2\right]=1$ for unitary Rayleigh fading.

Based on these probabilities, we define six different outcomes for the cases where both users transmit in the same time slot. 
These outcomes are ordered pairs $(o_1,o_2)$ where $o_u\in\{\mathcal{I},\mathcal{E},\mathcal{R}\}$ indicates the outcome of user $u$'s signal, described in the following. 
\begin{itemize}
    \item $\mathcal{I}$: Either 1) the signal of interest has sufficient \gls{sinr} to be immediately decoded or 2) the other signal has sufficient \gls{sinr} to be immediately decoded, its interference is removed through \gls{sic}, and the signal of interest has sufficient \gls{snr} and is decoded.
    \item $\mathcal{E}$: The signal has insufficient \gls{snr} to be decoded, even if the interference from the other signal is removed.
    \item $\mathcal{R}$: None of the signals has sufficient \gls{sinr} to be decoded immediately, but the signal of interest has sufficient \gls{snr} to be decoded if interference from the other is removed. 
\end{itemize} 
The probability of each of the outcomes, denoted as $\pi_{o_1 o_2}$, is derived in the following based on the operation of \gls{sic} and under Rayleigh fading.

\begin{itemize}
    \item $\left(\mathcal{I},\mathcal{I}\right)$: The signal with the highest \gls{sinr} is decoded and its interference is immediately removed through \gls{sic}. Then, the second signal is decoded. This occurs with probability
    \begin{IEEEeqnarray*}{rCl}
    \pi_{\mathcal{II}} &=&\Pr\left[\frac{\text{SNR}_{1,t}}{1+\text{SNR}_{2,t}}\geq\gamma_1~\land~ \text{SNR}_{2,t}\geq\gamma_2 \right]\\
    &&+\Pr\left[\frac{\text{SNR}_{2,t}}{1+\text{SNR}_{1,t}}\geq\gamma_2~\land~ \text{SNR}_{1,t}\geq\gamma_1 \right]\IEEEnonumber\\
    &=&\frac{\overline{\text{SNR}}_1}{\overline{\text{SNR}}_1+\gamma_1\overline{\text{SNR}}_2}e^{\frac{-\gamma_1}{\overline{\text{SNR}_1}}}e^{-\gamma_2\left(\frac{\gamma_1}{\overline{\text{SNR}}_1} +\frac{1}{\overline{\text{SNR}}_2} \right)}\\
    &&+
    \frac{\overline{\text{SNR}}_2}{\gamma_2\overline{\text{SNR}}_1+\overline{\text{SNR}}_2}e^{\frac{-\gamma_2}{\overline{\text{SNR}_2}}}e^{-\gamma_1\left(\frac{1}{\overline{\text{SNR}}_1}+\frac{\gamma_2}{\overline{\text{SNR}}_2} \right)}.\IEEEeqnarraynumspace
    \end{IEEEeqnarray*}
    
    \item $\left(\mathcal{I},\mathcal{E}\right)$ and $\left(\mathcal{E},\mathcal{I}\right)$: The signal with the higher \gls{sinr} is decoded and its interference is immediately removed through \gls{sic}. However, the second signal cannot be decoded due to the impact of noise, i.e., a low SNR. These outcomes occur with probabilities
    \begin{IEEEeqnarray}{rCl}
    \pi_{\mathcal{IE}} &=&\Pr\left[\frac{\text{SNR}_{1,t}}{1+\text{SNR}_{2,t}}\geq\gamma_1~\land~ \text{SNR}_{2,t}< \gamma_2 \right]\IEEEnonumber\\
    &=&\frac{\overline{\text{SNR}}_1 e^{\frac{-\gamma_1}{\overline{\text{SNR}_1}}}}{\overline{\text{SNR}}_1+\gamma_1\overline{\text{SNR}}_2}\left(1-e^{-\gamma_2\left(\frac{\gamma_1}{\overline{\text{SNR}}_1} +\frac{1}{\overline{\text{SNR}}_2} \right)}\right).
    \IEEEeqnarraynumspace
\end{IEEEeqnarray}
\begin{IEEEeqnarray}{rCl}
    \pi_{\mathcal{EI}} &=&\Pr\left[\frac{\text{SNR}_{2,t}}{1+\text{SNR}_{1,t}}\geq\gamma_2~\land~ \text{SNR}_{1,t}< \gamma_1 \right]\IEEEnonumber\\
    &=&\frac{\overline{\text{SNR}}_2 e^{\frac{-\gamma_2}{\overline{\text{SNR}_2}}}}{\gamma_2\overline{\text{SNR}}_1+\overline{\text{SNR}}_2}\left(1-e^{-\gamma_1\left(\frac{1}{\overline{\text{SNR}}_1} +\frac{\gamma_2}{\overline{\text{SNR}}_2} \right)}\right).
    \IEEEeqnarraynumspace
\end{IEEEeqnarray}

\item $\left(\mathcal{E},\mathcal{E}\right)$: The \gls{snr} of both signals is insufficient and, thus, neither can be decoded even if the interference from the other user were removed. This occurs with probability
\begin{IEEEeqnarray}{rCl}
    \pi_{\mathcal{EE}} &=&\Pr\left[\text{SNR}_{2,t}<\gamma_2~\land~ \text{SNR}_{1,t}< \gamma_1 \right]\\
    &=& \left(1-e^{\frac{-\gamma_1}{\overline{\text{SNR}}_1}}\right)\left(1-e^{\frac{-\gamma_2}{\overline{\text{SNR}}_2}}\right).
\end{IEEEeqnarray}

\item $\left(\mathcal{R},\mathcal{E}\right)$: The signal from user 2 has insufficient SNR, while the signal from user 2 has a sufficient SNR but insufficient SINR. Since the system cannot remove the interference from user 2 without decoding it first, both packets remain undecoded. This outcome occurs with probability
\begin{IEEEeqnarray*}{rCl}
    \pi_{\mathcal{RE}} &=&\Pr\left[\gamma_1\leq\text{SNR}_{1,t}<\gamma_1\left(1+\text{SNR}_{2,t}\right)\right.\\
    &&~\left.\land~ \text{SNR}_{2,t}< \gamma_2 \right]\\
    &=&\left( \frac{\overline{\text{SNR}}_1}{\overline{\text{SNR}}_1+\gamma_1\overline{\text{SNR}}_2}\left(e^{-\gamma_2\left(\frac{\gamma_1}{\overline{\text{SNR}}_1} +\frac{1}{\overline{\text{SNR}}_2} \right)}-1\right)\right.\\
    &&+\left.\left(1-e^{\frac{-\gamma_2}{\overline{\text{SNR}_2}}}\right)\right)e^{\frac{-\gamma_1}{\overline{\text{SNR}_1}}}\\
    \IEEEeqnarraynumspace
\end{IEEEeqnarray*}

\item $\left(\cdot,\mathcal{R}\right)$: In this case, none of the  signals can be immediately recovered but the signal from user 2 could be decoded if the interference from user 1 is removed via \gls{sic} after decoding the block of user 1. Therefore, this outcome includes the cases $\left(\mathcal{E},\mathcal{R}\right)$ and $\left(\mathcal{R},\mathcal{R}\right)$, and occurs with probability
\begin{IEEEeqnarray}{rCl}
    \pi_{\cdot \mathcal{R}}&=& 1- \pi_{\mathcal{II}}- \pi_{\mathcal{IE}}- \pi_{\mathcal{EI}}- \pi_{\mathcal{EE}}-\pi_{\mathcal{RE}}.
\end{IEEEeqnarray}
\end{itemize}

Note that the cases $\left(\mathcal{I},\mathcal{R}\right)$ and $\left(\mathcal{R},\mathcal{I}\right)$ are not feasible, as outcome $\mathcal{I}$ indicates that a signal is immediately decoded and that its interference to the other signal is removed.




\subsection{Key Performance Indicators}


The broadband user (user 1) is interested on maximizing its throughput $S$ under the constraint that the desired reliability $p_{s,1}$ must be greater than $1-\varepsilon_1$. Note that increasing the reliability of the broadband user entails a reduction in the coding rate $K/N$.

The intermittent user (user 2) is interested on the timeliness of its data, i.e., either \gls{lr} or \gls{paoi}, where we have selected their $90$th percentile as the main \glspl{kpi}. Let $T$ and $\Delta$ be the \glspl{rv} of \gls{lr} and \gls{paoi}, respectively. Then, the $90$th percentile of \gls{lr} is defined as
\begin{equation}
    T_{90}:=\min_{n}\{n : \Pr\left[T\leq n\right]>0.9\}
\end{equation}
and the $90$th percentile of \gls{paoi} $\Delta_{90}$ is defined analogously.
Note that the latter allows us to evaluate the tail distribution of the \gls{paoi} in a general scenario and can be used to compare the performance with different values of $\alpha$~\cite{Devassy2019}.

Since $S$ and the timeliness of the intermittent user are interlinked, we evaluate their trade-offs for a specific activation probability $\alpha$ and erasure probabilities $\varepsilon_u$, via the \emph{Pareto frontier} defined in the following. 
\begin{definition}
Let $\mathcal{C}$ be the set of feasible configurations for a specific access method and $f:\mathcal{C}\rightarrow \mathbb{R}^2$. Next, let \[Y=\{(S,\tau):(S,\tau)=f(c),c\in\mathcal{C}\},\]
where $S$ is the throughput of user 1 and $\tau$ is the timeliness of user 2; $\tau\in\{T_{90},\Delta_{90}\}$. The \emph{Pareto frontier} is the set 
\begin{IEEEeqnarray*}{rCl}
\mathcal{P}(Y)&=&\{(S,\tau)\in Y:\forall(S',\tau')\in Y:S> S'\vee\tau<\tau'\}.
\end{IEEEeqnarray*}
\end{definition}

Besides obtaining the Pareto frontiers, we evaluate the schemes by setting a minimum requirement for $S$, the throughput of user 2. Then, the optimal configuration of an access method is defined as the combination of parameters that minimizes the timing, either \gls{lr} or \gls{paoi} while maintaining $S$ above the minimum required. 

Table~\ref{tab:notation} summarizes the relevant notation introduced in this section.
To simplify the analytical expressions in the rest of the paper, we define the binomial function $\text{Bin}(K;N,p)$ as
\begin{equation}
    \text{Bin}(K;N,p)=\binom{N}{K}p^K(1-p)^{N-K}
\end{equation}
and the multinomial function $\text{Mult}(\mathbf{K};N,\mathbf{p})$ as
\begin{equation}
    \text{Mult}(\mathbf{K};N,\mathbf{p})=\frac{N!\prod_{i=1}^{|\mathbf{p}|}p_i^{K_i}(1-\sum_{i=1}^{|\mathbf{p}|}p_i)^{N-\sum_{i=1}^{|\mathbf{p}|}K_i} }{(N-\sum_{i=1}^{|\mathbf{p}|}K_i)!\prod_{i=1}^{|\mathbf{p}|} K_i!},
\end{equation}
where $|\mathbf{p}|$ is the length of vector $\mathbf{p}$.
Finally, we denote $\delta(x)$ as the delta function, which is equal to 1 if $x=0$ and 0 otherwise, and $[x]^+=\max(x,0)$.

\begin{table*}[t]
\centering
\caption{Notation summary}
\renewcommand{\arraystretch}{1.2}
\begin{tabular}{@{}llcll@{}}
\toprule
Symbol & Description & & Symbol & Description\\ \bottomrule
 $\mathcal{U}=\{1,2\}$    & Set of users; $u=1$ is the broadband user &&  $P_u$ & Transmission power of user $u$\\
& and $u=2$ is the intermittent user && $\overline{\text{SNR}}_u$ & Expected \gls{snr} for user $u$ \\
$K$    &  Size of the source block for user 1 &&$\text{SNR}_{u,t}$ & \gls{snr} of user $u$ at slot $t$\\
 $N$ & Size of the coded block for user 1 && $\text{SINR}_{u,t}$ & \gls{sinr} of user $u$ at slot $t$ \\
 $Q$ & Maximum queue length for user 2 && $\varepsilon_u$ & Erasure probability of user $u$\\
 $T_\text{int}$ & Period between slots allocated to user 2 in \gls{tdma} && $o_u\in\{\mathcal{I,R,E}\}$ & Outcome for user $u$ when signals overlap\\
 $t\in\mathbb{Z}$ & Time slot index && $\left(o_1,o_2\right)$ & Outcome when signals overlap\\
 $q_t$ & Length of the queue for user 2 at $t$ && $\pi_{o_1o_2}$ & Probability of outcome $\left(o_1,o_2\right)$\\
 $\mathcal{A}_t\subseteq \mathcal{U}$ & Allocation of time slot $t$  && $p_{s,u}$ & Success probability of user $u$\\
 $a_{u,t}$ & Activity indicator for user $u$ at $t$; 1 if active && $S$ & Throughput of user 1\\
 $h_{u,t}$ & Fading envelope for user $u$ at $t$ && $T$ & \Gls{rv} of \gls{lr} for user 2\\
 $\sigma^2$ & Noise power  &&$\Delta$ & \Gls{rv} of \gls{aoi} for user 2\\
  $\ell_u$ & Path loss of user $u$ && $T_{90}$, $\Delta_{90}$& $90$th percentile of \gls{lr} and \gls{aoi}\\
  $r$ & Distance between user 2 and the \gls{bs}&&$\delta(x)$ & Delta function, equal to 1 if $x=0$ and 0 otherwise \\
 \bottomrule
 
\end{tabular}
\label{tab:notation}
\end{table*}

\section{Performance with TDMA}
\label{sec:oma}

Here we derive the \glspl{kpi} for the \gls{tdma} system, for a \gls{lr}- or \gls{paoi}-oriented intermittent user.
For \gls{lr}, the length of the intermittent user's queue is assumed to be fixed to some $Q \geq 1$. On the other hand, for \gls{paoi}, the length of the intermittent user's queue is set to $Q=1$. This is because transmitting the newest packet is the optimal strategy to minimize \gls{paoi} but packet retransmissions are not allowed. 

In the assumed \gls{tdma} system, the broadband user has frames of $N$ slots, each of which contains $K$ data packets and $N-K$ redundancy packets, while the intermittent user has one reserved slot every $T_{\text{int}}$.
The success probability for user 1 is easy to compute
\begin{equation}
  p_{s,1}=\sum_{m=K}^N \text{Bin}(m;N,1-\varepsilon_1).
  \label{eq:ps1}
\end{equation}
The expected throughput of user 1 is
\begin{equation}
  S=p_{s,1}\frac{(T_{\text{int}}-1)K}{T_{\text{int}}N}.
  \label{eq:throughput}
\end{equation}
That is, the throughput measures the rate of innovative (i.e., non-redundant) packets received at the \gls{bs} from user 1 per time slot.
As the broadband user can only use $T_{\text{int}}-1$ slots for each $T_{\text{int}}$, setting up more frequent transmission opportunities for the intermittent user reduces the throughput. 

\subsection{Latency-reliability (LR)}

In order to derive the \gls{pmf} of \gls{lr} for the intermittent user, without loss of generality, we take the origin of time to be a slot in which a transmission occurs.
We define a Markov chain representing the state of the queue $q_{t}$ for the intermittent user, i.e., the number of packets in the queue at time $t$.
The transition matrix of the chain is $\textbf{P}^{(0)}$, whose elements $P_{ij}^{(0)}$ represent the probability of transitioning from state $i$ to state $j$ in the queue of the intermittent user at the end of such slot \cite{kleinrock1996queueing}.
The elements $P_{ij}^{(0)}$ are obtained as
\begin{equation}
  P^{(0)}_{ij}=\begin{cases}
           0 &\text{if } j<i-1;\\
           \text{Bin}(j-i+1;T_{\text{int}},\alpha)&\text{if } i-1\leq j<Q; \\
           \sum_{m=Q-i+1}^{T_{\text{int}}} \text{Bin}(m;T_{\text{int}},\alpha)&\text{if } j=Q.
         \end{cases}\label{eq:trans_queue}
\end{equation}
Let $\boldsymbol{\varphi}^{(0)}=\left[\varphi^{(0)}_0,\varphi^{(0)}_1,\dotsc,\varphi^{(0)}_Q\right]$ be the steady-state distribution vector of the queue immediately after a transmission. From the transition matrix computed in \eqref{eq:trans_queue}, we can easily derive $\boldsymbol{\varphi}^{(0)}$ as the left-eigenvector of $P^{(0)}$ with eigenvalue 1, normalized to sum to 1 to be a valid probability metric
\begin{equation}
\boldsymbol{\varphi}^{(0)}(\mathbf{I}-\mathbf{P}^{(0)})=\mathbf{0} \, \wedge \, \sum_{q=0}^Q\varphi^{(0)}_q=1.
\label{eq:steady_mat}
\end{equation}
It is easy to derive the steady-state distribution of the queue $q_{n}$ (i.e., $n$ slots after a transmission) as
\begin{equation}
  \varphi^{(n)}_q=\begin{cases}
              \sum_{s=0}^q \varphi^{(0)}_s \text{Bin}(q-s;n\alpha)&\text{if } q<Q;\\
              \sum_{s=0}^Q\sum_{m=Q-s}^n \varphi^{(0)}_s \text{Bin}(m;n,\alpha)&\text{if } q=Q,
            \end{cases}\label{eq:steady_slot_oma}
\end{equation}
where $\varphi_q^{(0)}$ is the $q$-th element of vector $\boldsymbol{\varphi}^{(0)}$.
If a packet is queued behind $q$ others, it will be transmitted at the $(q+1)$-th opportunity, unless new arrivals make the system drop some of the packets ahead of it in the queue: we remind the reader that, if the queue is full, the oldest packet (i.e., the first in the queue) is dropped. Let $g_i\in\{0,1,\ldots,T_\text{int}\}$ for $i\geq1$ be the number of packets generated by user 2 between the $i$-th and $(i+1)$-th intermittent slots after the current one. Further, let $g_0$ be the number of packets generated between the current time slot and the next intermittent slot. We define
\begin{equation}
    \mathcal{G}_{\ell}^{(n)}=\big\{\left[g_0\in\{0,1,\ldots,T_\text{int}-n\},g_1,\ldots,g_\ell\right]\big\}
\end{equation} 
to be the set of possible vectors for the number of packets generated by user 1 given that there are $T_\text{int}-n$ slots until the next intermittent slot.

The probability of occurrence of each element $\mathbf{g}\in\mathcal{G}_{\ell}^{(n)}$ is
\begin{equation}
  p_{\text{gen}}(\mathbf{g};\ell,n)=\text{Bin}(g_{0};T_{\text{int}}-n,\alpha)\prod_{i=1}^\ell \text{Bin}(g_i;T_{\text{int}},\alpha).
\end{equation}
At each intermittent slot, up to one packet is transmitted and, hence, removed from the queue. Other packets are removed if the number of generated packets exceeds the number of remaining spaces in the queue. For a given vector $\mathbf{g}\in\mathcal{G}_{\ell}^{(n)}$, the considered packet is transmitted at the $\ell$-th transmission opportunity, where $\ell$ is the first index that satisfies condition $\psi_k^{(\mathbf{g},q)}$ if the packet has $q$ others ahead of it in the queue
\begin{equation}
  \psi_k^{(\mathbf{g},q)}=\delta\left(\sum_{i=1}^k\left[q+1-Q+\sum_{j=1}^i g_j\right]^+\!+k-(q+1)\right).
\end{equation}
We now define the set $\mathcal{S}_{\ell}^{(n,q)}$, which contains the elements $\mathbf{g}\in\mathcal{G}_{\ell}^{(n)}$ for which the considered packet is transmitted at the $\ell$-th opportunity
\begin{equation}
  \mathcal{S}_{\ell}^{(n,q)}=\Big\{\mathbf{g}\in\mathcal{G}_{\ell}^{(n)}: \psi_\ell^{(\mathbf{g},q)}-\sum_{k=1}^{\ell-1}\psi_k^{(\mathbf{g},q)}=1\Big\}.
\end{equation}
Since the packet is either transmitted within $q+1$ transmission attempts or discarded, the conditioned success probability $p_{s,2}(n,q;T_{\text{int}})$ for the intermittent user is given by
\begin{align}
  p_{s,2}(n,q)=\sum_{\ell=1}^{q+1}\sum_{\mathbf{g}\in\mathcal{S}_{\ell}^{(n,q)}}p_{\text{gen}}(\mathbf{g};\ell,n)(1-\varepsilon_2).
\end{align}
We can now compute the latency \gls{pmf} $p_T(t)$, considering the fact that it takes 1 slot to transmit the packet
\begin{IEEEeqnarray}{rCl}
 p_T(t)&=&\sum_{n=1}^{T_{\text{int}}}\sum_{q=0}^{Q}\frac{\sum_{\mathbf{g}\in\mathcal{S}_{\ell}^{(n,\min(q,Q-1))}} p_{\text{gen}}\left(\mathbf{g};\frac{t+n-1}{T_{\text{int}}},n\right)}{T_{\text{int}}p_{s,2}(n,\min(q,Q-1))}\IEEEnonumber\\
 &&\times\varphi^{(n-1)}_q\delta\left(\frac{t+n-1}{T_{\text{int}}}-\left\lfloor\frac{t+n-1}{T_{\text{int}}}\right\rfloor\right).
\end{IEEEeqnarray}
The success probability of the intermittent user is
\begin{equation}
  p_{s,2}=\sum_{n=1}^{T_{\text{int}}}\sum_{q=0}^{Q}\frac{\varphi^{(n-1)}_q p_{s,2}(n,q)}{T_{\text{int}}}.
\end{equation}

\subsection{Peak age}

In the \gls{paoi}-oriented case, the \gls{pmf} is given by the sum of the waiting time $W$ and the inter-update interval $Z$~\cite{Kaul2012}.

Since $Q=1$, the generated packets are always sent at the first available transmission opportunity.
The \gls{pmf} of the waiting time $W$ for a successful transmission is given by:
\begin{equation}
  p_{W}(w)=\frac{\alpha(1-\alpha)^{w-1}}{1-(1-\alpha)^{T_{\text{int}}}},\ w\in\{1,\ldots,T_{\text{int}}\}.
\end{equation}
We now compute the \gls{pmf} of $Z$. Since exactly one slot every $T_{\text{int}}$ is reserved for the intermittent user, $Z$ is $T_{\text{int}}$ times the number of reserved slots between consecutive successful transmissions. This is a geometric random variable, whose probability of successful transmission is given by:
\begin{equation}
\xi=(1-(1-\alpha)^{T_{\text{int}}})(1-\varepsilon_2).
\label{eq:xi}
\end{equation}
The \gls{pmf} of $Z$ is then
\begin{equation}
 p_Z(z)=(1-\xi)^{\frac{z}{T_{\text{int}}}-1}\xi\delta\left(\text{mod}(z,T_{\text{int}})\right).
 \label{eq:pmfZ}
\end{equation}
The \gls{pmf} of the \gls{paoi} is
\begin{equation}
 p_{\Delta}(t+1)=p_{Z}\left(t-\text{mod}(t,T_{\text{int}})\right)p_{W}\left(1+\text{mod}(t,T_{\text{int}})\right).
\end{equation}

\section{Performance with NOMA}
\label{sec:noma}

We now derive the distributions of the \glspl{kpi} in the \gls{noma} case, in which the broadband user has frames of $N$ slots, all of which are mixed, i.e., allocated both to the intermittent and broadband user.

First, we define $p_1$ as the probability that a packet from the broadband user is received in a given slot, which is given by
\begin{equation}
 p_1=\left((1-\alpha)(1-\varepsilon_1)+\alpha (\pi_{\mathcal{I}\mathcal{I}}+\pi_{\mathcal{I}\mathcal{E}})\right).
\end{equation}
The probability that the block from the broadband user is decoded in the $d$-th slot of the frame, denoted as $p_D(d)$, is
\begin{equation}
\begin{aligned}
 p_D(d)=&p_1\text{Bin}(K-1;d-1,p_1),
\end{aligned}
\end{equation}
The \gls{cdf} of the decoding instant $D$, $P_D(d)$, is
\begin{equation}
  P_D(d)=\sum_{m=K}^{d}\text{Bin}(m;d,p_1).
\end{equation}
We then simply have $p_{s,1}=P_D(N)$.
The average throughput for the broadband user is
\begin{equation}
 S=\frac{K P_D(N)}{N}.
\end{equation}

\subsection{Latency-reliability (LR)}
We now analyze the latency distribution for the intermittent user. All the intermittent packets transmitted after decoding slot $d$ -- once the block from the broadband user has been decoded -- can be either decoded immediately or lost. Specifically, these packets are lost with probability $\varepsilon_2=\pi_{\mathcal{I}\mathcal{E}}+\pi_{\mathcal{E}\mathcal{E}}+\pi_{\mathcal{R}\mathcal{E}}$. On the other hand, if the intermittent user packet is sent before the decoding slot $d$, it is decoded instantly with probability $\pi_{\mathcal{I}\mathcal{I}}+\pi_{\mathcal{E}\mathcal{I}}$, while it can be decoded after \gls{sic} with probability $\pi_{\cdot\mathcal{R}}$. In order to compute $p_{s,2}$, we need to compute the conditioned probability of having $a_b$ collisions between the two users before the broadband user block is decoded in slot $d$, $i_b$ of which result in an immediate decoding, while $v_b$ are decoded after \gls{sic}, for the intermittent user. This is denoted as $p_{A_b,I_b,V_b|D}(a_b,i_b,v_b|d)$, and given by:
\begin{equation}
\begin{aligned}
 p_{A_b,I_b,V_b|D}(a_b,i_b,v_b|d)=\sum_{\mathclap{\ell=K-d+a_b}}^{\mathclap{\min(a_b,K-1)}}\text{Bin}(a_b;d-1,\alpha)\\
 \times\frac{p_1}{p_D(d)}\sum_{m=0}^{\mathclap{\min(i_b,\ell)}}\text{Bin}(K-1-\ell;d-1-a_b,1-\varepsilon_1)\\
 \times \text{Mult}\left([m,i_b-m,v_b,\ell-m];a_b,[\pi_{\mathcal{I}\mathcal{I}},\pi_{\mathcal{E}\mathcal{I}},\pi_{\cdot\mathcal{R}},\pi_{\mathcal{I}\mathcal{E}}]\right).
 \end{aligned}\label{eq:pnoma_before}
\end{equation}
We can then simply take the four cases for packets from the intermittent user (transmitted before slot $d$, in slot $d$, after slot $d$, or in lost frames), and compute $p_{s,2}$.
We can compute $p_{A_d,I_d}(a_d,i_d)$, the probability that a packet from user 2 is sent and correctly decoded in the same slot as the broadband user block decoding:
\begin{equation}
 p_{A_d,I_d}(a_d,i_d)=\begin{cases}\frac{\alpha \pi_{\mathcal{I}\mathcal{I}}}{p_1},&a_d=1,i_d=1;\\
                         \frac{\alpha \pi_{\mathcal{I}\mathcal{E}}}{p_1},&a_d=1,i_d=0;\\
                         \frac{(1-\alpha)(1-\varepsilon_1)}{p_1},&a_d=0,i_d=0;\\
                         0, &\text{otherwise}.
                        \end{cases}
 \label{eq:pnoma_same}
\end{equation}
We then give the probability of having $A_a$ packets after the decoding of the broadband user block in slot $d$, $I_a$ of which are correctly received:
\begin{equation}
 \begin{aligned}
 p_{A_a,I_a|D}(a_a,i_a|d)=&\frac{\text{Bin}(i_a;a_a,\pi_{\mathcal{I}\mathcal{I}}+\pi_{\mathcal{E}\mathcal{I}}+\pi_{\cdot\mathcal{R}})}{p_D(d)}\\
 &\times\text{Bin}(a_a;N-d,\alpha).\label{eq:pnoma_after}
 \end{aligned}
\end{equation}
Finally, we can consider the case in which the broadband user frame is not decoded: in this case, the only intermittent packets that are decoded are immediate captures. We can then compute the probability $p_{A_z,I_z|\tilde{D}}(a_z,i_z)$:
\begin{equation}
\begin{aligned}
 p_{A_z,I_z|\tilde{D}}(a_z,i_z)=\sum_{c=0}^{\mathclap{\min(K-1,N-a)}}\text{Bin}(c;N-a,1-)\sum_{\ell=0}^{\mathclap{\min(a_z,K-1-c)}}\text{Bin}(a_z;N,\alpha)\\
 \times\sum_{m=0}^{\min(\ell,i_z)}\frac{\text{Mult}([m,i_z-m,\ell-m];a_z,[\pi_{\mathcal{I}\mathcal{I}},\pi_{\mathcal{E}\mathcal{I}},\pi_{\mathcal{I}\mathcal{E}}])}{1-p_{s,1}}.
 \end{aligned}
\end{equation}
We now know that all packets transmitted by the intermittent user at or after the decoding of the broadband block, or in frames for which the broadband block is not decoded, are either lost or decoded immediately. To compute the latency distribution, we then only need to distinguish the case in which a packet transmitted before $d$ is decoded instantly or after \gls{sic}. The probability of a packet from the intermittent user being decoded instantly is then $p_T(1)$:
\begin{equation}
\begin{aligned}
p_{T}(1)=&(1-p_{s,1})\sum_{a_z=1}^N\sum_{i_z=0}^{a_z}\frac{i_z p_{A_z,I_z|\tilde{D}}(a_z,i_z)}{(1-\text{Bin}(0;N,\alpha))a_z}\\
&+\sum_{d=K}^N \sum_{a_b=0}^{d-1}\sum_{a_d=0}^1\sum_{a_a=0}^{N-d}\sum_{i_b=0}^a\sum_{i_d=0}^1\sum_{i_a=0}^m\sum_{v_b=0}^{a_b-i_b}p_D(d) \\
&\times\frac{(i_b+i_d+i_a)p_{A_d,I_d}(a_d,i_d)p_{A_a,I_a|D}(a_a,i_a|d)}{(1-\text{Bin}(0;N,\alpha))(a_b+a_d+a_a)}\\
&\times p_{A_b,I_b,V_b|D}(a_b,i_b,v_b|d).\label{eq:pnoma_lat_0}
\end{aligned}
\end{equation}
As the delay from any packet decoded after \gls{sic} is distributed uniformly between 2 and $d+1$, we can easily compute $p_T(t)$:
\begin{equation}
\begin{aligned}
p_{T}(t)=&\sum_{\mathclap{d=\min(K,t-1)}}^N p_D(d)\sum_{a_b=0}^{d-1}\sum_{a_d=0}^1\sum_{a_a=0}^{N-d}\sum_{i_b=0}^{a_b}\sum_{v_b=0}^{a_b-i_b} \frac{v_b}{d}\\
&\times\sum_{i_d=0}^1\sum_{i_a=0}^m \frac{p_{A_d,I_d}(a_d,i_d)p_{A_a,I_a|D}(a_a,i_a|d)}{(a_b+a_d+a_a)(1-\text{Bin}(0;N,\alpha))}\\
&\times p_{A_b,I_b,V_b|D}(a_b,i_b,v_b|d),\ t\in\{2,\ldots,N\}.\label{eq:pnoma_lat_t}
\end{aligned}
\end{equation}
The combination of~\eqref{eq:pnoma_lat_0} and~\eqref{eq:pnoma_lat_t} is the latency-reliability \gls{pmf} for the intermittent user. We then have:
\begin{equation}
 p_{s,2}=\sum_{t=1}^{N} p_T(t).
\end{equation}

\subsection{Peak age}

\begin{figure*}
\begin{minipage}{\linewidth}
\begin{equation}
 \begin{aligned}
  p_F(f)=&\sum_{a=0}^{f-1}\left[\alpha \varepsilon_C(\cdot,I)\sum_{\ell=0}^{\mathclap{\min(K-1,a)}}\text{Bin}(a;f-1,\alpha)\sum_{m=0}^{\mathclap{\min(K-1-\ell,f-a-1)}}\text{Mult}([0,0,\ell];a,[\pi_{\mathcal{I}\mathcal{I}},\pi_{\mathcal{E}\mathcal{I}},\pi_{\mathcal{I}\mathcal{E}}])\times\text{Bin}(m;f-a-1,1-\varepsilon_1)\right]
  \\
  &+p_D(f)\sum_{a_d=0}^1 p_{A_b,I_b,V_b|D}(a_b,0,v_b|f)p_{A_d,I_d}(a_d,0)
  +\sum_{\mathclap{d=K}}^{\mathclap{f-1}}p_D(d)\times\sum_{\mathclap{a_b=0}}^f p_{A_b,I_b,V_b|D}(a_b,0,0|d)\\
  &\times\sum_{\mathclap{a_d=0}}^1 p_{A_d,I_d}(a_d,0)\text{Bin}(0;f-d-1,\alpha \pi_{\cdot\mathcal{R}}\alpha \pi_{\cdot\mathcal{R}}.
 \end{aligned}\label{eq:pff}
\end{equation}
\end{minipage}
\end{figure*}

In order to derive the \gls{pmf} of the \gls{paoi}, we first need to compute some auxiliary values. First, we derive the probability that the first decoded packet from the intermittent user in a frame is decoded in slot $f$, denoted as $p_F(f)$ and given in~\eqref{eq:pff}.

It is then easy to get $p_e$, the probability of decoding no new intermittent packets in a frame:
\begin{equation}
\begin{aligned}
 p_e=1-\sum_{f=1}^N p_F(f).
\end{aligned}
\end{equation}
The \gls{pmf} of the number of slots $Y$ from the beginning of a given frame until the first decoded packet from the intermittent user is
\begin{equation}
 p_Y(y)=p_e^{\left\lfloor\frac{y}{N}\right\rfloor}p_F(\text{mod}(y,N)),
\end{equation}
where $\text{mod}(m,n)$ is the integer modulo function. 

We now consider the probability $p_U(u)$ of receiving an update from the intermittent user, i.e., a packet with newer information than the one already available. We have the following \gls{pmf}, conditioned on the decoding slot $d$ of the broadband block. First, we consider the case in which $d<u$
\begin{equation}
 p_{U|D}(u|d)=\sum\limits_{a_b=1}^{d-1}\sum\limits_{i_b=1}^{a_b}\sum\limits_{v_b=0}^{\mathclap{a_b-i_b}} \frac{i_b p_{A_b,I_b,V_b|D}(a_b,i_b,v_b|d)}{d-1}, d<u.
\end{equation}
Next, for $d>u$
\begin{equation}
 p_{U|D}(u|d)=\sum\limits_{a_a=1}^{N-d}\sum\limits_{i_a=1}^{a_a} \frac{i_a}{N-d} p_{A_a,I_a|D}(a_a,i_a|d), d>u.
\end{equation}
Finally, for $d=u$
\begin{equation}
\begin{aligned}
 p_{U|D}(d|d)=&\sum\limits_{a_b=1}^{d-1}\sum\limits_{i_b=0}^{a_b-1}\sum\limits_{v_b=1}^{a_b-i_b}\sum_{m=i_b+v_b}^{d-1} \sum\limits_{a_d=0}^1\frac{v_b p_{A_d,I_d}(a_d,0) }{d-1}\\
&\times \mathcal{H}_{m-1,d-1}(i_b+v_b-1,i_b+v_b-1)\\
&\times p_{A_b,I_b,V_b(a_b,i_b,v_b|d)|D}+p_{A_d,I_d}(1,1),
\end{aligned}
\end{equation}
where $\mathcal{H}_{M,N}(m,n)$ is the hypergeometric distribution, whose \gls{pmf} is given by
\begin{equation}
 \mathcal{H}_{M,N}(m,n)=\frac{\binom{M}{m}\binom{N-M}{n-m}}{\binom{N}{n}}.
\end{equation}
We also consider the probability $p_{U|\tilde{D}}(\mathtt{u})$, i.e., the probability of receiving an update in slot $\mathtt{u}$ if the broadband user block is not decoded
\begin{equation}
  p_{U|\tilde{D}}(\mathtt{u})=\sum\limits_{a_z=1}^{N}\sum\limits_{i_z=1}^{a_z}\frac{i_z p_{A_z,I_z|\tilde{D}}(a_z,i_z)}{N}.
\end{equation}
By applying the law of total probability, we obtain $p_U(\mathtt{u})$
\begin{equation}
  p_U(\mathtt{u})=\sum_{d=K}^N p_D(d)p_{U|D}(\mathtt{u}|d)+(1-p_{s,1})p_{U|\tilde{D}}(\mathtt{u}).
\end{equation}

We now compute the probability that a given update is the last in the frame, given that the decoding happens in slot $d$, denoted as $p_{L|D}(\ell|d)$. Again, we distinguish three cases, starting from $\ell<d$:
\begin{equation}
 \begin{aligned}
  p_{L|D}(\ell|d)=&\sum_{a_b=1}^{d-1}\sum_{i_b=1}^\ell\sum_{v_b=0}^{\ell-i_b}\sum_{a_d=0}^1\sum_{a_a=0}^{N-d}\frac{i_b p_{A_d,I_d}(a_d,0)}{d-1}\\
  &\times p_{A_b,I_b,V_b|D}(a_b,i_b,v_b|d)p_{A_a,I_a|D}(a_a,0|d)\\
  &\times \frac{\mathcal{H}_{\ell-1,d-2}(v_b+i_b-1,v_b+i_b-1)}{p_{U|D}(\ell|d)}, \ell<d.
 \end{aligned}
\end{equation}
If $\ell=d$, we have:
\begin{equation}
 p_{L|D}(d|d)=\sum_{a_a=0}^{N-d}p_{A_a,I_a|D}(a_a,0|d).
\end{equation}
Finally, if $\ell>d$ the probability is
\begin{equation}
 \begin{aligned}
  p_{L|D}(\ell|d)=&\sum_{a_a=1}^{\ell-d}\sum_{i_a=1}^{a_a}\frac{i_a p_{A_a,I_a|D}(a_a,i_a|d)}{(N-d)p_{U|D}(\ell|d)}\\
  &\times\mathcal{H}_{\ell-d-1,N-d-1}(i_a-1,i_a-1), \ell>d.
 \end{aligned}
\end{equation}
The probability that an update in slot $\ell$ is the last in the frame, given that the broadband user frame is lost, $p_{L|\tilde{D}}(\ell)$, is
\begin{equation}
  p_{L|\tilde{D}}(\ell)=\sum\limits_{a_z=1}^{\ell}\sum\limits_{i_z=1}^{a_z}\frac{p_{A_z,I_z}(a_z,i_z)i_z\mathcal{H}_{N-\ell,N-1}(0,i_z-1)}{N p_{U|\tilde{D}}(\ell)}.
\end{equation}
Combining the expressions derived above, we get
\begin{equation}
  p_L(\ell)=\sum_{d=K}^N p_D(d)p_{L|D}(\ell|d)+(1-p_{s,1})p_{L|\tilde{D}}(\ell).
\end{equation}
If the update is not the last in the frame, we can compute the conditioned \gls{pmf} $p_{Z|U,D,\tilde{L}}(z|\mathtt{u},d)$ of the inter-update interval $Z$. We first consider the case in which $z+\mathtt{u}<d$
\begin{equation}
\begin{aligned}
 p_{Z|U,D,\tilde{L}}(z|\mathtt{u},d)=\sum\limits_{a_b=2}^{d-1}\sum\limits_{i_b=2}^{a_b} \frac{i_b(i_b-1)\mathcal{H}_{z-1,d-3}(0,i_b-2)}{p_{U|D}(\mathtt{u}|d)}\\
 \times\sum\limits_{\mathclap{v_b=0}}^{\mathclap{a_b-i_b}}\frac{p_{A_b,I_b,V_b|D}(a_b,i_b,v_b|d)}{(d-1)(d-2)(1-p_{L|D}(\mathtt{u}|d))},\ \mathtt{u}+z<d.
 \end{aligned}
\end{equation}
In this case, the only possibility to have another update after $z$ is to have two immediate captures in slots $\mathtt{u}$ and $\mathtt{u}+z$, without any immediate captures in between.
Further, for $\mathtt{u}> d$
\begin{equation}
\begin{aligned}
 p_{Z|U,D,\tilde{L}}(z|\mathtt{u},d)= \sum_{a_a=2}^{N-d-z+1}
                    \sum_{i_a=2}^{a_a}\frac{p_{A_a,I_a|D}(a_a,i_a|d)}{(N-d)(N-d-1)}\\
                    \times\frac{i_a(i_a-1)\mathcal{H}_{z-1,N-d-2}(0,i_a-2)}{(1-p_{L|D}(\mathtt{u}|d))p_{U|D}(\mathtt{u}|d)},\ \mathtt{u}> d\wedge \mathtt{u}+z\leq N.
\end{aligned}
\end{equation}
Next, for $\mathtt{u}=d$:
\begin{equation}
\begin{aligned}
 p_{Z|U,D,\tilde{L}}(z|d,d)=& 
                    \sum_{a_a=1}^{N-d-z+1}\sum_{i_a=1}^{a_a}\frac{i_a p_{A_a,I_a|D}(a_a,i_a|d)}{(N-d)(1-p_{L|D}(d|d))}\\
                    &\times\mathcal{H}_{z-1,N-d-1}(0,i_a-1),\  d+z\leq N.
\end{aligned}
\end{equation}
We then consider the case that $\mathtt{u}+z=d$ 
\begin{equation}
\begin{aligned}
 p_{Z|U,D,\tilde{L}}(d-\mathtt{u}|\mathtt{u},d)=\sum\limits_{a_b=1}^{d-1}\sum\limits_{i_b=1}^{a_b}\sum\limits_{v_b=0}^{\mathclap{a_b-i_b}}p_{A_b,I_b,V_b|D}(a_b,i_b,v_b|d)\\
 \times\frac{\mathcal{H}_{\mathtt{u}-1,d-2}(i_b-1,i_b-1)}{(d-1)p_{U|D}(\mathtt{u}|d)(1-p_{L|D}(\mathtt{u}|d))}\Bigg(p_{A_d,I_D}(1,1)\\
 +\mathbbm{1}(v_b-1)\sum_{a_d=0}^1 p_{A_d,I_d}(a_d,0)(1-\mathcal{H}_{\mathtt{u}-i_b,d-i_b-1}(v_b,v_b))\Bigg),
\end{aligned}
\end{equation}
where $\mathbbm{1}(x)$ is the step function, equal to 1 if $x\geq0$ and 0 otherwise.
Finally, we can derive $p_{Z|U,\tilde{D},\tilde{L}}(z|\mathtt{u})$, if the broadband user frame is not decoded:
\begin{equation}
\begin{aligned}
 p_{Z|U,\tilde{D},\tilde{L}}(z|\mathtt{u})=&\sum\limits_{a_z=2}^{N}\sum\limits_{i_z=2}^{a_z}\frac{\mathcal{H}_{z-1,N-2}(0,i_z-2)}{N(N-1)(1-p_{L|\tilde{D}}(\mathtt{u})p_{U|\tilde{D}}(\mathtt{u})}\\
 &\times p_{A_z,I_z}(a_z,i_z),\ \mathtt{u}+z\leq N.
 \end{aligned}
\end{equation}

\begin{table*}[t]
\centering
\caption{Parameter settings}
\renewcommand{\arraystretch}{1.2}
\begin{tabular}{@{}lllclll@{}}
\toprule
Parameter & Symbol & Setting && Parameter & symbol& Setting\\
\midrule
Coded block length for user 1 & $N$ & $\{2,3,\dotsc,32\}$ && Source block length for user 1 & $K$ & $<N$ \\
Erasure probability of user 1 & $\varepsilon_1$ & $0.1$ && Transmission power of user 2 & $P_2$ & $23$\,dBm  \\
Activation probability for user 2 & $\alpha$ & $\{0.01,0.05,0.1\}$ && Period between intermittent slots in \gls{tdma}&  $T_\text{int}$ & $\{1,2,\dotsc,40\}$\\  
\gls{sinr} threshold to decode a packet &  $\gamma_1=\gamma_2$ & $3$\,dB && Noise power & $\sigma^2$ & $-127.216$\,dBm \\
Distance from user 2 to the \gls{bs} & $r$ & $\left\{50,100,\ldots,400\right\}$\,m&&  Carrier frequency & $f_c$ & $2$\,GHz\\
Path loss exponent & $\eta$ & $\{2.6,3\}$ && Queue length in TDMA & $Q$ & 4 packets \\
\bottomrule
\end{tabular}
\label{tab:param}
\end{table*}

We now compute the \gls{pmf} of the inter-update interval $Z$ if the next packet is in the same frame
\begin{equation}
\begin{aligned}
  p_Z(z|\mathtt{u})=&\sum_{d=K}^N p_{Z|U,D,\tilde{L}}(z|\mathtt{u},d)p_D(d)
            +(1-p_{s,1})\\
            &\times p_{Z|U,\tilde{D},\tilde{L}}(z|\mathtt{u}),\ z\leq N-\mathtt{u}.
\end{aligned}
\end{equation}
On the other hand, if $z>N-\mathtt{u}$, we have
\begin{equation}
\begin{aligned}  
            p_Z(z|\mathtt{u})=p_L(\mathtt{u})p_e^{\left\lfloor\frac{z-(N-\mathtt{u})}{N}\right\rfloor}p_F(\text{mod}(z-(N-\mathtt{u}),N)),\\
            \forall z>N-\mathtt{u}.
\end{aligned}
\end{equation}

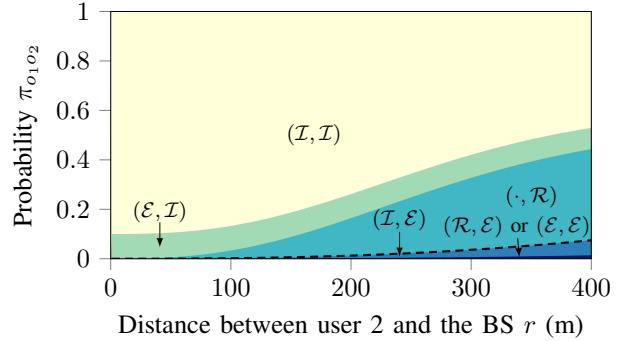
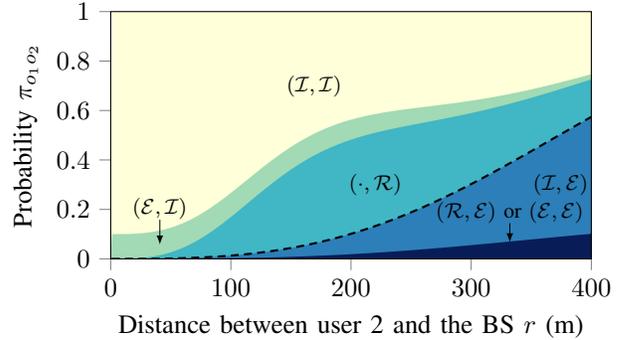
\begin{figure}
    \centering
    \begin{subfigure}[b]{\columnwidth}
    \centering
    \begin{tikzpicture}
\begin{axis}[
width=\fwidth,
height=\fheight,
tick align=outside,
tick pos=left,
legend style={ at={(0.03, 0.33)}, anchor=west},
ylabel={Probability $\pi_{o_1o_2}$},
xlabel={Distance between user 2 and the BS $r$~(m)},
xmin=0,
xmax=400,
ymin=0,
ymax=1,
reverse legend,
axis on top=true
]
\path[name path=floor] (axis cs:0,0) -- (axis cs:400,0);
\addplot[draw=none, name path=II, forget plot] table [x index=0, y index=7] {figs/outcomes/capture_probs_seq_pl_exponent_2p6_gamma_3_dB.txt};
\addplot[draw=none, name path=EI, forget plot] table [x index=0, y index=6] {figs/outcomes/capture_probs_seq_pl_exponent_2p6_gamma_3_dB.txt}node [pos=0.1, inner sep=0pt, yshift=-6pt, pin={[pin distance=0.13in]90, align=center, font=\footnotesize, inner sep=0pt:{$\left(\mathcal{E,I}\right)$}}]{};
\addplot[draw=none, name path=R, forget plot] table [x index=0, y index=5] {figs/outcomes/capture_probs_seq_pl_exponent_2p6_gamma_3_dB.txt};
\addplot[draw=black, densely dashed, thick, name path=IE, forget plot] table [x index=0, y index=4] {figs/outcomes/capture_probs_seq_pl_exponent_2p6_gamma_3_dB.txt}node [pos=0.60, inner sep=0pt, yshift=-2pt, pin={[pin distance=0.13in]90, align=center, font=\footnotesize, inner sep=0pt:{$\left(\mathcal{I,E}\right)$}}]{};
\addplot[draw=none, name path=RE, forget plot] table [x index=0, y index=3] {figs/outcomes/capture_probs_seq_pl_exponent_2p6_gamma_3_dB.txt}node [pos=0.85, inner sep=0pt, yshift=-1pt, pin={[pin distance=0.1in]90, xshift=-1pt, align=center, font=\footnotesize, inner sep=0pt:{$\left(\mathcal{R,E}\right)$ or $\left(\mathcal{E,E}\right)$}}]{};
\addplot[draw=none, name path=EE, forget plot] table [x index=0, y index=2] {figs/outcomes/capture_probs_seq_pl_exponent_2p6_gamma_3_dB.txt};

\addplot [fill=YlGnBu-A] fill between[ of=II and floor, soft clip second={domain=0:400}];
\addplot [thick, draw=none, fill=YlGnBu-E] fill between[ of=EI and floor, soft clip second={domain=0:400}];
\addplot [thick, draw=none, fill=YlGnBu-G] fill between[ of=R and floor, soft clip second={domain=0:400}];
\addplot [thick, draw=none, fill=YlGnBu-I] fill between[ of=IE and floor, soft clip second={domain=0:400}];
\addplot [thick, draw=none, fill=YlGnBu-M] fill between[ of=RE and floor, soft clip second={domain=0:400}];

\node[anchor=east, font=\footnotesize] at (axis cs: 200,0.5){$\left(\mathcal{I,I}\right)$};
\node[anchor=east, font=\footnotesize] at (axis cs: 380,0.25){$\left(\cdot,\mathcal{R}\right)$};

\end{axis}
 \end{tikzpicture}
    \caption{$\eta=2.6$}
    \end{subfigure}
    \begin{subfigure}[b]{\columnwidth}
    \centering
 \begin{tikzpicture}
\begin{axis}[
width=\fwidth,
height=\fheight,
tick align=outside,
tick pos=left,
legend style={ at={(0.03, 0.33)}, anchor=west},
ylabel={Probability $\pi_{o_1o_2}$},
xlabel={Distance between user 2 and the BS $r$~(m)},
xmin=0,
xmax=400,
ymin=0,
ymax=1,
reverse legend,
axis on top=true
]
\path[name path=floor] (axis cs:0,0) -- (axis cs:400,0);
\addplot[draw=none, name path=II, forget plot] table [x index=0, y index=7] {figs/outcomes/capture_probs_seq_pl_exponent_3_gamma_3_dB.txt};
\addplot[draw=none, name path=EI, forget plot] table [x index=0, y index=6] {figs/outcomes/capture_probs_seq_pl_exponent_3_gamma_3_dB.txt}node [pos=0.1, inner sep=0pt, yshift=-6pt, pin={[pin distance=0.13in]90, align=center, font=\footnotesize, inner sep=0pt:{$\left(\mathcal{E,I}\right)$}}]{};
\addplot[draw=none, name path=R, forget plot] table [x index=0, y index=5] {figs/outcomes/capture_probs_seq_pl_exponent_3_gamma_3_dB.txt};
\addplot[draw=black, densely dashed, thick, name path=IE, forget plot] table [x index=0, y index=4] {figs/outcomes/capture_probs_seq_pl_exponent_3_gamma_3_dB.txt};
\addplot[draw=none, name path=RE, forget plot] table [x index=0, y index=3] {figs/outcomes/capture_probs_seq_pl_exponent_3_gamma_3_dB.txt}node [pos=0.83, inner sep=0pt, yshift=-1pt, pin={[pin distance=0.1in]90, align=center, font=\footnotesize, inner sep=0pt:{$\left(\mathcal{R,E}\right)$ or $\left(\mathcal{E,E}\right)$}}]{};
\addplot[draw=none, name path=EE, forget plot] table [x index=0, y index=2] {figs/outcomes/capture_probs_seq_pl_exponent_3_gamma_3_dB.txt};

\addplot [fill=YlGnBu-A] fill between[ of=II and floor, soft clip second={domain=0:400}];
\addplot [thick, draw=none, fill=YlGnBu-E] fill between[ of=EI and floor, soft clip second={domain=0:400}];
\addplot [thick, draw=none, fill=YlGnBu-G] fill between[ of=R and floor, soft clip second={domain=0:400}];
\addplot [thick, draw=none, fill=YlGnBu-I] fill between[ of=IE and floor, soft clip second={domain=0:400}];
\addplot [thick, draw=none, fill=YlGnBu-M] fill between[ of=RE and floor, soft clip second={domain=0:400}];

\node[anchor=east, font=\footnotesize] at (axis cs: 200,0.7){$\left(\mathcal{I,I}\right)$};
\node[anchor=east, font=\footnotesize, inner sep=1pt] at (axis cs: 400,0.3){$\left(\mathcal{I,E}\right)$};
\node[anchor=east, font=\footnotesize] at (axis cs: 250,0.3){$\left(\cdot,\mathcal{R}\right)$};

\end{axis}
 \end{tikzpicture}
 \caption{$\eta=3$}
    \end{subfigure}
    \caption{Area plot for the probabilities of the different outcomes $\left(o_1,o_2\right)$ when the signals of both users collide for (a) $\eta=2.6$ and (b) $\eta=3$. The dashed line indicates the value of $\varepsilon_2$.}
    \label{fig:outcomes}
\end{figure}

The decoding delay $W$ component of \gls{paoi} applies only if the update transmitted before the decoding slot $d$ and decoded with \gls{sic} only after the decoding of the broadband user block. We then give $p_{W|U,D}(w|\mathtt{u},d)$, the \gls{pmf} of $W$ for an update in the same slot $d$ which the broadband user block is decoded in
\begin{equation}
\begin{aligned}
  p_{W|U,D}(w|d,d)=\frac{1}{p_{U|D}(\mathtt{u}|d)}\Bigg(p_{A_d,I_d}(1,1)\delta(w-1)+\\\quad\times\sum\limits_{v_b=1}^{\mathclap{\min(a_b,d-w+1)}}\mathcal{H}_{w-2,d-2}(0,v_b-1) \sum\limits_{i_b=0}^{\mathclap{\min(a_b,d-w+1)-v_b}}\mathcal{H}_{w-2,d-v_b-1}(0,i_b)\\
  \times\frac{p_{A_b,I_b,V_b|D}(a_b,i_b,v_b|d)}{(d-1)}\sum\limits_{a_d=0}^1 p_{A_d,I_d}(a_d,0)\Bigg),\ \mathtt{u}=d.
\end{aligned}
\end{equation}
In all other cases, the packet is captured instantaneously, and we simply have
\begin{equation}
p_{W|U,D}(w|\mathtt{u},d)=\delta(w-1),\ \mathtt{u}\neq d.
\end{equation}
If the broadband user frame is not decoded, the decoding delay is always 1, as the only updates are due to immediate capture
\begin{equation}
  p_{W|U,\tilde{D}}(w|\mathtt{u})=\delta(w-1).
\end{equation}
By applying the law of total probability, we get
\begin{equation}
\begin{aligned}
  p_{W|U}(w|\mathtt{u})=&\sum_{d=K}^N p_D(d)p_{W|U,D}(w|\mathtt{u},d)\\& + (1-p_{s,2})p_{W|U,\tilde{D}}(w|\mathtt{u}).
\end{aligned}
\end{equation}
Finally, we get the \gls{paoi} as the convolution of $Z$ and $W$ and removing the condition on $U$
\begin{equation}
  p_{\Delta}(t)=\sum_{\mathtt{u}=1}^N p_U(\mathtt{u})\sum_{w=1}^{\mathclap{\min(\mathtt{u},t-1)}}p_{W|U}(w|\mathtt{u})p_{Z|U}(t-w|\mathtt{u}).
\end{equation}

\section{Benchmark: Performance with FDMA}
\label{sec:fdma}

In case of \gls{fdma}, the two users are occupying a dedicated bandwidth part each, and their \glspl{kpi} are independent.
The success probability for user 1 is equal to that in \gls{tdma}, given by~\eqref{eq:ps1}. 
The throughput of user 1 can be computed by setting $T_\text{int}\to \infty$ in~\eqref{eq:throughput}, which gives
\begin{equation}
  S=\frac{Kp_{s,1}}{N}.
  \label{eq:th_fdma}
\end{equation}
For user 2, the latency for all successfully decoded packets is 1. 
Further, $p_{s,2}=1-\epsilon_2$ and the \gls{pmf} of \gls{lr} is simply $p_L(t)=\delta(t-1)$.

The \gls{paoi} for user 2 can be obtained as the latency $T=1$ plus the inter-decoding time $Z$ when setting $T_\text{int}=1$ in \eqref{eq:xi} and \eqref{eq:pmfZ}. Hence, it is simply a function of the inter-arrival time and $\epsilon_2$. Namely,
\begin{equation}
    p_\Delta(t)=\left(1-\alpha\left(1-\varepsilon_2\right)\right)^{t-2}\alpha\left(1-\varepsilon_2\right),\ t\geq2.
\end{equation}



\section{Evaluation}\label{sec:results}

We assume that user 1 (the broadband user) selects its transmission power to achieve $\varepsilon_1=0.1$.
On the other hand,  user 2 (the intermittent user) transmits infrequently, and thus cannot get up-to-date information on the channel state. The best possible strategy for it is then to always transmit at maximum power; in this case, $\varepsilon_2$ depends on its distance from the \gls{bs} $r$ and the erasure probability $\varepsilon_2$ is minimized. 

For performance evaluation, we set parameters that represent a typical 5G urban scenario~\cite{3GPPTR38913}.
Namely, the carrier frequency is $2$\,GHz, the path loss exponent is $\eta\in\{2.6,3\}$\,dB, the noise power $\sigma^2$ is determined by the noise temperature and the subcarrier spacing, set to a typical $\Delta f=15$\,kHz, plus a noise figure of $5$\,dB.
The resulting noise power and other relevant parameter settings are listed in Table~\ref{tab:param}.
For simplicity's sake, the \gls{sinr} thresholds for decoding both users are set to the same value $\gamma_1=\gamma_2=3$\,dB. As a reference, the \gls{sinr} threshold when calculating the maximum coverage in 5G is  $0$\,dB~\cite{3GPPTR38913}. Fig.~\ref{fig:outcomes} show the area plots for the probability of the outcomes when both users transmit in the same slot for $\eta \in \{2.6,3\}$. The figure shows that a high reliability for the intermittent user is only achievable when it is close to the base station, particularly when $\eta=3$. On the other hand, recovering packets after decoding the broadband user block is crucial, as case $(\cdot,\mathcal{R})$ occurs with a relatively high probability for both values of $\eta$ and is critical to achieve high reliability for the intermittent user.

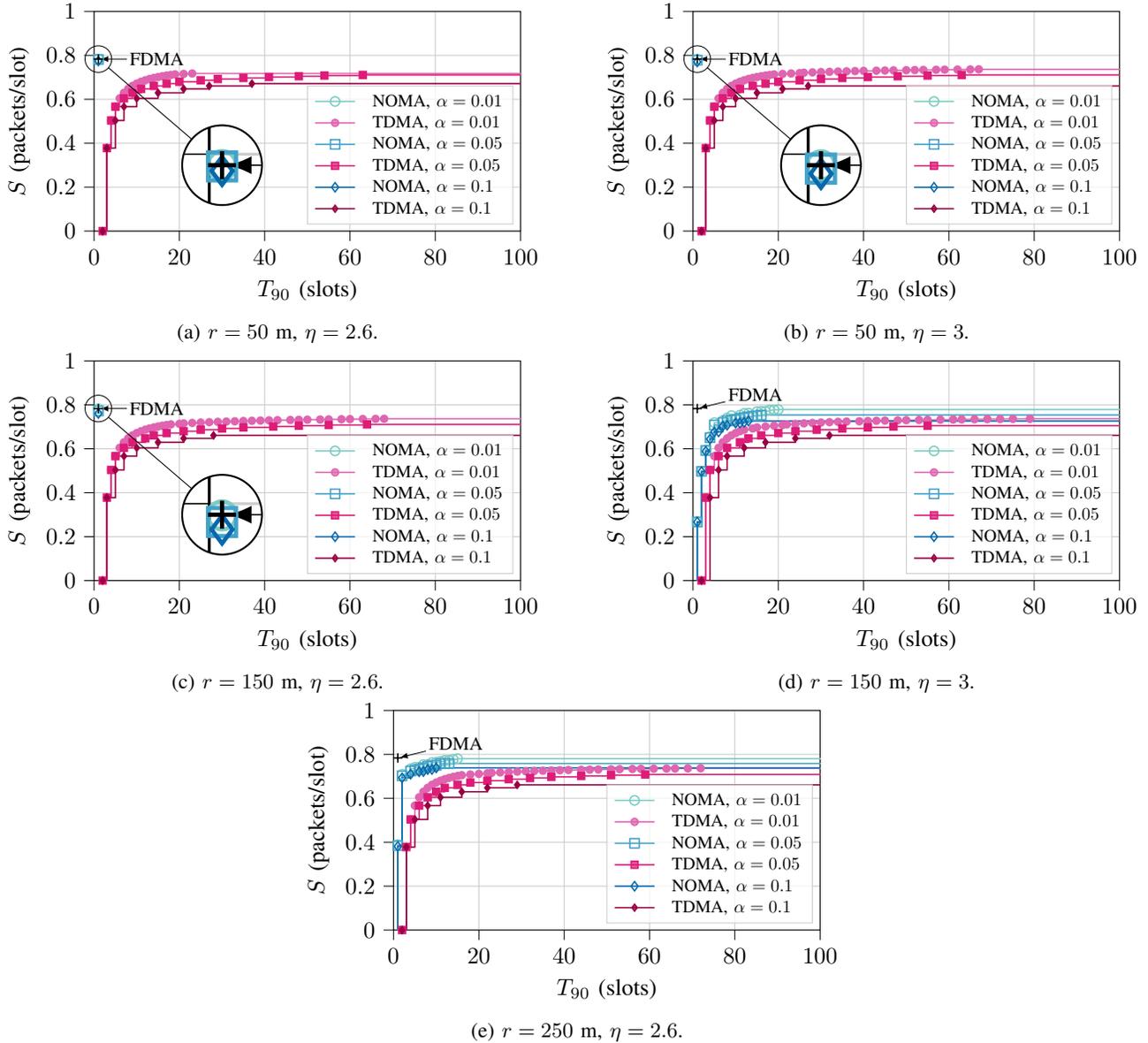
\begin{figure*}
    \centering
	\begin{subfigure}[b]{.49\linewidth}
	    \centering
\pgfdeclarelayer{fg}    
\pgfsetlayers{main,fg}
\begin{tikzpicture}
[spy using outlines={circle, magnification=3, connect spies}]
\begin{axis}[
legend cell align={left},
width=\fwidth,
height=\fheight,
legend style={
  nodes={scale=0.75, transform shape},
  fill opacity=0.75,
  draw opacity=1,
  text opacity=1,
  legend columns=1,
  at={(0.983,0.03)},
  anchor=south east,
  draw=white!80!black,
  column sep=0.2cm
},
xmajorgrids,
ymajorgrids,
tick align=outside,
tick pos=left,
x grid style={white!80!black},
xlabel={$T_{90}$ (slots)},
xmin=0, xmax=100,
xtick style={color=black},
y grid style={white!80!black},
ylabel style={align=center},
ylabel={$S$ (packets/slot)},
ymin=0, ymax=1,
ytick style={color=black}
]
\addplot [semithick, color2, const plot mark left, mark=o, mark options={solid}]
table {%
1 0.783299233952704
};
\addlegendentry{NOMA, $\alpha=0.01$}
\addplot [semithick, color3, const plot mark left, mark=*, mark options={solid}, mark size=1.5pt]
table {%
2 -1
2 0
3 0.377737057213186
4 0.503649409617581
5 0.566605585819778
6 0.604379291541097
7 0.629561762021976
8 0.647549240936889
9 0.661039850123075
10 0.671532546156774
11 0.679926702983734
12 0.686794649478519
13 0.692517938224174
14 0.697360721008958
15 0.70151167768163
16 0.705109173464613
17 0.708256982274723
18 0.711034460636584
19 0.713503330291573
21 0.715712318930246
23 0.717700408705052
201 0.717700408705052
};
\addlegendentry{TDMA, $\alpha=0.01$}
\addplot [semithick, color4, const plot mark left, mark=square, mark options={solid}]
table {%
1 0.781485894266509
};
\addlegendentry{NOMA, $\alpha=0.05$}
\addplot [semithick, color5, const plot mark left, mark=square*, mark options={solid}, mark size=1.5pt]
table {%
2 -1
2 0
3 0.377737057213186
4 0.503649409617581
5 0.566605585819778
7 0.604379291541097
9 0.629561762021976
11 0.647549240936889
14 0.661039850123075
17 0.671532546156774
20 0.679926702983734
25 0.686794649478519
29 0.692517938224174
35 0.697360721008958
41 0.70151167768163
48 0.705109173464613
54 0.708256982274723
63 0.711034460636584
201 0.711034460636584
};
\addlegendentry{TDMA, $\alpha=0.05$}
\addplot [semithick, color6, const plot mark left, mark=diamond, mark options={solid}]
table {%
1 0.774767402993376
};
\addlegendentry{NOMA, $\alpha=0.1$}
\addplot [semithick, color7, const plot mark left, mark=diamond*, mark options={solid}, mark size=1.5pt]
table {%
2 -1
2 0
3 0.377737057213186
5 0.503649409617581
7 0.566605585819778
10 0.604379291541097
15 0.629561762021976
21 0.647549240936889
27 0.661039850123075
37 0.671532546156774
201 0.671532546156774
};
\addlegendentry{TDMA, $\alpha=0.1$}

\addplot [semithick, black, const plot mark left, mark=+, mark options={solid}, mark size=2pt]coordinates{(1,0.7833)}node [inner sep=0pt, pin={[pin distance=0.4cm]0, inner sep=1pt,font=\footnotesize:{FDMA}}]{};
\coordinate (spypoint) at (axis cs:1,0.7833);
  \coordinate (magnifyglass) at (axis cs:30,0.3);
\end{axis}

\spy [black, size=1.2cm] on (spypoint)
   in node[fill=white] at (magnifyglass);
\end{tikzpicture}
        \caption{$r=50$ m, $\eta=2.6$.}
        \label{fig:ld50p26}
    \end{subfigure}	
	\centering
	\begin{subfigure}[b]{.49\linewidth}
	    \centering
\begin{tikzpicture}
[spy using outlines={circle, magnification=3, connect spies}]
\begin{axis}[
legend cell align={left},
width=\fwidth,
height=\fheight,
legend style={
  nodes={scale=0.75, transform shape},
  fill opacity=0.75,
  draw opacity=1,
  text opacity=1,
  legend columns=1,
  at={(0.983,0.03)},
  anchor=south east,
  draw=white!80!black,
  column sep=0.2cm
},
xmajorgrids,
ymajorgrids,
tick align=outside,
tick pos=left,
x grid style={white!80!black},
xlabel={$T_{90}$ (slots)},
xmin=0, xmax=100,
xtick style={color=black},
y grid style={white!80!black},
ylabel style={align=center},
ylabel={$S$ (packets/slot)},
ymin=0, ymax=1,
ytick style={color=black}
]
\addplot [semithick, color2, const plot mark left, mark=o, mark options={solid}]
table {%
1 0.783030369243897
};
\addlegendentry{NOMA, $\alpha=0.01$}
\addplot [semithick, color3, const plot mark left, mark=*, mark options={solid}, mark size=1.5pt]
table {%
2 -1
2 0
3 0.377737057213186
4 0.503649409617581
5 0.566605585819778
6 0.604379291541097
7 0.629561762021976
8 0.647549240936889
9 0.661039850123075
10 0.671532546156774
11 0.679926702983734
12 0.686794649478519
13 0.692517938224174
14 0.697360721008958
15 0.70151167768163
16 0.705109173464613
17 0.708256982274723
18 0.711034460636584
19 0.713503330291573
21 0.715712318930246
23 0.717700408705052
25 0.719499156596544
26 0.721134381952445
28 0.722627413799138
30 0.723996026325272
32 0.725255149849316
34 0.726417417717664
36 0.727493591669839
38 0.728492896054001
40 0.729423282894427
43 0.730291643945492
45 0.73110398170294
47 0.731865548350547
50 0.732580959443754
52 0.733254287531478
54 0.733889139728475
57 0.734488722358972
59 0.73505589511755
62 0.735593216678309
65 0.736102983287233
67 0.736587261565712
201 0.736587261565712
};
\addlegendentry{TDMA, $\alpha=0.01$}
\addplot [semithick, color4, const plot mark left, mark=square, mark options={solid}]
table {%
1 0.778057810938754
};
\addlegendentry{NOMA, $\alpha=0.05$}
\addplot [semithick, color5, const plot mark left, mark=square*, mark options={solid}, mark size=1.5pt]
table {%
2 -1
2 0
3 0.377737057213186
4 0.503649409617581
5 0.566605585819778
7 0.604379291541097
9 0.629561762021976
11 0.647549240936889
14 0.661039850123075
17 0.671532546156774
20 0.679926702983734
25 0.686794649478519
30 0.692517938224174
35 0.697360721008958
41 0.70151167768163
48 0.705109173464613
55 0.708256982274723
63 0.711034460636584
201 0.711034460636584
};
\addlegendentry{TDMA, $\alpha=0.05$}
\addplot [semithick, color6, const plot mark left, mark=diamond, mark options={solid}]
table {%
1 0.770327159705105
};
\addlegendentry{NOMA, $\alpha=0.1$}
\addplot [semithick, color7, const plot mark left, mark=diamond*, mark options={solid}, mark size=1.5pt]
table {%
2 -1
2 0
3 0.377737057213186
5 0.503649409617581
7 0.566605585819778
10 0.604379291541097
15 0.629561762021976
21 0.647549240936889
27 0.661039850123075
201 0.661039850123075
};
\addlegendentry{TDMA, $\alpha=0.1$}
\addplot [semithick, black, const plot mark left, mark=+, mark options={solid}, mark size=2pt]coordinates{(1,0.7833)}node [inner sep=0pt, pin={[pin distance=0.4cm]0, inner sep=1pt,font=\footnotesize:{FDMA}}]{};
\coordinate (spypoint) at (axis cs:1,0.7833);
  \coordinate (magnifyglass) at (axis cs:30,0.3);
\end{axis}

\spy [black, size=1.2cm] on (spypoint)
   in node[fill=white] at (magnifyglass);
\end{tikzpicture}
        \caption{$r=50$ m, $\eta=3$.}
        \label{fig:ld50p3}
    \end{subfigure}	
	\begin{subfigure}[b]{.49\linewidth}
	    \centering
\begin{tikzpicture}
[spy using outlines={circle, magnification=3, connect spies}]
\begin{axis}[
legend cell align={left},
width=\fwidth,
height=\fheight,
legend style={
  nodes={scale=0.75, transform shape},
  fill opacity=0.75,
  draw opacity=1,
  text opacity=1,
  legend columns=1,
  at={(0.983,0.03)},
  anchor=south east,
  draw=white!80!black,
  column sep=0.2cm
},
xmajorgrids,
ymajorgrids,
tick align=outside,
tick pos=left,
x grid style={white!80!black},
xlabel={$T_{90}$ (slots)},
xmin=0, xmax=100,
xtick style={color=black},
y grid style={white!80!black},
ylabel style={align=center},
ylabel={$S$ (packets/slot)},
ymin=0, ymax=1,
ytick style={color=black}
]
\addplot [semithick, color2, const plot mark left, mark=o, mark options={solid}]
table {%
1 0.782230210909973
};
\addlegendentry{NOMA, $\alpha=0.01$}
\addplot [semithick, color3, const plot mark left, mark=*, mark options={solid}, mark size=1.5pt]
table {%
2 -1
2 0
3 0.377737057213186
4 0.503649409617581
5 0.566605585819778
6 0.604379291541097
7 0.629561762021976
8 0.647549240936889
9 0.661039850123075
10 0.671532546156774
11 0.679926702983734
12 0.686794649478519
13 0.692517938224174
14 0.697360721008958
15 0.70151167768163
16 0.705109173464613
17 0.708256982274723
18 0.711034460636584
20 0.713503330291573
21 0.715712318930246
23 0.717700408705052
25 0.719499156596544
27 0.721134381952445
29 0.722627413799138
31 0.723996026325272
33 0.725255149849316
35 0.726417417717664
37 0.727493591669839
39 0.728492896054001
41 0.729423282894427
43 0.730291643945492
46 0.73110398170294
48 0.731865548350547
50 0.732580959443754
53 0.733254287531478
55 0.733889139728475
58 0.734488722358972
60 0.73505589511755
63 0.735593216678309
66 0.736102983287233
68 0.736587261565712
201 0.736587261565712
};
\addlegendentry{TDMA, $\alpha=0.01$}
\addplot [semithick, color4, const plot mark left, mark=square, mark options={solid}]
table {%
1 0.772196936016964
};
\addlegendentry{NOMA, $\alpha=0.05$}
\addplot [semithick, color5, const plot mark left, mark=square*, mark options={solid}, mark size=1.5pt]
table {%
2 -1
2 0
3 0.377737057213186
4 0.503649409617581
5 0.566605585819778
7 0.604379291541097
9 0.629561762021976
12 0.647549240936889
14 0.661039850123075
17 0.671532546156774
21 0.679926702983734
25 0.686794649478519
30 0.692517938224174
35 0.697360721008958
42 0.70151167768163
48 0.705109173464613
55 0.708256982274723
64 0.711034460636584
201 0.711034460636584
};
\addlegendentry{TDMA, $\alpha=0.05$}
\addplot [semithick, color6, const plot mark left, mark=diamond, mark options={solid}]
table {%
1 0.760896327260362
};
\addlegendentry{NOMA, $\alpha=0.1$}
\addplot [semithick, color7, const plot mark left, mark=diamond*, mark options={solid}, mark size=1.5pt]
table {%
2 -1
2 0
3 0.377737057213186
5 0.503649409617581
7 0.566605585819778
10 0.604379291541097
15 0.629561762021976
21 0.647549240936889
28 0.661039850123075
201 0.661039850123075
};
\addlegendentry{TDMA, $\alpha=0.1$}
\addplot [semithick, black, const plot mark left, mark=+, mark options={solid}, mark size=2pt]coordinates{(1,0.7833)}node [inner sep=0pt, pin={[pin distance=0.4cm]0, inner sep=1pt,font=\footnotesize:{FDMA}}]{};
\coordinate (spypoint) at (axis cs:1,0.7833);
  \coordinate (magnifyglass) at (axis cs:30,0.3);
\end{axis}

\spy [black, size=1.2cm] on (spypoint)
   in node[fill=white] at (magnifyglass);

\end{tikzpicture}
        \caption{$r=150$ m, $\eta=2.6$.}
        \label{fig:ld150p26}
    \end{subfigure}	
	\centering
	\begin{subfigure}[b]{.49\linewidth}
	    \centering
\begin{tikzpicture}

\begin{axis}[
legend cell align={left},
width=\fwidth,
height=\fheight,
legend style={
  nodes={scale=0.75, transform shape},
  fill opacity=0.75,
  draw opacity=1,
  text opacity=1,
  legend columns=1,
  at={(0.983,0.03)},
  anchor=south east,
  draw=white!80!black,
  column sep=0.2cm
},
xmajorgrids,
ymajorgrids,
tick align=outside,
tick pos=left,
x grid style={white!80!black},
xlabel={$T_{90}$ (slots)},
xmin=0, xmax=100,
xtick style={color=black},
y grid style={white!80!black},
ylabel style={align=center},
ylabel={$S$ (packets/slot)},
ymin=0, ymax=1,
ytick style={color=black}
]
\addplot [semithick, color2, const plot mark left, mark=o, mark options={solid}]
table {%
1 -1
1 0.269230769230769
2 0.497973793810734
3 0.59438908869768
4 0.655151864226289
5 0.719020566564817
7 0.728678945540877
8 0.7401676533059
9 0.751091492287062
11 0.75305689325747
12 0.761265703511682
13 0.762300168909553
15 0.769474447301091
16 0.771010914108836
17 0.771243234799801
18 0.775960546856772
19 0.777852404286572
20 0.778794699308648
201 0.778794699308648
};
\addlegendentry{NOMA, $\alpha=0.01$}
\addplot [semithick, color3, const plot mark left, mark=*, mark options={solid},mark size=1.5pt]
table {%
2 -1
2 0
3 0.377737057213186
4 0.503649409617581
5 0.566605585819778
6 0.604379291541097
7 0.629561762021976
8 0.647549240936889
9 0.661039850123075
10 0.671532546156774
11 0.679926702983734
12 0.686794649478519
14 0.692517938224174
15 0.697360721008958
17 0.70151167768163
19 0.705109173464613
21 0.708256982274723
22 0.711034460636584
24 0.713503330291573
26 0.715712318930246
28 0.717700408705052
30 0.719499156596544
32 0.721134381952445
34 0.722627413799138
36 0.723996026325272
39 0.725255149849316
41 0.726417417717664
43 0.727493591669839
45 0.728492896054001
48 0.729423282894427
50 0.730291643945492
53 0.73110398170294
55 0.731865548350547
58 0.732580959443754
61 0.733254287531478
63 0.733889139728475
66 0.734488722358972
69 0.73505589511755
72 0.735593216678309
75 0.736102983287233
79 0.736587261565712
201 0.736587261565712
};
\addlegendentry{TDMA, $\alpha=0.01$}
\addplot [semithick, color4, const plot mark left, mark=square, mark options={solid}]
table {%
1 -1
1 0.266666666162072
2 0.497161386517392
3 0.592219307037086
4 0.650859151830712
5 0.708302392230444
7 0.718905099787664
8 0.727277014935516
9 0.731048846455852
10 0.734834781418467
11 0.739605017781982
12 0.741737010777233
13 0.745701962086086
14 0.749097316002941
15 0.750749846487905
16 0.753973920608957
201 0.753973920608957
};
\addlegendentry{NOMA, $\alpha=0.05$}
\addplot [semithick, color5, const plot mark left, mark=square*, mark options={solid},mark size=1.5pt]
table {%
2 -1
2 0
3 0.377737057213186
4 0.503649409617581
6 0.566605585819778
8 0.604379291541097
11 0.629561762021976
13 0.647549240936889
16 0.661039850123075
20 0.671532546156774
24 0.679926702983734
29 0.686794649478519
35 0.692517938224174
41 0.697360721008958
47 0.70151167768163
55 0.705109173464613
201 0.705109173464613
};
\addlegendentry{TDMA, $\alpha=0.05$}
\addplot [semithick, color6, const plot mark left, mark=diamond, mark options={solid}]
table {%
1 -1
1 0.266666664431417
2 0.495884882634547
3 0.588864922228406
4 0.644329626044103
5 0.67644716481798
6 0.677135633218685
7 0.703409293706051
8 0.707168547559945
10 0.716813437531187
11 0.717376535964425
12 0.720696742570569
13 0.726463937056467
201 0.726463937056467
};
\addlegendentry{NOMA, $\alpha=0.1$}
\addplot [semithick, color7, const plot mark left, mark=diamond*, mark options={solid}, mark size=1.5pt]
table {%
2 -1
2 0
4 0.377737057213186
6 0.503649409617581
8 0.566605585819778
12 0.604379291541097
17 0.629561762021976
24 0.647549240936889
32 0.661039850123075
201 0.661039850123075
};
\addlegendentry{TDMA, $\alpha=0.1$}
\addplot [semithick, black, const plot mark left, mark=+, mark options={solid}, mark size=2pt]coordinates{(1,0.7833)}node [inner sep=0pt, pin={[pin distance=0.4cm]10, inner sep=1pt,font=\footnotesize:{FDMA}}]{};
\end{axis}

\end{tikzpicture}
        \caption{$r=150$ m, $\eta=3$.}
        \label{fig:ld150p3}
    \end{subfigure}	
    	\begin{subfigure}[b]{.49\linewidth}
	    \centering
\begin{tikzpicture}

\begin{axis}[
legend cell align={left},
width=\fwidth,
height=\fheight,
legend style={
  nodes={scale=0.75, transform shape},
  fill opacity=0.75,
  draw opacity=1,
  text opacity=1,
  legend columns=1,
  at={(0.983,0.03)},
  anchor=south east,
  draw=white!80!black,
  column sep=0.2cm
},
xmajorgrids,
ymajorgrids,
tick align=outside,
tick pos=left,
x grid style={white!80!black},
xlabel={$T_{90}$ (slots)},
xmin=0, xmax=100,
xtick style={color=black},
y grid style={white!80!black},
ylabel style={align=center},
ylabel={$S$ (packets/slot)},
ymin=0, ymax=1,
ytick style={color=black}
]
\addplot [semithick, color2, const plot mark left, mark=o, mark options={solid}]
table {%
1 -1
1 0.388888883679749
2 0.709117108616189
4 0.73419363549823
5 0.740924500899643
6 0.740988251049842
7 0.753083027537055
8 0.753884379813612
9 0.762505706635543
10 0.763777892900974
11 0.767533246501587
12 0.772343693769446
13 0.772794361412405
14 0.77697105684114
15 0.780160844719925
201 0.780160844719925
};
\addlegendentry{NOMA, $\alpha=0.01$}
\addplot [semithick, color3, const plot mark left, mark=*, mark options={solid}, mark size=1.5pt]
table {%
2 -1
2 0
3 0.377737057213186
4 0.503649409617581
5 0.566605585819778
6 0.604379291541097
7 0.629561762021976
8 0.647549240936889
9 0.661039850123075
10 0.671532546156774
11 0.679926702983734
12 0.686794649478519
13 0.692517938224174
14 0.697360721008958
15 0.70151167768163
16 0.705109173464613
18 0.708256982274723
20 0.711034460636584
22 0.713503330291573
23 0.715712318930246
25 0.717700408705052
27 0.719499156596544
29 0.721134381952445
31 0.722627413799138
33 0.723996026325272
35 0.725255149849316
37 0.726417417717664
39 0.727493591669839
41 0.728492896054001
44 0.729423282894427
46 0.730291643945492
48 0.73110398170294
51 0.731865548350547
53 0.732580959443754
56 0.733254287531478
58 0.733889139728475
61 0.734488722358972
64 0.73505589511755
66 0.735593216678309
69 0.736102983287233
72 0.736587261565712
201 0.736587261565712
};
\addlegendentry{TDMA, $\alpha=0.01$}
\addplot [semithick, color4, const plot mark left, mark=square, mark options={solid}]
table {%
1 -1
1 0.384615384588151
2 0.702204323747581
4 0.723961728024747
5 0.727270865526191
6 0.732246942970712
7 0.738559484815334
8 0.744883078363555
9 0.749143142851978
10 0.750036946521076
11 0.757044642124751
12 0.758151520090946
13 0.758222562711433
201 0.758222562711433
};
\addlegendentry{NOMA, $\alpha=0.05$}
\addplot [semithick, color5, const plot mark left, mark=square*, mark options={solid}, mark size=1.5pt]
table {%
2 -1
2 0
3 0.377737057213186
4 0.503649409617581
6 0.566605585819778
8 0.604379291541097
10 0.629561762021976
12 0.647549240936889
15 0.661039850123075
18 0.671532546156774
22 0.679926702983734
27 0.686794649478519
32 0.692517938224174
37 0.697360721008958
44 0.70151167768163
51 0.705109173464613
59 0.708256982274723
201 0.708256982274723
};
\addlegendentry{TDMA, $\alpha=0.05$}
\addplot [semithick, color6, const plot mark left, mark=diamond, mark options={solid}]
table {%
1 -1
1 0.380952377686781
2 0.692956479733128
4 0.709643237525829
6 0.721006144515513
7 0.721709746613257
8 0.730940226742546
9 0.731224350429976
10 0.738372928191143
201 0.738372928191143
};
\addlegendentry{NOMA, $\alpha=0.1$}
\addplot [semithick, color7, const plot mark left, mark=diamond*, mark options={solid}, mark size=1.5pt]
table {%
2 -1
2 0
3 0.377737057213186
5 0.503649409617581
8 0.566605585819778
11 0.604379291541097
16 0.629561762021976
22 0.647549240936889
29 0.661039850123075
201 0.661039850123075
};
\addlegendentry{TDMA, $\alpha=0.1$}
\addplot [semithick, black, const plot mark left, mark=+, mark options={solid}, mark size=2pt]coordinates{(1,0.7833)}node [inner sep=0pt, pin={[pin distance=0.4cm]10, inner sep=1pt, font=\footnotesize:{FDMA}}]{};
\end{axis}

\end{tikzpicture}
        \caption{$r=250$ m, $\eta=2.6$.}
        \label{fig:ld250p26}
    \end{subfigure}	
    \caption{Pareto frontiers for latency-reliability versus throughput with \gls{tdma} and \gls{noma}, with different values of $\alpha$. The cross marks indicate the performance with \gls{fdma} (benchmark).}
 \label{fig:latency}
\end{figure*}

An essential aspect of our analysis is to identify the values of $K$ and $N$ that maximize the throughput $S$ of user 1. These can be selected independently of user 2's parameters for \gls{tdma} and \gls{fdma} and, hence, represent the optimal configuration for user 1 with these schemes.

Note that implementing a longer coded block size $N$ would grant a greater throughput, bounded by $1-\varepsilon_1$ for $N\rightarrow\infty$, but would also lead to a longer decoding latency and complexity.
Hence, we limit the value of $N\leq 32$ to achieve an adequate balance between $S$ and decoding latency and complexity.
By restricting $N\leq32$, the optimal configuration for user 1 for both \gls{tdma} and \gls{fdma} is $K = 26$ and $N = 32$, which leads to $p_{s,1}=0.964$. With this configuration, \gls{fdma} achieves a throughput of $S=0.7833$ for all cases, as user 1 operates in a separate channel from user 2 and, hence, there is no trade-off between $S$ and the \gls{kpi} of user 2. On the other hand, the optimal configuration for \gls{tdma} and \gls{noma} depends on the desired performance trade-off and, hence, these are given at the end of this section.


\subsection{Pareto analysis}

We first present the Pareto frontier for throughput of user 1 $S$ and timing of user 2, for \gls{lr} $T_{90}$ or \gls{paoi} $\Delta_{90}$, which describes the best achievable trade-offs between these \glspl{kpi}. 
We consider three different distances (50, 150, and 250~m) for the intermittent user, with three different activation probabilities.
It is easy to see in Fig.~\ref{fig:latency} that \gls{noma} easily outperforms \gls{tdma} in terms of \gls{lr} and throughput in all scenarios.

Furthermore, \mbox{Fig.~\ref{fig:ld50p26}-\subref{fig:ld150p26}} show that $T_{90}=1$ can be achieved with \gls{noma} if the distance and path loss allow to immediately decode more than $90$\% of the packets from user 2 due to capture and the use of \gls{sic} in the same slot. In these cases, the throughput with \gls{noma} is only up to $2$\% lower than with \gls{fdma}. Therefore, \gls{noma} is the most efficient access scheme in these cases, as it achieves a similar performance to \gls{fdma} but with half the resources: one bandwidth part instead of two. 

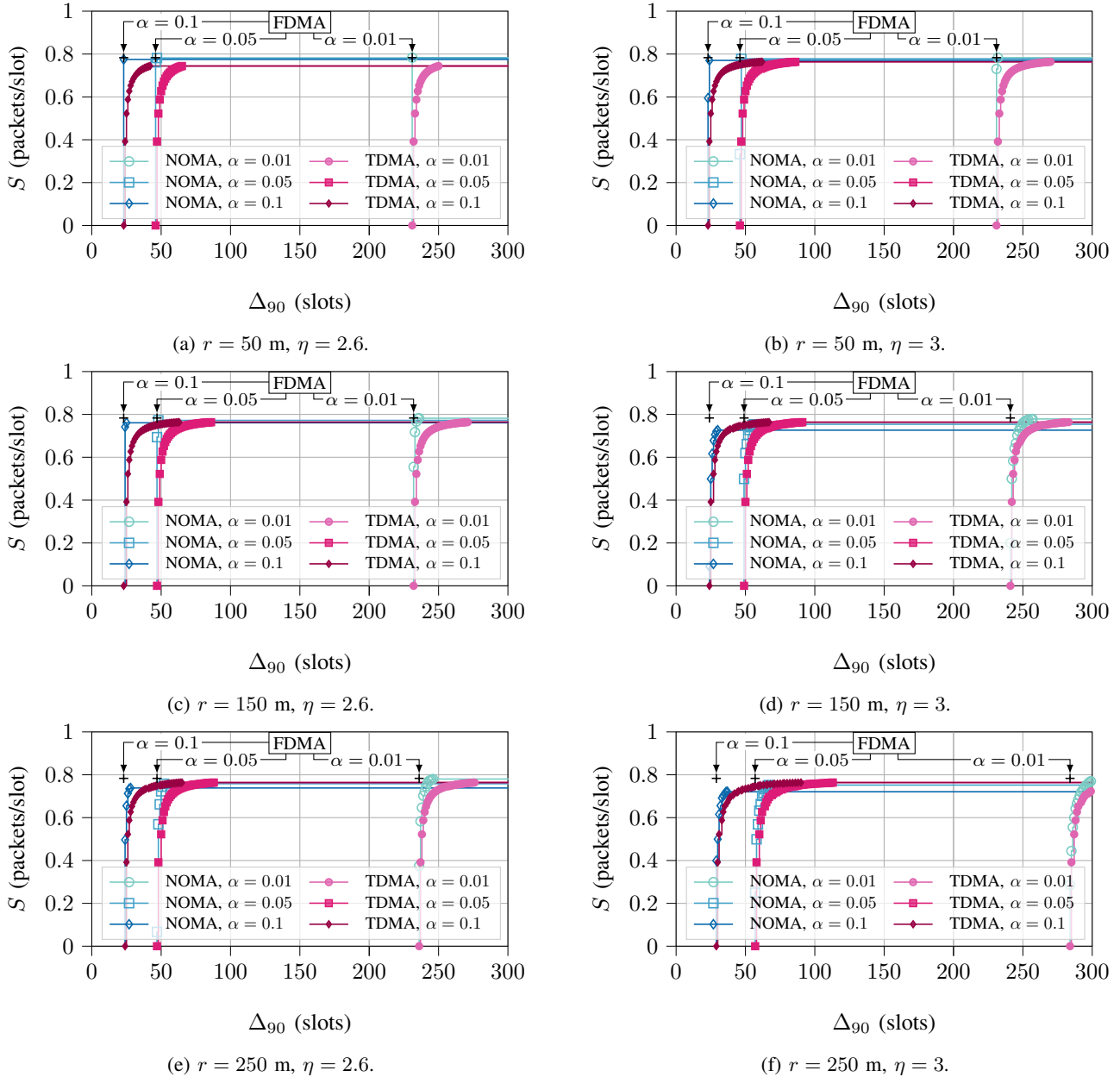
\begin{figure*}
    \centering
	\begin{subfigure}[b]{.49\linewidth}
	    \centering
\begin{tikzpicture}

\begin{axis}[
legend cell align={left},
width=\fwidth,
height=\fheight,
legend style={
  nodes={scale=0.75, transform shape},
  fill opacity=0.75,
  draw opacity=1,
  text opacity=1,
  legend columns=2,
  at={(0.983,0.03)},
  anchor=south east,
  draw=white!80!black,
  column sep=0.2cm
},
tick align=outside,
tick pos=left,
x grid style={white!69.0196078431373!black},
xlabel={$\Delta_{90}$ (slots)},
xmin=0, xmax=300,
ymajorgrids,
xmajorgrids,
xminorgrids,
xtick style={color=black},
extra x ticks={200},
extra x tick labels={},
y grid style={white!69.0196078431373!black},
ylabel style={align=center},
ylabel={$S$ (packets/slot)},
ymin=0, ymax=1,
ytick style={color=black}
]
\addplot [semithick, color2, const plot mark left, mark=o, mark options={solid}]
table {%
231 -1
231 0.783299233952704
1100 0.783299233952704
};
\addlegendentry{NOMA, $\alpha=0.01$}
\addplot [semithick, color3, const plot mark left, mark=*, mark options={solid}, mark size=1.5pt]
table {%
231 -1
231 0
232 0.391686397733124
233 0.522248530310832
234 0.587529596599686
235 0.626698236372998
236 0.65281066288854
237 0.671462396113927
238 0.685451196032967
239 0.696331373747776
240 0.705035515919623
241 0.712157086787498
242 0.718091729177394
243 0.723113349661152
244 0.727417595790087
245 0.731147942435165
246 0.734411995749607
247 0.737292042791763
248 0.739852084607012
249 0.742142648336445
250 0.744204155692935
1100 0.744204155692935
};
\addlegendentry{TDMA, $\alpha=0.01$}
\addplot [semithick, color4, const plot mark left, mark=square, mark options={solid}]
table {%
46 -1
46 0.761407121215056
47 0.781485894266509
1100 0.781485894266509
};
\addlegendentry{NOMA, $\alpha=0.05$}
\addplot [semithick, color5, const plot mark left, mark=square*, mark options={solid}, mark size=1.5pt]
table {%
46 -1
46 0
47 0.391686397733124
48 0.522248530310832
49 0.587529596599686
50 0.626698236372998
51 0.65281066288854
52 0.671462396113927
53 0.685451196032967
54 0.696331373747776
55 0.705035515919623
56 0.712157086787498
57 0.718091729177394
58 0.723113349661152
59 0.727417595790087
60 0.731147942435165
61 0.734411995749607
62 0.737292042791763
63 0.739852084607012
64 0.742142648336445
65 0.744204155692935
1100 0.744204155692935
};
\addlegendentry{TDMA, $\alpha=0.05$}
\addplot [semithick, color6, const plot mark left, mark=diamond, mark options={solid}]
table {%
23 -1
23 0.774767402993376
1100 0.774767402993376
};
\addlegendentry{NOMA, $\alpha=0.1$}
\addplot [semithick, color7, const plot mark left, mark=diamond*, mark options={solid}, mark size=1.5pt]
table {%
23 -1
23 0
24 0.391686397733124
25 0.522248530310832
26 0.587529596599686
27 0.626698236372998
28 0.65281066288854
29 0.671462396113927
30 0.685451196032967
31 0.696331373747776
32 0.705035515919623
33 0.712157086787498
34 0.718091729177394
35 0.723113349661152
36 0.727417595790087
37 0.731147942435165
38 0.734411995749607
39 0.737292042791763
40 0.739852084607012
41 0.742142648336445
42 0.744204155692935
1100 0.744204155692935
};
\addlegendentry{TDMA, $\alpha=0.1$}

\addplot [semithick, black, const plot mark left, mark=+, mark options={solid}, mark size=2pt]coordinates{(23,0.7833)};
\addplot [semithick, black, const plot mark left, mark=+, mark options={solid}, mark size=2pt]coordinates{(46,0.7833)};
\addplot [semithick, black, const plot mark left, mark=+, mark options={solid}, mark size=2pt]coordinates{(231,0.7833)};
\draw[->, yshift=2pt] (150,0.93)--(23,0.93)node[near end, fill=white,font=\footnotesize, inner sep=1pt]{$\alpha=0.1$}--(23,0.7833);
\draw[->, yshift=2pt](140,0.93)-- (140,0.85)--(46,0.85)node[midway, fill=white,font=\footnotesize, inner sep=1pt]{$\alpha=0.05$}--(46,0.7833);
\draw[->, yshift=2pt](160,0.93)-- (160,0.85)--(231,0.85)node[very near end, anchor=east, fill=white,font=\footnotesize, inner sep=1pt]{$\alpha=0.01$}--(231,0.7833);
\node[draw=black, fill=white, inner sep=2pt, font=\footnotesize, anchor=north] at (150,1){FDMA};
\end{axis}

\end{tikzpicture}
        \caption{$r=50$ m, $\eta=2.6$.}
        \label{fig:ad50p26}
    \end{subfigure}	
	\centering
	\begin{subfigure}[b]{.49\linewidth}
	    \centering
\begin{tikzpicture}

\begin{axis}[
legend cell align={left},
width=\fwidth,
height=\fheight,
legend style={
  nodes={scale=0.75, transform shape},
  fill opacity=0.75,
  draw opacity=1,
  text opacity=1,
  legend columns=2,
  at={(0.983,0.03)},
  anchor=south east,
  draw=white!80!black,
  column sep=0.2cm
},
tick align=outside,
tick pos=left,
x grid style={white!69.0196078431373!black},
xlabel={$\Delta_{90}$ (slots)},
xmin=0, xmax=300,
ymajorgrids,
xmajorgrids,
xminorgrids,
xtick style={color=black},
extra x ticks={200},
extra x tick labels={},
y grid style={white!69.0196078431373!black},
ylabel style={align=center},
ylabel={$S$ (packets/slot)},
ymin=0, ymax=1,
ytick style={color=black}
]
\addplot [semithick, color2, const plot mark left, mark=o, mark options={solid}]
table {%
231 -1
231 0.730434403277037
232 0.783030369243897
1100 0.783030369243897
};
\addlegendentry{NOMA, $\alpha=0.01$}
\addplot [semithick, color3, const plot mark left, mark=*, mark options={solid}, mark size=1.5pt]
table {%
231 -1
231 0
232 0.391686397733124
233 0.522248530310832
234 0.587529596599686
235 0.626698236372998
236 0.65281066288854
237 0.671462396113927
238 0.685451196032967
239 0.696331373747776
240 0.705035515919623
241 0.712157086787498
242 0.718091729177394
243 0.723113349661152
244 0.727417595790087
245 0.731147942435165
246 0.734411995749607
247 0.737292042791763
248 0.739852084607012
249 0.742142648336445
250 0.744204155692935
251 0.746069329015474
252 0.747764941126873
253 0.749313108706846
254 0.750732262321821
255 0.752037883647598
256 0.7532430725637
257 0.754358988226757
258 0.755395195628167
259 0.75635994045017
260 0.757260368950706
261 0.758102705289917
262 0.758892395607927
263 0.759634225906664
264 0.760332419129005
265 0.760990715595783
266 0.76161244003663
267 0.762200557750944
268 0.762757721901346
269 0.763286313531216
270 0.763788475579592
1100 0.763788475579592
};
\addlegendentry{TDMA, $\alpha=0.01$}
\addplot [semithick, color4, const plot mark left, mark=square, mark options={solid}]
table {%
46 -1
46 0.332987436796696
47 0.778057810938754
1100 0.778057810938754
};
\addlegendentry{NOMA, $\alpha=0.05$}
\addplot [semithick, color5, const plot mark left, mark=square*, mark options={solid}, mark size=1.5pt]
table {%
46 -1
46 0
47 0.391686397733124
48 0.522248530310832
49 0.587529596599686
50 0.626698236372998
51 0.65281066288854
52 0.671462396113927
53 0.685451196032967
54 0.696331373747776
55 0.705035515919623
56 0.712157086787498
57 0.718091729177394
58 0.723113349661152
59 0.727417595790087
61 0.731147942435165
62 0.734411995749607
63 0.737292042791763
64 0.739852084607012
65 0.742142648336445
66 0.744204155692935
67 0.746069329015474
68 0.747764941126873
69 0.749313108706846
70 0.750732262321821
71 0.752037883647598
72 0.7532430725637
73 0.754358988226757
74 0.755395195628167
75 0.75635994045017
76 0.757260368950706
77 0.758102705289917
78 0.758892395607927
79 0.759634225906664
80 0.760332419129005
81 0.760990715595783
82 0.76161244003663
83 0.762200557750944
84 0.762757721901346
85 0.763286313531216
86 0.763788475579592
1100 0.763788475579592
};
\addlegendentry{TDMA, $\alpha=0.05$}
\addplot [semithick, color6, const plot mark left, mark=diamond, mark options={solid}]
table {%
23 -1
23 0.594494141537652
24 0.770327159705105
1100 0.770327159705105
};
\addlegendentry{NOMA, $\alpha=0.1$}
\addplot [semithick, color7, const plot mark left, mark=diamond*, mark options={solid}, mark size=1.5pt]
table {%
23 -1
23 0
24 0.391686397733124
25 0.522248530310832
26 0.587529596599686
27 0.626698236372998
28 0.65281066288854
29 0.671462396113927
30 0.685451196032967
31 0.696331373747776
32 0.705035515919623
33 0.712157086787498
34 0.718091729177394
35 0.723113349661152
36 0.727417595790087
37 0.731147942435165
38 0.734411995749607
39 0.737292042791763
40 0.739852084607012
41 0.742142648336445
42 0.744204155692935
43 0.746069329015474
44 0.747764941126873
45 0.749313108706846
46 0.750732262321821
47 0.752037883647598
48 0.7532430725637
49 0.754358988226757
50 0.755395195628167
51 0.75635994045017
52 0.757260368950706
53 0.758102705289917
54 0.758892395607927
55 0.759634225906664
56 0.760332419129005
57 0.760990715595783
58 0.76161244003663
59 0.762200557750944
60 0.762757721901346
61 0.763286313531216
62 0.763788475579592
1100 0.763788475579592
};
\addlegendentry{TDMA, $\alpha=0.1$}
\addplot [semithick, black, const plot mark left, mark=+, mark options={solid}, mark size=2pt]coordinates{(23,0.7833)};
\addplot [semithick, black, const plot mark left, mark=+, mark options={solid}, mark size=2pt]coordinates{(46,0.7833)};
\addplot [semithick, black, const plot mark left, mark=+, mark options={solid}, mark size=2pt]coordinates{(231,0.7833)};

\draw[->, yshift=2pt] (150,0.93)--(23,0.93)node[near end, fill=white,font=\footnotesize, inner sep=1pt]{$\alpha=0.1$}--(23,0.7833);
\draw[->, yshift=2pt](140,0.93)-- (140,0.85)--(46,0.85)node[midway, fill=white,font=\footnotesize, inner sep=1pt]{$\alpha=0.05$}--(46,0.7833);
\draw[->, yshift=2pt](160,0.93)-- (160,0.85)--(231,0.85)node[very near end, anchor=east, fill=white,font=\footnotesize, inner sep=1pt]{$\alpha=0.01$}--(231,0.7833);
\node[draw=black, fill=white, inner sep=2pt, font=\footnotesize, anchor=north] at (150,1){FDMA};
\end{axis}

\end{tikzpicture}
        \caption{$r=50$ m, $\eta=3$.}
        \label{fig:ad50p3}
    \end{subfigure}	
	\begin{subfigure}[b]{.49\linewidth}
	    \centering
\begin{tikzpicture}

\begin{axis}[
legend cell align={left},
width=\fwidth,
height=\fheight,
legend style={
  nodes={scale=0.75, transform shape},
  fill opacity=0.75,
  draw opacity=1,
  text opacity=1,
  legend columns=2,
  at={(0.983,0.03)},
  anchor=south east,
  draw=white!80!black,
  column sep=0.2cm
},
tick align=outside,
tick pos=left,
x grid style={white!69.0196078431373!black},
xlabel={$\Delta_{90}$ (slots)},
xmin=0, xmax=300,
ymajorgrids,
xmajorgrids,
xminorgrids,
xtick style={color=black},
extra x ticks={200},
extra x tick labels={},
y grid style={white!69.0196078431373!black},
ylabel style={align=center},
ylabel={$S$ (packets/slot)},
ymin=0, ymax=1,
ytick style={color=black}
]
\addplot [semithick, color2, const plot mark left, mark=o, mark options={solid}]
table {%
232 -1
232 0.555041481472368
233 0.717399817932802
234 0.76640878972901
235 0.780819040440188
236 0.782230210909973
1100 0.782230210909973
};
\addlegendentry{NOMA, $\alpha=0.01$}
\addplot [semithick, color3, const plot mark left, mark=*, mark options={solid}, mark size=1.5pt]
table {%
232 -1
232 0
233 0.391686397733124
234 0.522248530310832
235 0.587529596599686
236 0.626698236372998
237 0.65281066288854
238 0.671462396113927
239 0.685451196032967
240 0.696331373747776
241 0.705035515919623
242 0.712157086787498
243 0.718091729177394
244 0.723113349661152
245 0.727417595790087
246 0.731147942435165
247 0.734411995749607
248 0.737292042791763
249 0.739852084607012
250 0.742142648336445
251 0.744204155692935
252 0.746069329015474
253 0.747764941126873
254 0.749313108706846
255 0.750732262321821
256 0.752037883647598
257 0.7532430725637
258 0.754358988226757
259 0.755395195628167
260 0.75635994045017
261 0.757260368950706
262 0.758102705289917
263 0.758892395607927
264 0.759634225906664
265 0.760332419129005
266 0.760990715595783
267 0.76161244003663
268 0.762200557750944
269 0.762757721901346
270 0.763286313531216
271 0.763788475579592
1100 0.763788475579592
};
\addlegendentry{TDMA, $\alpha=0.01$}
\addplot [semithick, color4, const plot mark left, mark=square, mark options={solid}]
table {%
47 -1
47 0.693851255923948
48 0.772196936016964
1100 0.772196936016964
};
\addlegendentry{NOMA, $\alpha=0.05$}
\addplot [semithick, color5, const plot mark left, mark=square*, mark options={solid}, mark size=1.5pt]
table {%
47 -1
47 0
48 0.391686397733124
49 0.522248530310832
50 0.587529596599686
51 0.626698236372998
52 0.65281066288854
53 0.671462396113927
54 0.685451196032967
55 0.696331373747776
56 0.705035515919623
57 0.712157086787498
58 0.718091729177394
59 0.723113349661152
60 0.727417595790087
61 0.731147942435165
62 0.734411995749607
63 0.737292042791763
64 0.739852084607012
65 0.742142648336445
66 0.744204155692935
67 0.746069329015474
68 0.747764941126873
69 0.749313108706846
70 0.750732262321821
71 0.752037883647598
72 0.7532430725637
73 0.754358988226757
74 0.755395195628167
75 0.75635994045017
76 0.757260368950706
77 0.758102705289917
78 0.758892395607927
79 0.759634225906664
80 0.760332419129005
81 0.760990715595783
82 0.76161244003663
83 0.762200557750944
84 0.762757721901346
85 0.763286313531216
86 0.763788475579592
1100 0.763788475579592
};
\addlegendentry{TDMA, $\alpha=0.05$}
\addplot [semithick, color6, const plot mark left, mark=diamond, mark options={solid}]
table {%
24 -1
24 0.741671915190466
25 0.760896327260362
1100 0.760896327260362
};
\addlegendentry{NOMA, $\alpha=0.1$}
\addplot [semithick, color7, const plot mark left, mark=diamond*, mark options={solid}, mark size=1.5pt]
table {%
23 -1
23 0
25 0.391686397733124
26 0.522248530310832
27 0.587529596599686
28 0.626698236372998
29 0.65281066288854
30 0.671462396113927
31 0.685451196032967
32 0.696331373747776
33 0.705035515919623
34 0.712157086787498
35 0.718091729177394
36 0.723113349661152
37 0.727417595790087
38 0.731147942435165
39 0.734411995749607
40 0.737292042791763
41 0.739852084607012
42 0.742142648336445
43 0.744204155692935
44 0.746069329015474
45 0.747764941126873
46 0.749313108706846
47 0.750732262321821
48 0.752037883647598
49 0.7532430725637
50 0.754358988226757
51 0.755395195628167
52 0.75635994045017
53 0.757260368950706
54 0.758102705289917
55 0.758892395607927
56 0.759634225906664
57 0.760332419129005
58 0.760990715595783
59 0.76161244003663
60 0.762200557750944
61 0.762757721901346
62 0.763286313531216
63 0.763788475579592
1100 0.763788475579592
};
\addlegendentry{TDMA, $\alpha=0.1$}
\addplot [semithick, black, const plot mark left, mark=+, mark options={solid}, mark size=2pt]coordinates{(23,0.7833)};
\addplot [semithick, black, const plot mark left, mark=+, mark options={solid}, mark size=2pt]coordinates{(47,0.7833)};
\addplot [semithick, black, const plot mark left, mark=+, mark options={solid}, mark size=2pt]coordinates{(232,0.7833)};

\draw[->, yshift=2pt] (150,0.93)--(23,0.93)node[near end, fill=white,font=\footnotesize, inner sep=1pt]{$\alpha=0.1$}--(23,0.7833);
\draw[->, yshift=2pt](140,0.93)-- (140,0.85)--(47,0.85)node[midway, fill=white,font=\footnotesize, inner sep=1pt]{$\alpha=0.05$}--(47,0.7833);
\draw[->, yshift=2pt](160,0.93)-- (160,0.85)--(232,0.85)node[very near end, anchor=east, fill=white,font=\footnotesize, inner sep=1pt]{$\alpha=0.01$}--(232,0.7833);
\node[draw=black, fill=white, inner sep=2pt, font=\footnotesize, anchor=north] at (150,1){FDMA};
\end{axis}

\end{tikzpicture}
        \caption{$r=150$ m, $\eta=2.6$.}
        \label{fig:ad150p26}
    \end{subfigure}	
	\centering
	\begin{subfigure}[b]{.49\linewidth}
	    \centering
\begin{tikzpicture}

\begin{axis}[
legend cell align={left},
width=\fwidth,
height=\fheight,
legend style={
  nodes={scale=0.75, transform shape},
  fill opacity=0.75,
  draw opacity=1,
  text opacity=1,
  legend columns=2,
  at={(0.983,0.03)},
  anchor=south east,
  draw=white!80!black,
  column sep=0.2cm
},
tick align=outside,
tick pos=left,
x grid style={white!69.0196078431373!black},
xlabel={$\Delta_{90}$ (slots)},
xmin=0, xmax=300,
ymajorgrids,
xmajorgrids,
xminorgrids,
xtick style={color=black},
extra x ticks={200},
extra x tick labels={},
y grid style={white!69.0196078431373!black},
ylabel style={align=center},
ylabel={$S$ (packets/slot)},
ymin=0, ymax=1,
ytick style={color=black}
]
\addplot [semithick, color2, const plot mark left, mark=o, mark options={solid}]
table {%
241 -1
241 0.199999997597879
242 0.499749918807375
243 0.582959312884089
244 0.641741003555561
245 0.665726039090688
246 0.702034679335965
247 0.72347368774895
248 0.737264937487078
249 0.745512339992716
250 0.759496149311342
251 0.765299572991503
252 0.770342586742156
253 0.773101267775198
254 0.775960546856772
256 0.777852404286572
257 0.778794699308648
1100 0.778794699308648
};
\addlegendentry{NOMA, $\alpha=0.01$}
\addplot [semithick, color3, const plot mark left, mark=*, mark options={solid}, mark size=1.5pt]
table {%
241 -1
241 0
242 0.391686397733124
243 0.522248530310832
244 0.587529596599686
245 0.626698236372998
246 0.65281066288854
248 0.671462396113927
249 0.685451196032967
250 0.696331373747776
251 0.705035515919623
252 0.712157086787498
253 0.718091729177394
254 0.723113349661152
255 0.727417595790087
256 0.731147942435165
257 0.734411995749607
258 0.737292042791763
259 0.739852084607012
260 0.742142648336445
261 0.744204155692935
262 0.746069329015474
263 0.747764941126873
264 0.749313108706846
266 0.750732262321821
267 0.752037883647598
268 0.7532430725637
269 0.754358988226757
270 0.755395195628167
271 0.75635994045017
272 0.757260368950706
273 0.758102705289917
274 0.758892395607927
275 0.759634225906664
276 0.760332419129005
277 0.760990715595783
278 0.76161244003663
279 0.762200557750944
280 0.762757721901346
281 0.763286313531216
283 0.763788475579592
1100 0.763788475579592
};
\addlegendentry{TDMA, $\alpha=0.01$}
\addplot [semithick, color4, const plot mark left, mark=square, mark options={solid}]
table {%
49 -1
49 0.49888781452211
50 0.619647166549726
51 0.661217269476682
52 0.70518937462193
53 0.727201102545928
54 0.74048254758966
55 0.749913277821643
56 0.753973920608957
1100 0.753973920608957
};
\addlegendentry{NOMA, $\alpha=0.05$}
\addplot [semithick, color5, const plot mark left, mark=square*, mark options={solid}, mark size=1.5pt]
table {%
49 -1
49 0
50 0.391686397733124
51 0.522248530310832
52 0.587529596599686
53 0.626698236372998
54 0.65281066288854
55 0.671462396113927
56 0.685451196032967
57 0.696331373747776
58 0.705035515919623
59 0.712157086787498
60 0.718091729177394
61 0.723113349661152
62 0.727417595790087
63 0.731147942435165
65 0.734411995749607
66 0.737292042791763
67 0.739852084607012
68 0.742142648336445
69 0.744204155692935
70 0.746069329015474
71 0.747764941126873
72 0.749313108706846
73 0.750732262321821
74 0.752037883647598
75 0.7532430725637
77 0.754358988226757
78 0.755395195628167
79 0.75635994045017
80 0.757260368950706
81 0.758102705289917
82 0.758892395607927
83 0.759634225906664
84 0.760332419129005
85 0.760990715595783
86 0.76161244003663
88 0.762200557750944
89 0.762757721901346
90 0.763286313531216
91 0.763788475579592
1100 0.763788475579592
};
\addlegendentry{TDMA, $\alpha=0.05$}
\addplot [semithick, color6, const plot mark left, mark=diamond, mark options={solid}]
table {%
24 -1
24 0.0909090908903347
25 0.499216911483954
26 0.616544256679387
27 0.678211572309266
28 0.705496621577429
29 0.720696742570569
30 0.726463937056467
1100 0.726463937056467
};
\addlegendentry{NOMA, $\alpha=0.1$}
\addplot [semithick, color7, const plot mark left, mark=diamond*, mark options={solid}, mark size=1.5pt]
table {%
24 -1
24 0
25 0.391686397733124
27 0.522248530310832
28 0.587529596599686
29 0.626698236372998
30 0.65281066288854
31 0.671462396113927
32 0.685451196032967
33 0.696331373747776
34 0.705035515919623
35 0.712157086787498
36 0.718091729177394
37 0.723113349661152
38 0.727417595790087
39 0.731147942435165
41 0.734411995749607
42 0.737292042791763
43 0.739852084607012
44 0.742142648336445
45 0.744204155692935
46 0.746069329015474
48 0.747764941126873
49 0.749313108706846
50 0.750732262321821
52 0.752037883647598
53 0.7532430725637
54 0.754358988226757
55 0.755395195628167
56 0.75635994045017
57 0.757260368950706
58 0.758102705289917
59 0.758892395607927
60 0.759634225906664
61 0.760332419129005
62 0.760990715595783
63 0.76161244003663
64 0.762200557750944
65 0.762757721901346
66 0.763286313531216
67 0.763788475579592
1100 0.763788475579592
};
\addlegendentry{TDMA, $\alpha=0.1$}
\addplot [semithick, black, const plot mark left, mark=+, mark options={solid}, mark size=2pt]coordinates{(24,0.7833)};
\addplot [semithick, black, const plot mark left, mark=+, mark options={solid}, mark size=2pt]coordinates{(49,0.7833)};
\addplot [semithick, black, const plot mark left, mark=+, mark options={solid}, mark size=2pt]coordinates{(241,0.7833)};

\draw[->, yshift=2pt] (150,0.93)--(24,0.93)node[near end, fill=white,font=\footnotesize, inner sep=1pt]{$\alpha=0.1$}--(24,0.7833);
\draw[->, yshift=2pt](140,0.93)-- (140,0.85)--(49,0.85)node[midway, fill=white,font=\footnotesize, inner sep=1pt]{$\alpha=0.05$}--(49,0.7833);
\draw[->, yshift=2pt](160,0.93)-- (160,0.85)--(241,0.85)node[very near end, anchor=east, fill=white,font=\footnotesize, inner sep=1pt]{$\alpha=0.01$}--(241,0.7833);
\node[draw=black, fill=white, inner sep=2pt, font=\footnotesize, anchor=north] at (150,1){FDMA};
\end{axis}

\end{tikzpicture}
        \caption{$r=150$ m, $\eta=3$.}
        \label{fig:ad150p3}
    \end{subfigure}	
    	\begin{subfigure}[b]{.49\linewidth}
	    \centering
\begin{tikzpicture}

\begin{axis}[
legend cell align={left},
width=\fwidth,
height=\fheight,
legend style={
  nodes={scale=0.75, transform shape},
  fill opacity=0.75,
  draw opacity=1,
  text opacity=1,
  legend columns=2,
  at={(0.983,0.03)},
  anchor=south east,
  draw=white!80!black,
  column sep=0.2cm
},
tick align=outside,
tick pos=left,
x grid style={white!69.0196078431373!black},
xlabel={$\Delta_{90}$ (slots)},
xmin=0, xmax=300,
ymajorgrids,
xmajorgrids,
xminorgrids,
xtick style={color=black},
extra x ticks={200},
extra x tick labels={},
y grid style={white!69.0196078431373!black},
ylabel style={align=center},
ylabel={$S$ (packets/slot)},
ymin=0, ymax=1,
ytick style={color=black}
]
\addplot [semithick, color2, const plot mark left, mark=o, mark options={solid}]
table {%
236 -1
236 0.374990007048749
237 0.582976928703455
238 0.646478519446309
239 0.70220074571922
240 0.723656835124541
241 0.745747356576288
242 0.759860385202844
243 0.770879619239518
244 0.773957232446898
245 0.77697105684114
246 0.780160844719925
1100 0.780160844719925
};
\addlegendentry{NOMA, $\alpha=0.01$}
\addplot [semithick, color3, const plot mark left, mark=*, mark options={solid}, mark size=1.5pt]
table {%
236 -1
236 0
237 0.391686397733124
238 0.522248530310832
239 0.587529596599686
240 0.626698236372998
241 0.65281066288854
242 0.671462396113927
243 0.685451196032967
244 0.696331373747776
245 0.705035515919623
246 0.712157086787498
247 0.718091729177394
248 0.723113349661152
249 0.727417595790087
250 0.731147942435165
251 0.734411995749607
252 0.737292042791763
253 0.739852084607012
254 0.742142648336445
255 0.744204155692935
257 0.746069329015474
258 0.747764941126873
259 0.749313108706846
260 0.750732262321821
261 0.752037883647598
262 0.7532430725637
263 0.754358988226757
264 0.755395195628167
265 0.75635994045017
266 0.757260368950706
267 0.758102705289917
268 0.758892395607927
269 0.759634225906664
270 0.760332419129005
271 0.760990715595783
272 0.76161244003663
273 0.762200557750944
274 0.762757721901346
275 0.763286313531216
276 0.763788475579592
1100 0.763788475579592
};
\addlegendentry{TDMA, $\alpha=0.01$}
\addplot [semithick, color4, const plot mark left, mark=square, mark options={solid}]
table {%
47 -1
47 0.0666666666666663
48 0.569107794666075
49 0.662079429141682
50 0.722689672821832
51 0.744883078363555
52 0.758222562711433
1100 0.758222562711433
};
\addlegendentry{NOMA, $\alpha=0.05$}
\addplot [semithick, color5, const plot mark left, mark=square*, mark options={solid}, mark size=1.5pt]
table {%
47 -1
47 0
48 0.391686397733124
50 0.522248530310832
51 0.587529596599686
52 0.626698236372998
53 0.65281066288854
54 0.671462396113927
55 0.685451196032967
56 0.696331373747776
57 0.705035515919623
58 0.712157086787498
59 0.718091729177394
60 0.723113349661152
61 0.727417595790087
62 0.731147942435165
63 0.734411995749607
64 0.737292042791763
65 0.739852084607012
66 0.742142648336445
67 0.744204155692935
68 0.746069329015474
69 0.747764941126873
70 0.749313108706846
71 0.750732262321821
72 0.752037883647598
73 0.7532430725637
74 0.754358988226757
75 0.755395195628167
77 0.75635994045017
78 0.757260368950706
79 0.758102705289917
80 0.758892395607927
81 0.759634225906664
82 0.760332419129005
83 0.760990715595783
84 0.76161244003663
85 0.762200557750944
86 0.762757721901346
87 0.763286313531216
88 0.763788475579592
1100 0.763788475579592
};
\addlegendentry{TDMA, $\alpha=0.05$}
\addplot [semithick, color6, const plot mark left, mark=diamond, mark options={solid}]
table {%
24 -1
24 0.496657683240354
25 0.655301041392013
26 0.712794096756109
27 0.734515178141104
28 0.738372928191143
1100 0.738372928191143
};
\addlegendentry{NOMA, $\alpha=0.1$}
\addplot [semithick, color7, const plot mark left, mark=diamond*, mark options={solid}, mark size=1.5pt]
table {%
24 -1
24 0
25 0.391686397733124
26 0.522248530310832
27 0.587529596599686
28 0.626698236372998
29 0.65281066288854
30 0.671462396113927
31 0.685451196032967
32 0.696331373747776
33 0.705035515919623
34 0.712157086787498
35 0.718091729177394
36 0.723113349661152
37 0.727417595790087
38 0.731147942435165
39 0.734411995749607
41 0.737292042791763
42 0.739852084607012
43 0.742142648336445
44 0.744204155692935
45 0.746069329015474
46 0.747764941126873
47 0.749313108706846
49 0.750732262321821
50 0.752037883647598
51 0.7532430725637
52 0.754358988226757
53 0.755395195628167
54 0.75635994045017
55 0.757260368950706
56 0.758102705289917
57 0.758892395607927
58 0.759634225906664
59 0.760332419129005
60 0.760990715595783
61 0.76161244003663
62 0.762200557750944
63 0.762757721901346
64 0.763286313531216
65 0.763788475579592
1100 0.763788475579592
};
\addlegendentry{TDMA, $\alpha=0.1$}
\addplot [semithick, black, const plot mark left, mark=+, mark options={solid}, mark size=2pt]coordinates{(23,0.7833)};
\addplot [semithick, black, const plot mark left, mark=+, mark options={solid}, mark size=2pt]coordinates{(47,0.7833)};
\addplot [semithick, black, const plot mark left, mark=+, mark options={solid}, mark size=2pt]coordinates{(236,0.7833)};

\draw[->, yshift=2pt] (150,0.93)--(23,0.93)node[near end, fill=white,font=\footnotesize, inner sep=1pt]{$\alpha=0.1$}--(23,0.7833);
\draw[->, yshift=2pt](140,0.93)-- (140,0.85)--(47,0.85)node[midway, fill=white,font=\footnotesize, inner sep=1pt]{$\alpha=0.05$}--(47,0.7833);
\draw[->, yshift=2pt](160,0.93)-- (160,0.85)--(236,0.85)node[very near end, anchor=east, fill=white,font=\footnotesize, inner sep=1pt]{$\alpha=0.01$}--(236,0.7833);
\node[draw=black, fill=white, inner sep=2pt, font=\footnotesize, anchor=north] at (150,1){FDMA};
\end{axis}

\end{tikzpicture}
        \caption{$r=250$ m, $\eta=2.6$.}
        \label{fig:ad250p26}
    \end{subfigure}	    	
    \begin{subfigure}[b]{.49\linewidth}
	    \centering
\begin{tikzpicture}

\begin{axis}[
legend cell align={left},
width=\fwidth,
height=\fheight,
legend style={
  nodes={scale=0.75, transform shape},
  fill opacity=0.75,
  draw opacity=1,
  text opacity=1,
  legend columns=2,
  at={(0.983,0.03)},
  anchor=south east,
  draw=white!80!black,
  column sep=0.2cm
},
tick align=outside,
tick pos=left,
x grid style={white!69.0196078431373!black},
xlabel={$\Delta_{90}$ (slots)},
xmin=0, xmax=300,
ymajorgrids,
xmajorgrids,
xminorgrids,
xtick style={color=black},
extra x ticks={200},
extra x tick labels={},
y grid style={white!69.0196078431373!black},
ylabel style={align=center},
ylabel={$S$ (packets/slot)},
ymin=0, ymax=1,
ytick style={color=black}
]
\addplot [semithick, color2, const plot mark left, mark=o, mark options={solid}]
table {%
284 -1
284 0.285712039593399
285 0.44440960818842
286 0.554973924291573
287 0.599768469300783
288 0.641719474907495
289 0.665705472870941
290 0.684796208910295
291 0.701964449941417
292 0.717968729878743
293 0.72805865646525
294 0.737186266826094
295 0.746873560292265
296 0.752784903230309
297 0.759341403933312
298 0.765110724176068
299 0.77011475551736
301 0.772739874571506
303 0.775534209529247
305 0.777354653500214
307 0.77821914434722
1100 0.77821914434722
};
\addlegendentry{NOMA, $\alpha=0.01$}
\addplot [semithick, color3, const plot mark left, mark=*, mark options={solid}, mark size=1.5pt]
table {%
284 -1
284 0
285 0.391686397733124
287 0.522248530310832
288 0.587529596599686
289 0.626698236372998
290 0.65281066288854
292 0.671462396113927
293 0.685451196032967
294 0.696331373747776
295 0.705035515919623
297 0.712157086787498
298 0.718091729177394
299 0.723113349661152
301 0.727417595790087
302 0.731147942435165
303 0.734411995749607
304 0.737292042791763
306 0.739852084607012
307 0.742142648336445
308 0.744204155692935
310 0.746069329015474
311 0.747764941126873
312 0.749313108706846
314 0.750732262321821
315 0.752037883647598
316 0.7532430725637
317 0.754358988226757
319 0.755395195628167
320 0.75635994045017
321 0.757260368950706
323 0.758102705289917
324 0.758892395607927
325 0.759634225906664
326 0.760332419129005
328 0.760990715595783
329 0.76161244003663
331 0.762200557750944
332 0.762757721901346
333 0.763286313531216
334 0.763788475579592
1100 0.763788475579592
};
\addlegendentry{TDMA, $\alpha=0.01$}
\addplot [semithick, color4, const plot mark left, mark=square, mark options={solid}]
table {%
57 -1
57 0.249999437841791
58 0.498817521433611
59 0.568569666289782
60 0.632767244038619
61 0.663222887107467
62 0.701355399984985
63 0.71531472806777
64 0.732088426568191
65 0.738904989767772
66 0.748312524675918
67 0.752044080480033
1100 0.752044080480033
};
\addlegendentry{NOMA, $\alpha=0.05$}
\addplot [semithick, color5, const plot mark left, mark=square*, mark options={solid}, mark size=1.5pt]
table {%
57 -1
57 0
58 0.391686397733124
60 0.522248530310832
61 0.587529596599686
62 0.626698236372998
64 0.65281066288854
65 0.671462396113927
66 0.685451196032967
67 0.696331373747776
69 0.705035515919623
70 0.712157086787498
72 0.718091729177394
73 0.723113349661152
74 0.727417595790087
76 0.731147942435165
77 0.734411995749607
78 0.737292042791763
80 0.739852084607012
81 0.742142648336445
83 0.744204155692935
85 0.746069329015474
86 0.747764941126873
87 0.749313108706846
88 0.750732262321821
89 0.752037883647598
91 0.7532430725637
92 0.754358988226757
94 0.755395195628167
96 0.75635994045017
97 0.757260368950706
99 0.758102705289917
101 0.758892395607927
103 0.759634225906664
105 0.760332419129005
107 0.760990715595783
109 0.76161244003663
110 0.762200557750944
111 0.762757721901346
112 0.763286313531216
113 0.763788475579592
1100 0.763788475579592
};
\addlegendentry{TDMA, $\alpha=0.05$}
\addplot [semithick, color6, const plot mark left, mark=diamond, mark options={solid}]
table {%
29 -1
29 0.39939979974267
30 0.499603383088237
31 0.615657880298104
32 0.656152616264933
33 0.691706932770932
34 0.702905070621189
35 0.711730133297427
36 0.72009979689574
37 0.720869389205455
1100 0.720869389205455
};
\addlegendentry{NOMA, $\alpha=0.1$}
\addplot [semithick, color7, const plot mark left, mark=diamond*, mark options={solid}, mark size=1.5pt]
table {%
29 -1
29 0
30 0.391686397733124
31 0.522248530310832
33 0.587529596599686
34 0.626698236372998
35 0.65281066288854
37 0.671462396113927
38 0.685451196032967
39 0.696331373747776
41 0.705035515919623
42 0.712157086787498
44 0.718091729177394
45 0.723113349661152
47 0.727417595790087
48 0.731147942435165
50 0.734411995749607
52 0.737292042791763
54 0.739852084607012
55 0.742142648336445
56 0.744204155692935
57 0.746069329015474
59 0.747764941126873
60 0.749313108706846
61 0.750732262321821
63 0.752037883647598
65 0.7532430725637
66 0.754358988226757
68 0.755395195628167
70 0.75635994045017
71 0.757260368950706
73 0.758102705289917
75 0.758892395607927
77 0.759634225906664
78 0.760332419129005
80 0.760990715595783
82 0.76161244003663
84 0.762200557750944
86 0.762757721901346
88 0.763286313531216
90 0.763788475579592
1100 0.763788475579592
};
\addlegendentry{TDMA, $\alpha=0.1$}
\addplot [semithick, black, const plot mark left, mark=+, mark options={solid}, mark size=2pt]coordinates{(29,0.7833)};
\addplot [semithick, black, const plot mark left, mark=+, mark options={solid}, mark size=2pt]coordinates{(57,0.7833)};
\addplot [semithick, black, const plot mark left, mark=+, mark options={solid}, mark size=2pt]coordinates{(284,0.7833)};

\draw[->, yshift=2pt] (150,0.93)--(29,0.93)node[near end, fill=white,font=\footnotesize, inner sep=1pt]{$\alpha=0.1$}--(29,0.7833);
\draw[->, yshift=2pt](140,0.93)-- (140,0.85)--(57,0.85)node[midway, fill=white,font=\footnotesize, inner sep=1pt]{$\alpha=0.05$}--(57,0.7833);
\draw[->, yshift=2pt](160,0.93)-- (160,0.85)--(284,0.85)node[very near end, anchor=east, fill=white,font=\footnotesize, inner sep=1pt]{$\alpha=0.01$}--(284,0.7833);
\node[draw=black, fill=white, inner sep=2pt, font=\footnotesize, anchor=north] at (150,1){FDMA};
\end{axis}

\end{tikzpicture}
        \caption{$r=250$ m, $\eta=3$.}
        \label{fig:ad250p3}
    \end{subfigure}	
    \caption{Pareto frontiers for \gls{paoi} versus throughput with \gls{tdma} and \gls{noma}, with different values of $\alpha$. The cross marks indicate the performance with \gls{fdma} (benchmark).}
 \label{fig:age}
\end{figure*}

On the other hand, there is a strict trade-off between \gls{lr} and throughput for all cases with \gls{tdma}, as the only way to reduce the latency is to decrease the period between intermittent slot $T_\text{int}$, which decreases the resources assigned to the broadband user. The same trade-off appears with \gls{noma} for the cases where $\pi_{\mathcal{I},\mathcal{I}}+\pi_{\mathcal{E},\mathcal{I}}<0.9$ due to an increase in path loss, shown in Fig.~\ref{fig:ld150p3}-\subref{fig:ld250p26}. In these, reducing the latency also requires reducing the efficiency of the code. However, the Pareto frontier for \gls{noma} is always above and to the left of the curve for the equivalent scenario with \gls{tdma}, showing that \gls{noma} is clearly the best choice in this scenario. The frontiers for $r=250$~m and path loss exponent $\eta=3$ are not shown, as in this case it is impossible for the intermittent user to deliver 90\% of packets at any finite latency (i.e., $T_{90}=\infty$).

\gls{noma} also achieves a lower \gls{paoi} with high throughput for all cases in Fig.~\ref{fig:ad50p26}-\subref{fig:ad150p26}. In these, the Pareto frontier increases abruptly and reaches its maximum $S\approx0.78$, which is close to the one achieved with \gls{fdma} of $0.7833$. This occurs at exactly or only a few time slots later than the minimum $\Delta_{90}$.
Thus, the resource efficiency of \gls{noma} is much greater than that of \gls{fdma} and achieves similar trade-offs. 

\begin{figure*}
    \centering
    \begin{subfigure}[b]{.49\linewidth}
	    \centering
\begin{tikzpicture}

\begin{axis}[
legend cell align={left},
width=\fwidth,
height=\fheight,
legend style={
  nodes={scale=0.75, transform shape},
  fill opacity=0.75,
  draw opacity=1,
  text opacity=1,
  legend columns=2,
  at={(0.017,0.97)},
  anchor=north west,
  draw=white!80!black,
  column sep=0.2cm
},
xmajorgrids,
ymajorgrids,
tick align=outside,
tick pos=left,
x grid style={white!80!black},
xlabel={$r$ (m)},
xmin=50, xmax=400,
xtick style={color=black},
y grid style={white!80!black},
ylabel style={align=center},
ylabel={$T_{90}$ (slots)},
ymin=0, ymax=100,
ytick style={color=black}
]
\addplot [semithick, color2, mark=*, mark options={solid}]
table {%
50 1
100 1
150 1
200 1
250 2
300 5
350 7
400 9
};
\addlegendentry{NOMA}
\addplot [semithick, color6, mark=square, mark options={solid}]
table {%
50 12
100 12
150 12
200 12
250 13
300 14
350 15
400 17
};
\addlegendentry{NOMA (no capture)}
\addplot [semithick, color5, mark=diamond*, mark options={solid}]
table {%
50 15
100 15
150 15
200 15
250 15
300 16
350 18
400 22
};
\addlegendentry{TDMA}
\addplot [semithick, black, mark=+,  mark options={solid}]
table {%
50 1 
100 1 
150 1 
200 1 
250 1 
300 1 
350 1 
400 1
};
\addlegendentry{FDMA}
\end{axis}

\end{tikzpicture}
        \caption{$\alpha=0.01$, $\eta=2.6$.}
        \label{fig:la1p26}
    \end{subfigure}	
    \begin{subfigure}[b]{.49\linewidth}
	    \centering
\begin{tikzpicture}

\begin{axis}[
legend cell align={left},
width=\fwidth,
height=\fheight,
legend style={
  nodes={scale=0.75, transform shape},
  fill opacity=0.75,
  draw opacity=1,
  text opacity=1,
  legend columns=2,
  at={(0.017,0.97)},
  anchor=north west,
  draw=white!80!black,
  column sep=0.2cm
},
xmajorgrids,
ymajorgrids,
tick align=outside,
tick pos=left,
x grid style={white!80!black},
xlabel={$r$ (m)},
xmin=50, xmax=400,
xtick style={color=black},
y grid style={white!80!black},
ylabel style={align=center},
ylabel={$T_{90}$ (slots)},
ymin=0, ymax=100,
ytick style={color=black}
]
\addplot [semithick, color2, mark=*, mark options={solid}]
table {%
50 1
100 1
150 5
200 121
250 121
300 121
350 121
400 121
};
\addlegendentry{NOMA}
\addplot [semithick, color6, mark=square, mark options={solid}]
table {%
50 12
100 12
150 14
200 121
250 121
300 121
350 121
400 121
};
\addlegendentry{NOMA (no capture)}
\addplot [semithick, color5, mark=diamond*, mark options={solid}]
table {%
50 15
100 15
150 17
200 121
250 121
300 121
350 121
400 121
};
\addlegendentry{TDMA}
\addplot [semithick, black, mark=+,  mark options={solid}]
table {%
50 1 
100 1 
150 1 
200 121 
};
\addlegendentry{FDMA}
\end{axis}

\end{tikzpicture}
        \caption{$\alpha=0.01$, $\eta=3$.}
        \label{fig:la1p3}
    \end{subfigure}	
	\centering
    \begin{subfigure}[b]{.49\linewidth}
	    \centering
\begin{tikzpicture}

\begin{axis}[
legend cell align={left},
width=\fwidth,
height=\fheight,
legend style={
  nodes={scale=0.75, transform shape},
  fill opacity=0.75,
  draw opacity=1,
  text opacity=1,
  legend columns=2,
  at={(0.017,0.97)},
  anchor=north west,
  draw=white!80!black,
  column sep=0.2cm
},
xmajorgrids,
ymajorgrids,
tick align=outside,
tick pos=left,
x grid style={white!80!black},
xlabel={$r$ (m)},
xmin=50, xmax=400,
xtick style={color=black},
y grid style={white!80!black},
ylabel style={align=center},
ylabel={$T_{90}$ (slots)},
ymin=0, ymax=100,
ytick style={color=black}
]
\addplot [semithick, color2, mark=*, mark options={solid}]
table {%
50 1
100 1
150 1
200 1
250 2
300 5
350 7
400 9
};
\addlegendentry{NOMA}
\addplot [semithick, color6, mark=square, mark options={solid}]
table {%
50 20
100 21
150 21
200 21
250 22
300 23
350 26
400 30
};
\addlegendentry{NOMA (no capture)}
\addplot [semithick, color5, mark=diamond*, mark options={solid}]
table {%
50 41
100 41
150 42
200 43
250 44
300 46
350 49
400 121
};
\addlegendentry{TDMA}
\addplot [semithick, black, mark=+,  mark options={solid}]
table {%
50 1 
100 1 
150 1 
200 1 
250 1 
300 1 
350 1 
400 1
};
\addlegendentry{FDMA}
\end{axis}

\end{tikzpicture}
        \caption{$\alpha=0.05$, $\eta=2.6$.}
        \label{fig:la5p26}
    \end{subfigure}	
    \begin{subfigure}[b]{.49\linewidth}
	    \centering
\begin{tikzpicture}

\begin{axis}[
legend cell align={left},
width=\fwidth,
height=\fheight,
legend style={
  nodes={scale=0.75, transform shape},
  fill opacity=0.75,
  draw opacity=1,
  text opacity=1,
  legend columns=2,
  at={(0.017,0.97)},
  anchor=north west,
  draw=white!80!black,
  column sep=0.2cm
},
xmajorgrids,
ymajorgrids,
tick align=outside,
tick pos=left,
x grid style={white!80!black},
xlabel={$r$ (m)},
xmin=50, xmax=400,
xtick style={color=black},
y grid style={white!80!black},
ylabel style={align=center},
ylabel={$T_{90}$ (slots)},
ymin=0, ymax=100,
ytick style={color=black}
]
\addplot [semithick, color2, mark=*, mark options={solid}]
table {%
50 1
100 1
150 5
200 121
250 121
300 121
350 121
400 121
};
\addlegendentry{NOMA}
\addplot [semithick, color6, mark=square, mark options={solid}]
table {%
50 21
100 21
150 24
200 121
250 121
300 121
350 121
400 121
};
\addlegendentry{NOMA (no capture)}
\addplot [semithick, color5, mark=diamond*, mark options={solid}]
table {%
50 41
100 43
150 47
200 121
250 121
300 121
350 121
400 121
};
\addlegendentry{TDMA}
\addplot [semithick, black, mark=+,  mark options={solid}]
table {%
50 1 
100 1 
150 1 
200 121
};
\addlegendentry{FDMA}
\end{axis}

\end{tikzpicture}
        \caption{$\alpha=0.05$, $\eta=3$.}
        \label{fig:la5p3}
    \end{subfigure}	
    \caption{Minimum \gls{lr} with $S\geq 0.7$ as a function of the distance between user 2 and the \gls{bs} for different values of $\alpha$. }
 \label{fig:latdist}
\end{figure*}
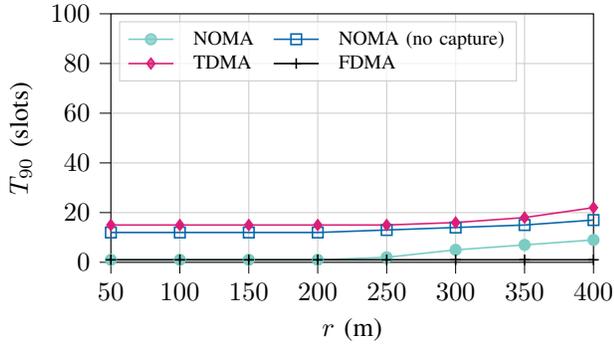
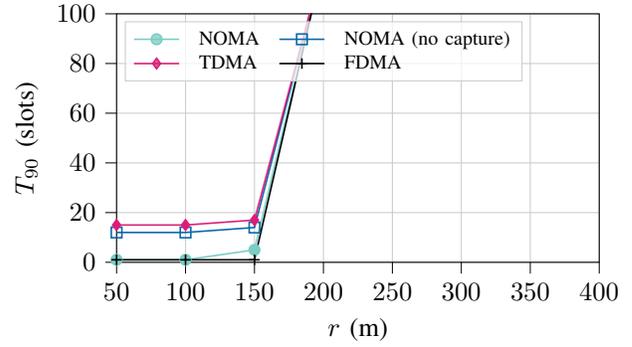
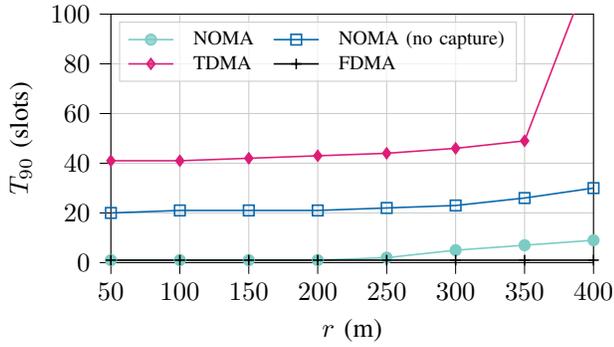
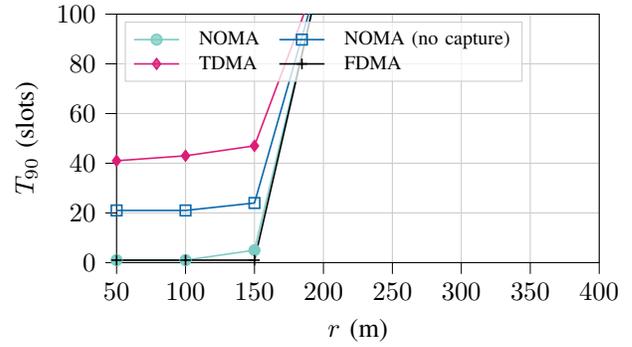

On the other hand, for $r\geq150$~m, \gls{tdma} achieves a higher throughput than \gls{noma} at the expense of an increase in \gls{paoi}. This is expected, as greater values of $T_\text{int}$ increase $S$ but also $\Delta$. Specifically, as described in Section~\ref{sec:fdma}, the throughput with \gls{tdma} for $T_\text{int}\to\infty$ is equal to that with \gls{fdma}. 

Note, however, that the activation rate $\alpha$ has the greatest impact on the \gls{paoi}.
This is because the interval time between consecutive packets with low values of $\alpha$ can be so long that reducing the latency for each individual packet has only a minor effect on the \gls{paoi}.

\subsection{Distance analysis}

We now investigate the performance of the schemes as a function of the distance $r$ between user 2 and the base station. In this case, we also consider the case for \gls{noma} with fully destructive interference and, hence, no capture, which was investigated in our previous work~\cite{chiariotti2021non}.
In this later case, we have $\pi_{\cdot,\mathcal{R}}=1-\varepsilon_2$ and $\pi_{\mathcal{E},\mathcal{E}}=\varepsilon_2$ for any slot in which the two users collide, eliminating the possibility of instantaneous \gls{sic}. This scenario is naturally a lower bound for \gls{noma}'s performance, as removing the possibility of capture makes the scheme perform significantly worse.

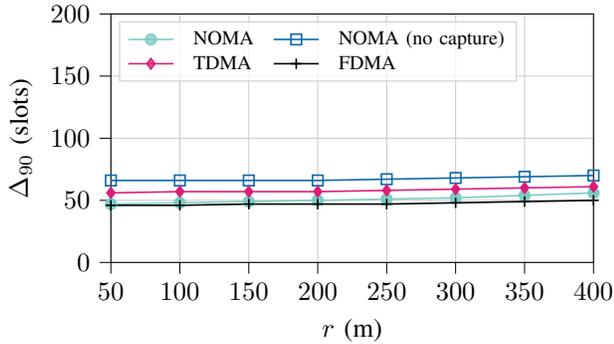
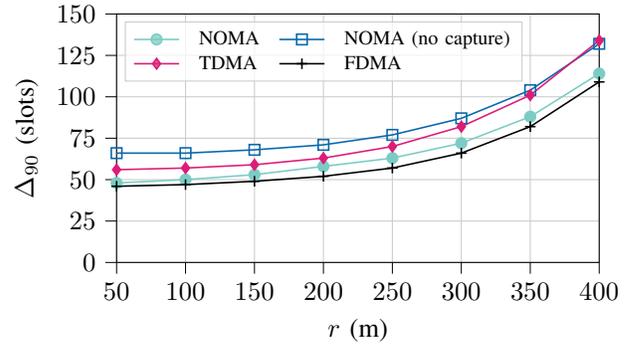
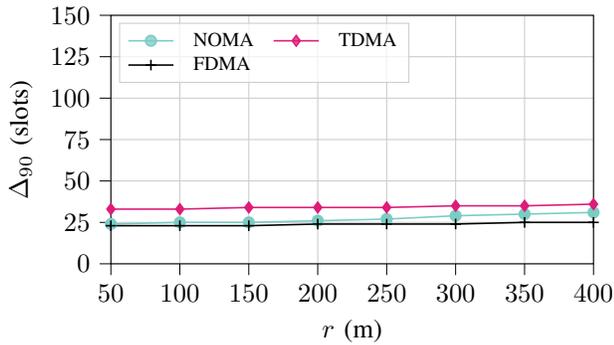
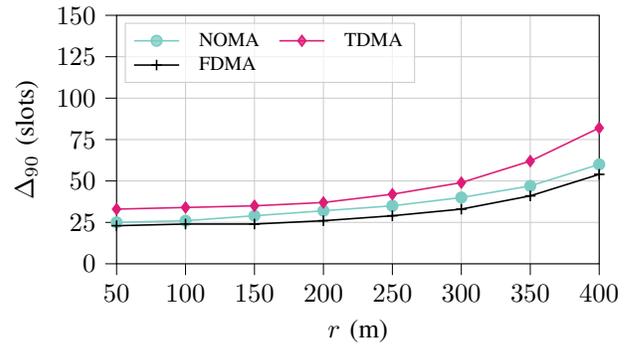
\begin{figure*}
    \centering
	\centering
    \begin{subfigure}[b]{.49\linewidth}
	    \centering
\begin{tikzpicture}

\begin{axis}[
legend cell align={left},
width=\fwidth,
height=\fheight,
legend style={
  nodes={scale=0.75, transform shape},
  fill opacity=0.75,
  draw opacity=1,
  text opacity=1,
  legend columns=2,
  at={(0.017,0.97)},
  anchor=north west,
  draw=white!80!black,
  column sep=0.2cm
},
xmajorgrids,
ymajorgrids,
tick align=outside,
tick pos=left,
x grid style={white!80!black},
xlabel={$r$ (m)},
xmin=50, xmax=400,
xtick style={color=black},
y grid style={white!80!black},
ylabel style={align=center},
ylabel={$\Delta_{90}$ (slots)},
ymin=0, ymax=200,
ytick style={color=black}
]
\addplot [semithick, color2, mark=*, mark options={solid}]
table {%
50 47
100 48
150 49
200 50
250 51
300 52
350 54
400 56
};
\addlegendentry{NOMA}
\addplot [semithick, color6, mark=square, mark options={solid}]
table {%
50 66
100 66
150 66
200 66
250 67
300 68
350 69
400 70
};
\addlegendentry{NOMA (no capture)}
\addplot [semithick, color5, mark=diamond*, mark options={solid}]
table {%
50 56
100 57
150 57
200 57
250 58
300 59
350 60
400 61
};
\addlegendentry{TDMA}
\addplot [semithick, black, mark=+,  mark options={solid}]
table {%
50 46 
100 46 
150 47 
200 47 
250 47 
300 48 
350 49 
400 50
};
\addlegendentry{FDMA}
\end{axis}

\end{tikzpicture}
        \caption{$\alpha=0.05$, $\eta=2.6$.}
        \label{fig:aa5p26}
    \end{subfigure}	
    \begin{subfigure}[b]{.49\linewidth}
	    \centering
\begin{tikzpicture}

\begin{axis}[
legend cell align={left},
width=\fwidth,
height=\fheight,
legend style={
  nodes={scale=0.75, transform shape},
  fill opacity=0.75,
  draw opacity=1,
  text opacity=1,
  legend columns=2,
  at={(0.017,0.97)},
  anchor=north west,
  draw=white!80!black,
  column sep=0.2cm
},
xmajorgrids,
ymajorgrids,
tick align=outside,
tick pos=left,
x grid style={white!80!black},
xlabel={$r$ (m)},
xmin=50, xmax=400,
xtick style={color=black},
y grid style={white!80!black},
ylabel style={align=center},
ylabel={$\Delta_{90}$ (slots)},
ymin=0, ymax=150, ytick distance=25,
ytick style={color=black}
]
\addplot [semithick, color2, mark=*, mark options={solid}]
table {%
50 48
100 50
150 53
200 58
250 63
300 72
350 88
400 114
};
\addlegendentry{NOMA}
\addplot [semithick, color6, mark=square, mark options={solid}]
table {%
50 66
100 66
150 68
200 71
250 77
300 87
350 104
400 132
};
\addlegendentry{NOMA (no capture)}
\addplot [semithick, color5, mark=diamond*, mark options={solid}]
table {%
50 56
100 57
150 59
200 63
250 70
300 82
350 101
400 134
};
\addlegendentry{TDMA}

\addplot [semithick, black, mark=+,  mark options={solid}]
table {%
50 46 
100 47 
150 49 
200 52 
250 57 
300 66 
350 82 
400 109
};
\addlegendentry{FDMA}
\end{axis}

\end{tikzpicture}
        \caption{$\alpha=0.05$, $\eta=3$.}
        \label{fig:aa5p3}
    \end{subfigure}	
    \begin{subfigure}[b]{.49\linewidth}
 	    \centering
\begin{tikzpicture}

\begin{axis}[
legend cell align={left},
width=\fwidth,
height=\fheight,
legend style={
  nodes={scale=0.75, transform shape},
  fill opacity=0.75,
  draw opacity=1,
  text opacity=1,
  legend columns=2,
  at={(0.017,0.97)},
  anchor=north west,
  draw=white!80!black,
  column sep=0.2cm
},
xmajorgrids,
ymajorgrids,
tick align=outside,
tick pos=left,
x grid style={white!80!black},
xlabel={$r$ (m)},
xmin=50, xmax=400,
xtick style={color=black},
y grid style={white!80!black},
ylabel style={align=center},
ylabel={$\Delta_{90}$ (slots)},
ymin=0, ymax=150, ytick distance=25,
ytick style={color=black}
]
\addplot [semithick, color2, mark=*, mark options={solid}]
table {%
50 24
100 25
150 25
200 26
250 27
300 29
350 30
400 31
};
\addlegendentry{NOMA}
\addplot [semithick, color5, mark=diamond*, mark options={solid}]
table {%
50 33
100 33
150 34
200 34
250 34
300 35
350 35
400 36
};
\addlegendentry{TDMA}
\addplot [semithick, black, mark=+,  mark options={solid}]
table {%
50 23 
100 23 
150 23 
200 24 
250 24 
300 24 
350 25 
400 25
};
\addlegendentry{FDMA}
\end{axis}

\end{tikzpicture}
         \caption{$\alpha=0.1$, $\eta=2.6$.}
         \label{fig:aa10p26}
     \end{subfigure}	
 	\centering
     \begin{subfigure}[b]{.49\linewidth}
 	    \centering
\begin{tikzpicture}

\begin{axis}[
legend cell align={left},
width=\fwidth,
height=\fheight,
legend style={
  nodes={scale=0.75, transform shape},
  fill opacity=0.75,
  draw opacity=1,
  text opacity=1,
  legend columns=2,
  at={(0.017,0.97)},
  anchor=north west,
  draw=white!80!black,
  column sep=0.2cm
},
xmajorgrids,
ymajorgrids,
tick align=outside,
tick pos=left,
x grid style={white!80!black},
xlabel={$r$ (m)},
xmin=50, xmax=400,
xtick style={color=black},
y grid style={white!80!black},
ylabel style={align=center},
ylabel={$\Delta_{90}$ (slots)},
ymin=0, ymax=150, ytick distance=25,
ytick style={color=black}
]
\addplot [semithick, color2, mark=*, mark options={solid}]
table {%
50 25
100 26
150 29
200 32
250 35
300 40
350 47
400 60
};
\addlegendentry{NOMA}
\addplot [semithick, color5, mark=diamond*, mark options={solid}]
table {%
50 33
100 34
150 35
200 37
250 42
300 49
350 62
400 82
};
\addlegendentry{TDMA}

\addplot [semithick, black, mark=+,  mark options={solid}]
table {%
50 23 
100 24 
150 24 
200 26 
250 29 
300 33 
350 41 
400 54
};
\addlegendentry{FDMA}
\end{axis}

\end{tikzpicture}
         \caption{$\alpha=0.1$, $\eta=3$.}
         \label{fig:aa10p3}
     \end{subfigure}	
    \caption{Minimum \gls{paoi} with $S\geq 0.7$ as a function of the distance between user 2 and the \gls{bs} for different values of $\alpha$.}
 \label{fig:agedist}
\end{figure*}

Fig.~\ref{fig:latdist} shows the performance of the schemes in terms of the minimum \gls{lr} $T_{90}$ that can be achieved while fulfilling a relatively high throughput requirement $S\geq0.7$ for $\alpha=0.01,0.05$. In general, \gls{noma} can outperform \gls{tdma} in most cases, but it is interesting to observe the behavior of the schemes when $\alpha$ is high. In these cases, we notice a performance drop for both \gls{tdma}, which has to allocate more slots to the intermittent user, and \gls{noma} without capture, which has to increase the robustness of the packet-level code to protect the transmission from the additional intermittent user packets. On the other hand, capture allows \gls{noma} to be more robust to the increased activation probability, maintaining a performance that is close to \gls{fdma}. In fact, while not shown in the figures, \gls{noma} and, naturally, \gls{fdma} are the only schemes that can achieve $S\geq0.7$ with $\alpha=0.1$, while the other schemes do not achieve the required throughput for any configuration.

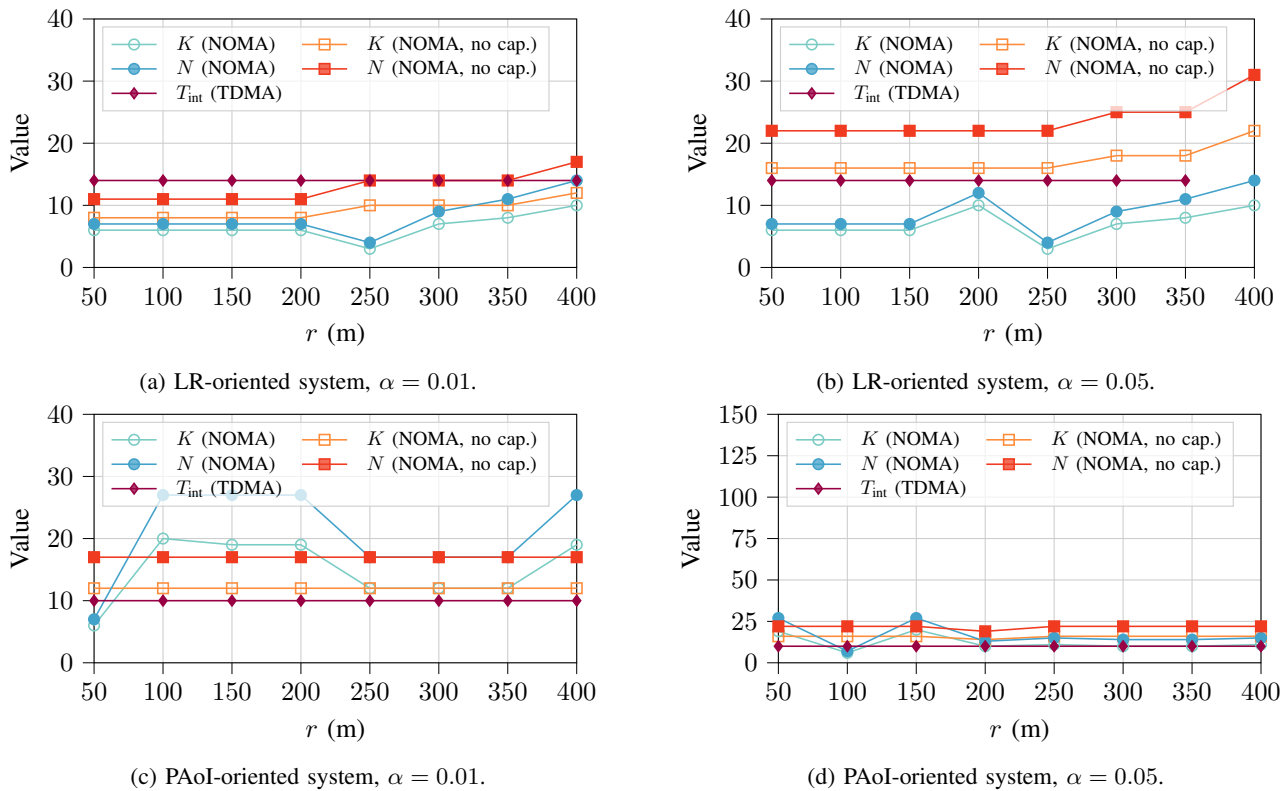
\begin{figure*}
	\centering
    \begin{subfigure}[b]{.49\linewidth}
	    \centering
\begin{tikzpicture}

\begin{axis}[
legend cell align={left},
width=\fwidth,
height=\fheight,
legend style={
  nodes={scale=0.75, transform shape},
  fill opacity=0.75,
  draw opacity=1,
  text opacity=1,
  legend columns=2,
  at={(0.017,0.97)},
  anchor=north west,
  draw=white!80!black,
  column sep=0.2cm
},
xmajorgrids,
ymajorgrids,
tick align=outside,
tick pos=left,
x grid style={white!80!black},
xlabel={$r$ (m)},
xmin=50, xmax=400,
xtick style={color=black},
y grid style={white!80!black},
ylabel style={align=center},
ylabel={Value},
ymin=0, ymax=40,
ytick style={color=black}
]
\addplot [semithick, color2, mark=o, mark options={solid}]
table {%
50 6
100 6
150 6
200 6
250 3
300 7
350 8
400 10
};
\addlegendentry{$K$ (NOMA)}
\addplot [semithick, color0, mark=square, mark options={solid}]
table {%
50 8
100 8
150 8
200 8
250 10
300 10
350 10
400 12
};
\addlegendentry{$K$ (NOMA, no cap.)}
\addplot [semithick, color4, mark=*, mark options={solid}]
table {%
50 7
100 7
150 7
200 7
250 4
300 9
350 11
400 14
};
\addlegendentry{$N$ (NOMA)}
\addplot [semithick, color1, mark=square*, mark options={solid}]
table {%
50 11
100 11
150 11
200 11
250 14
300 14
350 14
400 17
};
\addlegendentry{$N$ (NOMA, no cap.)}
\addplot [semithick, color7, mark=diamond*, mark options={solid}]
table {%
50 14
100 14
150 14
200 14
250 14
300 14
350 14
400 14
};
\addlegendentry{$T_{\text{int}}$ (TDMA)}
\end{axis}

\end{tikzpicture}
        \caption{\gls{lr}-oriented system, $\alpha=0.01$.}
        \label{fig:lp1}
    \end{subfigure}	
    \begin{subfigure}[b]{.49\linewidth}
	    \centering
\begin{tikzpicture}

\begin{axis}[
legend cell align={left},
width=\fwidth,
height=\fheight,
legend style={
  nodes={scale=0.75, transform shape},
  fill opacity=0.75,
  draw opacity=1,
  text opacity=1,
  legend columns=2,
  at={(0.017,0.97)},
  anchor=north west,
  draw=white!80!black,
  column sep=0.2cm
},
xmajorgrids,
ymajorgrids,
tick align=outside,
tick pos=left,
x grid style={white!80!black},
xlabel={$r$ (m)},
xmin=50, xmax=400,
xtick style={color=black},
y grid style={white!80!black},
ylabel style={align=center},
ylabel={Value},
ymin=0, ymax=40,
ytick style={color=black}
]
\addplot [semithick, color2, mark=o, mark options={solid}]
table {%
50 6
100 6
150 6
200 10
250 3
300 7
350 8
400 10
};
\addlegendentry{$K$ (NOMA)}
\addplot [semithick, color0, mark=square, mark options={solid}]
table {%
50 16
100 16
150 16
200 16
250 16
300 18
350 18
400 22
};
\addlegendentry{$K$ (NOMA, no cap.)}
\addplot [semithick, color4, mark=*, mark options={solid}]
table {%
50 7
100 7
150 7
200 12
250 4
300 9
350 11
400 14
};
\addlegendentry{$N$ (NOMA)}
\addplot [semithick, color1, mark=square*, mark options={solid}]
table {%
50 22
100 22
150 22
200 22
250 22
300 25
350 25
400 31
};
\addlegendentry{$N$ (NOMA, no cap.)}
\addplot [semithick, color7, mark=diamond*, mark options={solid}]
table {%
50 14
100 14
150 14
200 14
250 14
300 14
350 14
};
\addlegendentry{$T_{\text{int}}$ (TDMA)}
\end{axis}

\end{tikzpicture}
        \caption{\gls{lr}-oriented system, $\alpha=0.05$.}
        \label{fig:lp5}
    \end{subfigure}	
        \begin{subfigure}[b]{.49\linewidth}
	    \centering
\begin{tikzpicture}

\begin{axis}[
legend cell align={left},
width=\fwidth,
height=\fheight,
legend style={
  nodes={scale=0.75, transform shape},
  fill opacity=0.75,
  draw opacity=1,
  text opacity=1,
  legend columns=2,
  at={(0.017,0.97)},
  anchor=north west,
  draw=white!80!black,
  column sep=0.2cm
},
xmajorgrids,
ymajorgrids,
tick align=outside,
tick pos=left,
x grid style={white!80!black},
xlabel={$r$ (m)},
xmin=50, xmax=400,
xtick style={color=black},
y grid style={white!80!black},
ylabel style={align=center},
ylabel={Value},
ymin=0, ymax=40,
ytick style={color=black}
]
\addplot [semithick, color2, mark=o, mark options={solid}]
table {%
50 6
100 20
150 19
200 19
250 12
300 12
350 12
400 19
};
\addlegendentry{$K$ (NOMA)}
\addplot [semithick, color0, mark=square, mark options={solid}]
table {%
50 12
100 12
150 12
200 12
250 12
300 12
350 12
400 12
};
\addlegendentry{$K$ (NOMA, no cap.)}
\addplot [semithick, color4, mark=*, mark options={solid}]
table {%
50 7
100 27
150 27
200 27
250 17
300 17
350 17
400 27
};
\addlegendentry{$N$ (NOMA)}
\addplot [semithick, color1, mark=square*, mark options={solid}]
table {%
50 17
100 17
150 17
200 17
250 17
300 17
350 17
400 17
};
\addlegendentry{$N$ (NOMA, no cap.)}
\addplot [semithick, color7, mark=diamond*, mark options={solid}]
table {%
50 10
100 10
150 10
200 10
250 10
300 10
350 10
400 10
};
\addlegendentry{$T_{\text{int}}$ (TDMA)}
\end{axis}

\end{tikzpicture}
        \caption{\gls{paoi}-oriented system, $\alpha=0.01$.}
        \label{fig:ap1}
    \end{subfigure}	
	\centering
    \begin{subfigure}[b]{.49\linewidth}
	    \centering
\begin{tikzpicture}

\begin{axis}[
legend cell align={left},
width=\fwidth,
height=\fheight,
legend style={
  nodes={scale=0.75, transform shape},
  fill opacity=0.75,
  draw opacity=1,
  text opacity=1,
  legend columns=2,
  at={(0.017,0.97)},
  anchor=north west,
  draw=white!80!black,
  column sep=0.2cm
},
xmajorgrids,
ymajorgrids,
tick align=outside,
tick pos=left,
x grid style={white!80!black},
xlabel={$r$ (m)},
xmin=50, xmax=400,
xtick style={color=black},
y grid style={white!80!black},
ylabel style={align=center},
ylabel={Value},
ymin=0, ymax=150, ytick distance=25,
ytick style={color=black}
]
\addplot [semithick, color2, mark=o, mark options={solid}]
table {%
50 19
100 6
150 20
200 10
250 11
300 10
350 10
400 11
};
\addlegendentry{$K$ (NOMA)}
\addplot [semithick, color0, mark=square, mark options={solid}]
table {%
50 16
100 16
150 16
200 14
250 16
300 16
350 16
400 16
};
\addlegendentry{$K$ (NOMA, no cap.)}
\addplot [semithick, color4, mark=*, mark options={solid}]
table {%
50 27
100 7
150 27
200 13
250 15
300 14
350 14
400 15
};
\addlegendentry{$N$ (NOMA)}
\addplot [semithick, color1, mark=square*, mark options={solid}]
table {%
50 22
100 22
150 22
200 19
250 22
300 22
350 22
400 22
};
\addlegendentry{$N$ (NOMA, no cap.)}
\addplot [semithick, color7, mark=diamond*, mark options={solid}]
table {%
50 10
100 10
150 10
200 10
250 10
300 10
350 10
400 10
};
\addlegendentry{$T_{\text{int}}$ (TDMA)}
\end{axis}

\end{tikzpicture}
        \caption{\gls{paoi}-oriented system, $\alpha=0.05$.}
        \label{fig:ap5}
    \end{subfigure}	
    \caption{Optimal settings for the three schemes with $S\geq 0.7$ as a function of the distance between user 2, with $\eta=2.6$.}
 \label{fig:settings}
\end{figure*}

On the other hand, we can confirm the trend that we observed in Fig.~\ref{fig:age} for \gls{paoi} at different distances, as Fig.~\ref{fig:agedist} shows that \gls{noma} achieves a slightly lower $\Delta_{90}$ than \gls{tdma}. However, capture is essential for the \gls{noma} scheme with higher values of $\alpha$: without it, it performs slightly worse than \gls{tdma} for $\alpha=0.05$, and it never reaches the required throughput for $\alpha=0.1$. Finally, it can be seen that \gls{noma} achieves similar values of $\Delta_{90}$ than \gls{fdma} for (1) most values of $r$ with $\eta=2.6$ and (2) short distances $r\leq150$ with $\eta=3$. This demonstrates that, in the cases where the system can benefit from capture and \gls{sic}, \gls{noma} is nearly equivalent to \gls{fdma} in terms of performance, even when the latter utilizes twice the amount of resources. These cases occur, for example, when pairing the broadband user with an intermittent user located near the \gls{bs} and, hence, that achieves a high mean \gls{snr}.

\subsection{Parameter analysis}

We conclude by investigating the optimal configurations for the schemes as a function of the distance of user 2, under the constraint $S\geq 0.7$. Fig.~\ref{fig:settings} shows the optimal values of $K$ and $N$ for \gls{noma} and $T_{\text{int}}$ for \gls{tdma}, for both \gls{lr}-oriented and \gls{paoi}-oriented systems with $\eta=2.6$.
Fig.~\ref{fig:lp1}-\subref{fig:lp5}, which are related to \gls{lr}-oriented systems, show that the value of $T_{\text{int}}$ is always 14, independently from the distance.
On the other hand, \gls{lr}-oriented \gls{noma} systems tend to slightly increase both $K$ and $N$ as the distance increases. This occurs because the capture probability decreases as the distance from user 2 to the \gls{bs} increases. Increasing $N$ and $K$  then increases the robustness of the codes to errors in the transmission. This also implies that, when the capture probability is high, the \gls{noma} system can significantly reduce the frame size, which simplifies the decoding and reduces the latency, even for intermittent user packets that need to wait for \gls{sic}. 

On the other hand, if \gls{paoi} is the main objective, Fig.~\ref{fig:ap1}-\subref{fig:ap5} show a different picture: the value of $K$ and $N$ for \gls{noma} without capture is almost constant, as is the value of $T_{\text{int}}$ for \gls{tdma}, while the best possible values of $K$ and $N$ for \gls{noma} are higher at some distances and lower for others.
This phenomenon is likely due to the interplay between the different outcome probabilities and their effects on the \gls{paoi}.

\section{Conclusions}\label{sec:concl}
In this paper, we evaluated orthogonal and non-orthogonal slicing for heterogeneous services, namely, broadband and intermittent, in the \gls{ran}. Our model considered power control and packet-level coding for the broadband user and the use of \gls{sic} at the \gls{bs}. Our analyses and results highlighted the different characteristics and achievable performance of \gls{tdma} and \gls{noma} access schemes when compared to \gls{fdma}, which utilizes double the amount of resources -- two bandwidth parts instead of one for \gls{tdma} and \gls{noma}. In addition, we observed stark differences in terms of achievable trade-offs, impact of the inter-arrival times, and optimal configuration of the access schemes between the cases where the intermittent user aims to minimize either \gls{lr} and \gls{paoi}. Hence, our results highlight the importance of the considered performance indicator for the intermittent user and of its wireless conditions, which must be taken into account for an efficient user pairing in \gls{noma}.

In particular, our results showed that, with the considered schemes, \gls{noma} achieves a closely similar performance as \gls{fdma} if the intermittent user has a sufficiently high mean \gls{sinr} as a result of a relatively low path loss. Furthermore, \gls{noma} achieved better trade-offs between throughput and \gls{lr} than \gls{tdma} in every studied scenario. Since \gls{noma} utilized half of the resources of \gls{fdma} (the upper bound in performance), it achieved the best balance between resource efficiency and performance when the intermittent user aims to minimize \gls{lr}. Furthermore, even \gls{noma} without capture achieved a better performance than \gls{tdma} in terms of \gls{lr}.

Moreover, \gls{noma} achieved better throughput and \gls{paoi} trade-offs in most cases. In particular, \gls{tdma} only showed a superior performance when aiming for the highest throughput possible in exchange for a longer \gls{paoi}. However, the differences in \gls{paoi} were considerably smaller than those for the \gls{lr} cases, especially for short distances from user 2 to the \gls{bs}. Hence, \gls{tdma} may be preferred in the cases where the intermittent user is close to the \gls{bs} due to its simplicity. Finally, \gls{noma} without capture achieved the worst performance for \gls{paoi}.

Finally, is it important to note that, since the slicing is performed independently for each bandwidth part, our model and analyses can be easily extended to the case with multiple users and multiple bandwidth parts. This is the case with multiple broadband users, each with its own bandwidth part that can be shared with up to one of the intermittent users. Further, the \gls{fdma} scheme could be used to allocate multiple intermittent users in the same bandwidth part. However, the analysis of this scenario is out of the scope of this paper as \gls{fdma} is used as a benchmark for \gls{tdma} and \gls{noma} and it requires to define an access scheme for the intermittent users. 
\bibliographystyle{IEEEtran}
\bibliography{bib}
\end{document}